%% file: paper.tex
\documentstyle[12pt,iopfts,psfig]{ioplppt}

\begin{document}

\jl{1}

\title{Penumbra diffraction in the quantization of concave billiards}

\author{
Harel Primack\dag, 
Holger Schanz\ddag\ftnote{3}{Present address: Max-Planck-Institut f\"ur Physik komplexer Systeme, 01\,187 Dresden, Germany.}, 
Uzy Smilansky\dag~and \\
Iddo Ussishkin\dag\ftnote{4}{Present address: Department of Condensed Matter Physics, The Weizmann Institute of Science, Rehovot 76\,100, Israel}
}

\address{\dag Department of Physics of Complex Systems, 
The Weizmann Institute of Science, Rehovot 76\,100, Israel}
\address{\ddag Institut f\"ur Physik, Humboldt-Universit\"at, 10\,099 Berlin, Germany}

\begin{abstract}
The semiclassical description of billiard spectra is extended to
include the diffractive contributions from orbits which are nearly
tangent to a concave part of the boundary. The leading correction for
an unstable isolated orbit is of the same order as the standard
Gutzwiller expression itself. The importance of the diffraction
corrections is further emphasized by an estimate which shows that for
any large fixed $k$ almost all contributing periodic orbits are
affected. The theory is tested numerically using the annulus and the
Sinai billiard. For the Sinai billiard the investigation of the
spectral density is complemented by an analysis which is based on the
scattering approach to quantization. The merits of this approach as a
tool to investigate refined semiclassical theories are discussed and
demonstrated.
\end{abstract}



\section {Introduction} \label {sec:Introduction}
Beside the generic orbits which chaotic billiards support, there may
exist important families of non-generic orbits, which affect the
dynamics, and play a prominent r\^ole when the billiards are quantized
semiclassically. The bouncing ball orbits which reflect between
straight sections of the boundaries are an important example which was
studied extensively in the past \cite {Ber81a,SSCL93} and will not be
dealt with here. Another type exists in {\em convex} billiards (or
along convex sections), and it comprises of whispering gallery orbits:
these are classical trajectories which provide a hierarchy of
polygonal approximants to the boundary.  A detailed study of the
r\^ole of such orbits in the quantization of convex and smooth
boundaries is given in \cite{Laz93}. The particular case of the
stadium billiard, to which Lazutkin's theory does not apply, was
studied in \cite{SSCL93,Tan96}. The whispering gallery orbits occupy a
narrow strip in phase space, limited on one side by the boundary of
the phase space domain. In the case of a smooth, convex billiard,
every point on the phase space boundary is a fixed point of the bounce
map. We can therefore consider the boundary as a one-parameter family
of fixed points, which is the limit of the family of whispering
gallery orbits.

\input{FIG/fig-sinai}
In the present paper, we shall focus our attention on {\em concave}
billiards, where the whispering gallery modes do not appear. Instead,
concave billiards are characterized by the existence of {\em tangent}
orbits along the concave sections of the boundary.  They also form a
one parameter family, which belongs to the boundary of the phase space
domain. In this sense they are the counterparts of the whispering
gallery orbits in convex billiards.  In contrast to convex
billiards, however, tangency introduces discontinuities in the
classical map, and therefore the set (of zero measure) of tangent
orbits is excluded from the classical phase space when the ergodic
properties of the billiards are studied. Tangency is also responsible
for a special kind of bifurcation which can be best illustrated by
considering the standard Sinai billiard, or rather, its desymmetrized
quarter (see figure \ref{fig:sinai}). This will be the example we
shall use throughout this article and it can be generalized to other
chaotic concave billiards in a straightforward manner.  The chord
$AB$ in figure \ref {fig:sinai} is tangent to the arc of radius
$R_0$ and it is a classical trajectory which leads from $A$ to $B$.
Bifurcations due to tangency occur when one allows the radius of the
arc to vary.  When the radius is reduced to any value $R_- < R_0$, two
classical trajectories can serve to connect $A$ and $B$. One of them
goes directly, the other reflects specularly from the arc at a point
$C$. As $R_- \rightarrow R_0$ the two trajectories become closer,
until they coalesce when $R_-=R_0$. When the radius is increased to
values $R_+>R_0$, both the direct and the reflected trajectories from
$A$ to $B$ become classically forbidden. That is, $A$ and $B$ are
mutually shaded from each other by the geometrical shadow cast by the
arc.

Tangency affects also the quantum (wave) dynamics in the billiard. Due
to the finite wave length, the sharp geometrical shadow is replaced by
a transition region which smoothly interpolates between the strictly
illuminated and shaded regions. This is the {\it penumbra}
(Latin: almost shadow) domain. The standard semiclassical quantization
of billiards, which is restricted to the illuminated domain, expresses
the density of states $d(k)\equiv\sum_n \delta(k-k_n)$ ($E_n=k_n^2$
are the eigenvalues) in terms of classical periodic orbits via the
Gutzwiller trace formula \cite{Gut71}.  A semiclassical theory for
transitions to the strictly shaded domain can be formulated in terms
of non-classical orbits which are allowed to creep along a section of
the boundary.  This approach was first discussed by Keller and
co-workers who developed the concept of geometrical diffraction theory
in a systematic way \cite{LK59}.  Vattay et al.\ \cite{VWR94} have
recently generalized Gutzwiller's trace formula by including periodic
orbits with creeping sections that give exponentially small
corrections.  In a previous paper
\cite{PSSU96} we gave a preliminary account of a semiclassical theory
which is valid in the penumbra where the diffraction contributions are
not exponentially small. The purpose of the present paper is twofold -
to give a complete discussion of the penumbra effects, and to show how
they can be scrutinized within the scattering approach to quantization
\cite{DS92b,RS95,SS95,Smi94}.
 
Diffraction effects appear as corrections to the leading
semiclassical expressions which are based on Gutzwiller's trace
formula. To identify corrections to the contributions from individual
periodic orbits, it is advantageous to study the ``length spectrum"
\begin{equation}
D(x) =\int_{0}^{\infty} {\rm d}k\,\e^{\i kx} d(k) 
=\sum_{n=1}^{\infty} e^{\i k_n x}\,.
\label {eq:lengthspec}
\end{equation}
Strictly speaking, $D(x)$ is a tempered distribution whose singular support is
at lengths of periodic manifolds of the classical billiard \cite{AM77}. In the
Sinai billiard these are the continuous families of neutral orbits
(bouncing ball manifolds) and isolated unstable periodic orbits.  In
practice, we do not have the complete spectrum at our disposal, and we
study the truncated length spectrum
\begin{equation}
D_w(x) = \int_0^\infty {\rm d}k \, \e^{\i k x} w(k) d(k) \, 
=\sum_{n=1}^{\infty} w(k_n)\,\e^{\i k_n x}\,.
\label {eq:lengthspec_win}
\end{equation}
Here $w(k)$ is a smooth positive function with a finite support
centered at $k=\bar k$.
$D_w(x)$ is a smooth function with finite spikes
where $D(x)$ is singular. The leading semiclassical contributions to
(\ref{eq:lengthspec_win}) can be evaluated by substituting the
Gutzwiller trace formula, augmented with the special expressions
due to the neutral families.  The difference between the
semiclassical and the exact length spectra gives a measure of the
quality of the semiclassical approximation. We shall show below that
the corrections due to diffraction effects are responsible for the
largest deviations. The main difficulty of this approach is that the
length spectrum for a chaotic billiard is rather dense, and even with
spectral sections containing thousands of eigenvalues, the
contributions from neighbouring lengths overlap 
to the extent that a detailed investigation of contributions from individual
periodic orbits is impossible. However, the
scattering approach to quantization offers a method 
to disentangle some of this complexity which is due to the proliferation
of periodic orbits.
 
The scattering approach we shall employ here makes use of an
auxiliary scattering system which couples the billiard to a wave guide
in an appropriate manner \cite{DS92b,RS95,SS95,Smi94}.  In
a way, this method provides the quantum analogue of the classical
Poincar\'e section in terms of the scattering matrix, and it has much
in common with earlier \cite {Bog92} and later \cite
{Gut93,Pro94} attempts aiming at a similar goal. Since the method
is well documented and reviewed, we shall only mention its basic
ingredients for the particular application to the Sinai billiard.

The scattering problem which we use in the present context is
described in figure \ref {fig:sinai}.  We define two systems to
which we apply the Krein spectral shift theorem \cite{BY93}.  The
Hamiltonian $H_0$ of the first system is the kinetic energy ($-\Delta
$) in the space of functions which satisfy Dirichlet boundary
conditions on the parallel channel walls and on the section $\Gamma_{\rm S}$
which separates the original Sinai billiard from the wave guide.  In
the second system $H$ the wall at $\Gamma_{\rm S}$ is not present and the
functions under consideration have to satisfy Dirichlet boundary
conditions on the extended billiard boundary displayed with a full
line in figure (\ref{fig:sinai}).
One can now define the scattering matrix $S(E)$ for any energy
$E=k^2$, and at the energy eigenvalues $E_n$ of the billiard the
secular equation $\det(I+S(E_n))=0$ is satisfied. Krein's theorem is
expressed by the relation
\begin{equation}
\frac {1}{\pi}\lim _{\epsilon \downarrow 0} 
{\rm Im}\, {\rm Tr}\, \left[ G(E+i\epsilon) - G_0(E+i\epsilon) \right]
 =
\frac {1}{2\pi}\frac {\d \Theta (E)}{\d E}\,.
\label{eq:Krein}
\end{equation}
The trace in (\ref{eq:Krein}) is taken over the space of continuum
eigenstates of $H$.  $G_0$ and $G$ are the Green functions for the two
systems ($G_0$ is zero inside the billiard domain), and $\Theta(E)$ is
the total phase of the scattering matrix $S(E)$, defined by
$\Theta(E)= -\i\,\log \det S(E)$. Krein's theorem connects the
excess density of continuum states due to the introduction of the
scatterer, with the total phase of the S-matrix. Performing the trace
operation in the semiclassical approximation (see
e.g.\,\cite{GR89b,Gas93}), one finds that the left hand side in
(\ref{eq:Krein}) can be expressed as a sum of a smooth term and
contributions from {\em trapped} periodic orbits - orbits which do not
escape in spite of the fact that the billiard is opened. This is very
important in the context of the semiclassical quantization of
billiards by the scattering approach. There, one expresses the
spectral density of the closed billiard as
\begin{equation}
d(k) = 
\frac {1}{2\pi}\frac {\d \Theta (k)}{\d k}+\frac {1}{\pi} {\rm Im}\, {\d\over\d k}\sum_{n=1}^{\infty}
\frac{(-1)^n}{n}\Tr\,S^n(k)\,.
\label {eq:density}
\end{equation}  
In the semiclassical limit, each ${\rm Tr}\, S^n$ is expressed as a sum of
contributions from the periodic orbits of the closed billiard which
bounce on the section $\Gamma_{\rm S}$ exactly $n$ times.  However, there
may be periodic orbits which do not reflect from $\Gamma_{\rm S}$ at
all. They are trapped in the open billiard, and their contributions to
the level density come from the semiclassical expression of $\frac
{1}{2\pi}\frac {\d \Theta (k)}{\d k}$ via
(\ref{eq:Krein}), as explained above \cite{DS92b}.  $\frac
{1}{2\pi}\frac {\d \Theta (k)}{\d k}$ provides also the
leading term in the expression for the smooth part of the spectral
density. It contains higher order contributions as well, but not
necessarily all of them, as is discussed in \cite{SU96}.

Taking the Fourier transform of (\ref {eq:density}), we get an
expression of the length spectrum in terms of the Fourier transforms
of $\frac {1}{2\pi}\frac {\d \Theta (k)}{\d k}$ and of
$\frac {\d {\rm Tr}\, S^n(k)}{\d k}$ for all positive $n$. The
transform of the total phase will provide the length spectrum of
orbits which are trapped in the open billiard. In our system there
exists only one isolated and unstable trapped orbit. Another
contribution comes from the family of marginally stable orbits which
bounce perpendicularly between the straight sections of the billiard.
This is a very sparse set of lengths, and it leaves a lot of space to
observe diffractive orbits of various kinds. The Fourier transform of
$\frac {\d {\rm Tr}\, S^n(k)}{\d k}$ provides the length spectrum
of orbits which bounce $n$ times from the section $\Gamma_{\rm S}$. In this
way we can partition in a systematic way the contribution from orbits
with different $n$ to the total length spectrum and next to leading
order effects such as diffraction contributions can be better
observed.

To have an idea of the way how the theory which was described above
works in practice, let us consider the simple case of the square
billiard. Denoting the length of the square by $a$, we get a diagonal
scattering matrix
\begin{equation}
S_{l,l'}(k) = -\exp\left[2\pi \i \sqrt{\left(\frac {ak}{\pi}\right)^2-l^2 }\,\right] \delta_{ll'}\,.
\label{eq:Ssq}
\end{equation}
The subspace of conducting modes is of dimension $\Lambda =\left [ \frac{ak}{\pi} \right]$, where
the symbol $\left [ \cdot \right ]$ stands here for the integer part. In this case
\begin{equation} 
\frac {1}{2\pi} \Theta (k)= \sum _{l=1}^{\Lambda} 
\sqrt{\left(\frac {ak}{\pi}\right)^2-l^2 } - \frac{\Lambda}{2}
\label {eq:totphasesq}
\end{equation}
Using the Poisson summation formula and other standard relations, one
gets
\begin{equation}
\fl \frac {1}{2\pi} \Theta (k)= \frac{a^2k^2}{4\pi} - \frac{ak}{\pi} +\frac{1}{4} 
+\frac {ak}{2\pi} \sum_{m=1}^{\infty}\frac{1}{m} J_1(2mak) -\frac{1}{2\pi}\sum_{m=1}^{\infty}
\frac{1}{m}\sin(2mak)\,.
\label {eq:finphase}
\end{equation}
This is an exact equality which can be interpreted in the following
way. The expression appearing in the upper line of (\ref{eq:finphase})
is the smooth spectral counting function $\bar N(k)$ which consists of
three terms only - the area, circumference and corners terms \cite
{BH76}. The next infinite sum is the contribution of the (open)
manifold of periodic orbits which are parallel to the section
$\Gamma_{\rm S}$. The last sum is due to to the limiting periodic orbits
which run along the edge $\Gamma_{\rm S}$ and its counterpart on the other
side of the billiard. These two limiting orbits are the closure of the
manifold mentioned previously. We thus see that the oscillatory part
of (\ref {eq:finphase}) consists of contributions which exhaust all
possible {\em trapped} periodic motion in the open billiard. We would
like to emphasize again that this result is exact. In the sequel, when
we treat the more interesting Sinai billiard, we shall obtain
similar relationships which involve other possible trapped
orbits. This will be done, however, within the semiclassical
approximation and its refinements.

The paper is organized in the following way. In the next section we
shall present the semiclassical theory of quantization for billiards
which are exterior to a circle taking into account the leading
diffraction effects both in the penumbra and the deep shadow domains. Once
this is achieved, we demonstrate the success of the theory using
two numerical examples.  The first example in section 
\ref{sec:annulus} will be the annular sector which represents an
integrable billiard. In the following section \ref{sec:sb} we turn to
a detailed numerical study of the quantized Sinai billiard which is
chaotic and therefore necessitates the methods of analysis which were
outlined above. In section \ref{sec:penstat} we shall discuss the
general significance of diffraction corrections for the semiclassical
quantization of a concave and chaotic billiard. The results of the
paper will then be summarized in section \ref{sec:conc}.

\section{Theory of diffraction for dispersing billiards}
\label{sec:tod}
In this section, we consider billiards with a domain $\Omega$ that is
exterior to a circle (e.g.\,the Sinai billiard). We find expressions
for the contribution to the density of states of periodic orbits which
are nearly tangent to the circle (either by reflecting with a very
small angle or by passing very close to the circle). As will be shown,
the standard semiclassical expressions for contributions of such
orbits fail. In addition creeping orbits \cite {LK59} exist due to the
concave billiard boundary (the circle in our case).  The contribution
of periodic orbits which have a creeping part was studied by Vattay,
Wirzba and Rosenqvist \cite {VWR94}. However, their expressions for
creeping orbits also fail, if the orbit is too close to tangency
(i.e.\,if the creeping angle is too small). The expressions for
periodic orbits near tangency, derived in this section, extend the
semiclassical description of a general billiard exterior to a
circle. The methods we will use for nearly tangent periodic orbits are
adapted from the calculations of Nussenzveig
\cite{Nus65} for the problem of scattering off a three-dimensional
sphere.

The free Green function satisfies 
\begin{equation}
  (\Delta + k^2)\, G ({\bf r},{\bf r}') = -\delta ({\bf r}-{\bf r}')
  \label{gequation}
\end{equation}
for any ${\bf r},{\bf r'}$ and outgoing boundary conditions at
infinity. From Green's theorem one obtains that the eigenvalues $k_n$ of the
billiard are those values for which the boundary integral equation 
\begin{equation}
  u ({\bf r}_s) = 2 \int_\Gamma {\rm d}s' \, \frac{\partial G}{\partial
    \hat n_{s}} ({\bf r}_s,{\bf r}_{s'}) u ({\bf r}_{s'})
  \label{integeq}
\end{equation}
for the normal derivative of the wave function $u({\bf r}_s)=\partial
\psi / \partial \hat n ({\bf r_s})$ has a solution. The billiard
boundary is denoted by $\Gamma$, and the normal direction $\hat n_{s}$
in a point on the boundary ${\bf r_s}$ is pointing from $\Omega$
outwards.

Equation (\ref{integeq}) is used to obtain a secular equation, from
which the density of states $d(k)$ may be found by a multiple reflection expansion
containing terms as
\begin{equation}\fl
  {\rm Im} \frac {2^N}{\pi N} \frac {\d}{\d k} \int_\Gamma {\rm
    d} s_1 \ldots {\rm d} s_N \frac{\partial G}{\partial \hat n_1}
  ({\bf r}_1,{\bf r}_2) \cdots \frac{\partial G}{\partial \hat
    n_{N-1}} ({\bf r}_{N-1},{\bf r}_N) \frac{\partial G}{\partial \hat
    n_N} ({\bf r}_N,{\bf r}_1) \,.
  \label{nperiodic}
\end{equation}
When the integrals are evaluated in stationary phase approximation,
each saddle point in one of these terms yields
the contribution of a periodic orbit which reflects $N$ times off the boundary.
This method has often been used as a starting point to derive
semiclassical theories for billiards, e.g.\,Alonso and Gaspard
employed it for finding higher order corrections to the contributions
of periodic orbits \cite{AG93}.

For the purpose of considering nearly tangent or creeping orbits at
the circle, the free Green function is replaced by the circle Green
function, which will also be denoted by $G$. It satisfies
(\ref{gequation}) for ${\bf r}$ and $\bf r'$ exterior to the circle,
and moreover the prescribed boundary conditions on the circle. The
derivation of (\ref{nperiodic}) may still be followed and yields the
contribution of periodic orbits which reflect $N$ times off $\Gamma$,
which is now the billiard boundary excluding the circle. The
reflections off the circle are included in the Green function, and so
is the diffractive behaviour associated with creeping or nearly
tangent orbits. The problem is thus reduced to understanding the
behaviour of the Green function of the circle for different positions
of $\bf r$ and $\bf r'$. 

Although in principle the multiple reflection expansions based on
the free and the circle Green functions are equivalent and both exact,
the latter greatly simplifies the mathematical treatment close to
tangencies. A hint to this is already contained in the classical
analogue representing a Poincar\'e map from the boundary $\Gamma$ to
itself, which is discontinuous when the circle is a part of
$\Gamma$. The same system can be described by a mapping which is
$C^{(0)}$ when the circle is excluded from $\Gamma$ and the mapping
function includes the reflection the circle.

\subsection {The circle Green function}
\label{sec:circgreen}
In this section we consider a circle of radius $R$ centered at the
origin, with Dirichlet boundary conditions ($\psi({\bf r})=0$) on its
boundary. In 
\ref{sec:mixed} we indicate how the results
of this section modify for Neumann ($\partial_{\bf \hat n}\psi({\bf
r})=0$) or mixed boundary conditions ($\kappa\psi({\bf
r})+\partial_{\bf \hat n}\psi({\bf r})=0)$.

The Green function of the circle satisfies (\ref{gequation})
for any $\bf r$, $\bf r'$ outside the circle, outgoing boundary
conditions at infinity, and Dirichlet boundary conditions on the
circle.  The exact expression for the Green function is (in polar
coordinates)
\begin{equation}
  G ({\bf r},{\bf r'}) = \frac{\i}{8} \sum_{l=-\infty}^\infty \left[
  H_l^-(kr_<) + S_l(kR) H_l^+(kr_<) \right] H_l^+(kr_>)
  \e^{\i l\Delta\theta} \,,
  \label{gexact}
\end{equation}
where $r_>$ ($r_<$) is the larger (smaller) of $r$ and $r'$, and
$\Delta\theta=\theta-\theta'$. The elements of the scattering
matrix of the circle which is diagonal in the angular momentum representation are
given by
\begin{equation}
  S_l(kR) = -\frac{H_l^-(kR)}{H_l^+(kR)} \,.
  \label{scatmatrix}
\end{equation}
We continue the discussion assuming $r, r'\gtrsim [1 + (kR)^{-2/3}]R$,
so that the Debye approximation may be used for $H_l^+(kr)$ and
$H_l^+(kr')$ if $l\leq kR$. In addition we assume, without loss of
generality, that $0<\Delta\theta<\pi$.

\input{FIG/fig-regions}
The Poisson summation rule is used to express the Green function as
\begin{equation}
  G ({\bf r},{\bf r'}) = \sum_{m=-\infty}^\infty G^{(m)} ({\bf r},{\bf
    r'}) \,,
  \label{greensum}
\end{equation}
where
\begin{equation}\fl
  G^{(m)} ({\bf r},{\bf r'}) = \frac{\i}{8} \int_{-\infty}^{+\infty} {\rm
    d}l \, \left[ H_l^-(kr_<) + S_l(kR) H_l^+(kr_<) \right]
  H_l^+(kr_>) \e^{\i l\Delta\theta + 2\pi iml} \,.
  \label{greenint}
\end{equation}
We discuss separately three regions for {\bf r} and {\bf r'} (see
figure \ref{fig:regions}). The first is the lit region, in which the
usual semiclassical results hold. It is obtained from the $m=0$ term
in (\ref{greensum}), while the $m\neq0$ terms always describe the
contributions of creeping waves.  The second is the deep shadow
region, in which only creeping waves contribute and the third is the
penumbra, in which the leading order contribution, the $m=0$ term,
comes from nearly tangent orbits.  The lit and the deep shadow regions
lead to known contributions of periodic orbits and the corresponding derivations will only
briefly be described for the sake of completeness, and for understanding
where the employed approximations fail when approaching the transition
region.

\subsubsection {The lit region}
\label {sec:lit}
\input{FIG/fig-lit}
In the lit region the integral $G^{(0)}({\bf r},{\bf r'})$ gives the
usual semiclassical result. 
For the scattering matrix element $S_l(kR)$ the Debye approximation is used for $l<kR$, 
and for $l>kR$ it is approximated by $1$. 
The integral is evaluated in stationary phase approximation. 
There are two saddle points, which relate to the two
classical trajectories from $\bf r$ to $\bf r'$ (see figure \ref{fig:lit}). 
One is direct, and the other reflects specularly from the circle.
The Green function is then a sum of two terms. The contribution of the
direct path is given by
\begin{equation}
  G_{\rm d} ({\bf r},{\bf r'}) = \frac{1}{4} \left(\frac {2\i}{\pi
    kL}\right)^\frac{1}{2} \e^{\i kL} \,,
  \label{greenlitd}
\end{equation}
where $L=|{\bf r} - {\bf r'}|$, and the reflected path yields
\begin{equation}
  G_{\rm r} ({\bf r},{\bf r'}) = -\frac{1}{4} \left[\frac{2\i}{\pi
    k(L_{\rm r}+L_{\rm r'}+2L_{\rm r}L_{\rm r'}/a)}\right]^\frac{1}{2} \e^{\i k(L_{\rm r}+L_{\rm r'})}
  \,,
  \label{greenlitr}
\end{equation}
where $L_{\rm r}$ ($L_{\rm r'}$) is the distance from $\bf r$ ($\bf r'$) to the
point of reflection such that the length of the reflected trajectory is
$L_{\rm r}+L_{\rm r'}$. The impact parameter of the reflected path is $b_{\rm r}$, and
$a=\sqrt{R^2-b_{\rm r}^2}$.

The semiclassical result fails when the reflected trajectory becomes
nearly tangent to the circle. To be precise, the semiclassical
approximation holds provided that $b_{\rm r}\lesssim
[1-(kR)^{-2/3}]R$. In other words the angle of incidence $\theta$
should be smaller than $\pi/2-(kR)^{-1/3}$. This condition is
necessary for the semiclassical approximation of the scattering matrix
$S_l(kR)$ to hold at the saddle point, and thus defines the borderline
between the illuminated region and the penumbra.

\subsubsection {The deep shadow region}
\label{sec:deepshadow}
\input{FIG/fig-lplane}
In the deep shadow region the contour of the integral (\ref{greenint})
for $m=0$ may be closed
in the upper half plane. The integral is
then calculated by summing the contributions of the poles of the integrand which are
just the poles of the $S$ matrix.  The first few poles
are close to $kR$ (figure \ref{fig:lplane}) and give a
contribution which is evaluated using the transition region approximation
for the Hankel functions. The position of these poles is given by
\begin{equation}
  l_n \approx kR + \e^{\i\pi/3} x_n \left( \frac{kR}{2} \right)^\frac{1}{3}
  \,,
  \label{spoles}
\end{equation}
where $-x_n$ are the zeros of the Airy function ${\rm Ai}(x)$ \cite{AS65}. The
residues of the $S$ matrix at these poles are given by
\begin{equation}
  r_n \approx \frac {\e^{-i\pi/6}}{2\pi {\rm Ai}'(-x_n)^2} \left(
  \frac{kR}{2} \right)^\frac{1}{3} \,.
  \label{sresidues}
\end{equation}
The result is interpreted as the contribution of a creeping wave \cite
{LK59}. It starts from $\bf r$ along a path which is tangent to the
circle and creeps along the circle in one of the available creeping
modes $n$ until it leaves to $\bf r'$ along a path which is again tangent
to the circle (see figure \ref{fig:lit}). The contribution of a creeping wave to the
Green function is
\begin{equation}
  G_{\rm c}^{(0)} ({\bf r},{\bf r'}) = \frac{1}{4} \left( \frac{2\i}{\pi
    kL_{\rm r}L_{\rm r'}} \right)^\frac{1}{2} \e^{\i k(L_{\rm r}+L_{\rm r'})} \sum_n D_n^2
  \e^{\i l_n\gamma^{(0)}} \,,
  \label{greencreep}
\end{equation}
where $L_{\rm r}$ ($L_{\rm r'}$) is the distance from $\bf r$ ($\bf r'$) to the
corresponding point of tangency, and $\gamma^{(0)}$ is the creeping
angle. The length of the creeping trajectory is
$L_{\rm r}+L_{\rm r'}+\gamma^{(0)}R$. The diffraction coefficient $D_n^2$ 
is given for each creeping mode by
\begin{equation}
  D_n^2 = \e^{\i\pi/4} \left( \frac{2\pi}{k} \right)^\frac{1}{2} r_n \,.
  \label{creepcoeff}
\end{equation}
The angular momentum along the creeping orbit in a creeping mode $n$
is $l_n$. Its real part is slightly larger than $kR$. 
The angular momentum also has a
positive imaginary part, which is the result of the continuous
decrease of amplitude as waves leave the circle in a tangent direction
all along the way.

The same calculation holds for all $m\neq0$ terms of (\ref{greenint}),
for any location of $\bf r$ and $\bf r'$. For $m>0$,
$G_{\rm c}^{(m)} ({\bf r},{\bf r'})$ gives the contribution of creeping
waves going around the circle an additional $m$ full times in the
clockwise direction, such that $\gamma^{(m)}=\gamma^{(0)}+2\pi m$. For
$m<0$, the integration contour should be closed from below, and the
result describes the contribution of creeping waves going in the
anti-clockwise direction $|m|-1$ full times.

The Airy approximation (\ref{spoles}), (\ref{sresidues}) fails when
inserted into (\ref{greencreep}) if the creeping angle does not obey
$\gamma\gg(kR)^{-1/3}$. This condition guarantees that the
contribution of the poles rapidly decays with $n$ and defines the
borderline between the deep shadow region and the penumbra. If it is
not satisfied, the sum in (\ref{greencreep}) has to be extended to
poles which cannot be described by (\ref{spoles}) and (\ref{sresidues}).

\subsubsection {The penumbra region}
\label {sec:transition}
In the penumbra region, between the lit and the deep shadow regions, the
above expressions fail. In this section we obtain expressions valid
in the penumbra, following the methods used by Nussenzveig
\cite{Nus65} for studying the problem of scattering off a sphere. The
contour of the integral (\ref{greenint}) for $m = 0$ is first
deformed in the upper half complex plane (see figure \ref{fig:lplane}), 
so that it goes from $\sigma_1 \infty$, through
$kR$ and around the poles of the $S$ matrix to $\sigma_2 \infty$. The
limits $\sigma_1 \infty$ and $\sigma_2 \infty$ are defined by
demanding that the contour does not cross the lines asymptotically
defined by $l = \sigma |l|$, where
\begin{equation}
  \sigma = \exp \left[ \i \left( \frac {\pi}{2} + \epsilon \right)
\right]
  \label{defsigma1}
\end{equation}
and
\begin{equation}
  \eta = \epsilon \ln \left[ \frac {2 l}{\e k r_<} \right] .
  \label{defsigma2}
\end{equation}
The limits $l \rightarrow \sigma \infty$ and $\epsilon \rightarrow 0$
are taken simultaneously in such a way that $\eta \rightarrow \pm \pi /
2$. The deformation of the contour allows a separate treatment of the two
parts of the integral (\ref{greenint}). For the first part
we immediately obtain
\begin{equation}
  \frac {\i}{8} \int_{\sigma_1 \infty}^{\sigma_2 \infty} {\rm d} l \,
  H_l^-(k r_<) H_l^+(k r_>) \e^{\i l \Delta \theta} = 0 ,
  \label{vanishint}
\end{equation}
as the contour of the integral may be closed from above and the
integrand has no poles. The second part of the integral is split into
two parts, which we call the direct and glancing parts of the Green
function for reasons which will become clear. The result is
\begin{equation}
  G_{\rm d} ({\bf r},{\bf r}') = \frac{\i}{8} \int_{kR}^{\sigma_2\infty} {\rm
    d}l \, H_l^+(kr) H_l^+(kr') \e^{\i l\Delta\theta}
  \label{greentransid}
\end{equation}
and
\begin{eqnarray}\fl
  G_{\rm g} ({\bf r},{\bf r}') =  \frac{\i}{8} \int_{\sigma_1\infty}^{kR}
  {\rm d}l \, S_l(kR) H_l^+(kr) H_l^+(kr') \e^{\i l\Delta\theta} 
  \nonumber \\ + \frac{\i}{8} \int_{kR}^{\sigma_2\infty} {\rm d}l \,
  [S_l(kR)-1] H_l^+(kr) H_l^+(kr') \e^{\i l\Delta\theta} \,.
  \label{greentransit}
\end{eqnarray}
Each of these terms will now be treated separately.

\input{FIG/fig-glance}
In the direct part (\ref {greentransid}) the Hankel functions are
replaced by their Debye approximation, and the integrand is evaluated
around its saddle point.  The only difference in the contribution of
the direct term from that of the lit region expression is that now the
limit of the integration must be accounted for, as it is close to the
saddle point. The result is just the lit region expression
(\ref{greenlitd}) multiplied by a simple factor. It is given by
\begin{equation}
  G_{\rm d} ({\bf r},{\bf r}') = \left( \frac {F(\infty) -
    F(\nu)}{\sqrt{2\i}} \right) \frac {1}{4} \left[ \frac {2 \i}{\pi k
    L} \right]^\frac{1}{2} \e^{\i kL} \,,
  \label{greentransd}
\end{equation}
where $F(x)=C(x)+iS(x)$ is the Fresnel integral, and
\begin{equation}
  \nu = \left( \frac {kL}{\pi zz'} \right)^\frac{1}{2} (R-b_{\rm d}) \,.
  \label{transnu}
\end{equation}
The impact parameter is again denoted by $b_{\rm d}$, and we have
$z=\sqrt{r^2-b_{\rm d}^2}$, $z'=\sqrt{r'^2-b_{\rm d}^2}$ (see
figure \ref{fig:glance}). The Fresnel factor is in general a complex
number. It equals $\frac{1}{2}$ for exact tangency. Close to the
border of the illuminated region it approaches $1$, and it tends to
$0$ at the border of the deep shadow region. Although
this is the correct limiting behaviour, (\ref{greentransd}) is
restricted to the penumbra and does not necessarily represent the
correct interpolation between the illuminated and the deep shadow
regions.

In the glancing part (\ref {greentransit}) the main contribution of
both integrands comes from the vicinity of $kR$. The integrals are
evaluated using the following approximations: The $S$ matrix is approximated by the
transition region expression
\begin{equation}
  S_l (kR) = -\e^{2\pi \i/3} \frac {{\rm Ai}(x \e^{-2\pi \i/3})}{{\rm
      Ai}(x \e^{2\pi \i/3})} \,,
  \label{stransapp}
\end{equation}
where
\begin{equation}
  x = \left( \frac {2}{kR} \right)^\frac{1}{3} (l-kR) \,.
  \label{transarg}
\end{equation}
The rest of the integrand is taken at $kR$, using the Debye
approximation for the Hankel functions. The result of all these steps
is that the only $l$ dependence in the integrand is in the argument of
the transition region approximations.  By a change of the integration variable the
problem reduces to finding the value of the constant
\begin{equation}
  c = \frac {\e^{\pi \i/3}}{2^{1/3}} \int_{-\infty}^0 {\rm d}x \frac
  {{\rm Ai}(x\e^{-2\pi \i/3})}{{\rm Ai}(x\e^{2\pi \i/3})} + \frac
  {1}{2^{1/3}} \int_0^\infty {\rm d}x \frac {{\rm Ai}(x)}{{\rm
      Ai}(x\e^{2\pi \i/3})} \,.
  \label{transconst}
\end{equation}
The first term is the complex conjugate of the second, as was
shown by Rubinow and Wu \cite {RW56}, and after a numerical integration
over the second term, we have
\begin{displaymath}
  c \approx 0.996193019928 \,.
\end{displaymath}
Finally, the glancing term of the Green function is given by
\begin{equation}
  G_{\rm g} ({\bf r},{\bf r}') = -\frac{c}{4\pi}
  \frac{(kR)^{1/3}}{k\sqrt{L_{\rm r}L_{\rm r'}}}
  \e^{\i k(L_{\rm r}+L_{\rm r'}+R\gamma^{(0)})+i\pi/3} \,,
  \label{greentranst}
\end{equation}
where $L_{\rm r}$ ($L_{\rm r'}$) is the distance from $\bf r$ ($\bf r'$), along
a line which is tangent to the circle, to the corresponding point of
tangency, and
\begin{equation}
  \gamma^{(0)} = \Delta\theta - \arccos \left( \frac{R}{r} \right)
  -\arccos \left( \frac{R}{r'} \right) \,.
  \label{transgamma}
\end{equation}
When $b_{\rm d}<R$, the values for $L_{\rm r}$, $L_{\rm r'}$ and $\gamma^{(0)}$ are
equivalent to the ones in the creeping case. When $b_{\rm d}$ becomes larger
than $R$ (see figure \ref{fig:glance}), the points of tangency cross
and $\gamma^{(0)}$ becomes negative. Note that in contrast to
(\ref{greencreep}), equation (\ref{greentranst}) contains no exponential
damping.

\subsection {Diffraction corrections in the trace formula}
\label{sec:dos}
The contribution of periodic orbits to the density of states of the
billiard may now be calculated. As was explained above, the
contribution of orbits which bounce $N$ times on the exterior boundary
$\Gamma$ is found from the multiple reflection expansion
(\ref{nperiodic}) using the circle Green function and performing the
integrals over the billiard boundary excluding the circle in saddle
point approximation.  For each circle Green function the appropriate
expression is used, depending on the positions of ${\bf r}_i$ and
${\bf r}_{i+1}$. In the lit region we have $G=G_{\rm d}+G_{\rm r}$
(\ref{greenlitd}-\ref{greenlitr}), in the deep shadow region $G=G_{\rm
c}$ (\ref{greencreep}), and in the penumbra $G=G_{\rm d}+G_{\rm g}$
(\ref{greentransd},\ref{greentranst}).  In the deep shadow region the
deviation from $\i kL$ in the exponent in (\ref{greencreep}) is taken
as part of the prefactor, as it is proportional to $k^{1/3}$ and can
be considered a slowly varying function for large $k$. Then in all
cases the expressions for the Green function are of the form $A\e^{\i
kL}$ suitable for the saddle point approximation.  After the saddle
point integration a purely classical path yields the Gutzwiller
contribution of an ordinary unstable isolated periodic orbit corrected
with the product of the Fresnel factors for all segments traversing
the penumbra. An important difference for creeping orbits is,
that the action entering the Green function $G({\bf r},{\bf r}')$ can
be represented as a sum of two terms, one depending exclusively on
$\bf r$ and the other on $\bf r'$.  This is obvious from
(\ref{greencreep}), (\ref{greentranst}) and (\ref{transgamma}).  As a
consequence, the matrix of second derivatives entering the saddle
point approximation to (\ref{nperiodic}) decomposes into $K^{(\rm P)}$
independent blocks for an orbit (P) with $K^{(\rm P)}\ge 1$ creeping
segments. Finally one obtains an expression which is formally similar
to the one derived by Vattay et al. \cite{VWR94} for the contribution
of a creeping orbit in the deep shadow region
\begin{equation}
\label{nsc}
d_P^{\rm (scl)}(k)={1\over r\pi} \frac{\d}{\d k}{\rm Im}\,\prod\limits_{j=1}^{K^{(\rm P)}} G_j^{\rm
(scl)}\,G_j^{\rm (diff)}\,.
\end{equation}
The semiclassical Green function $G_j^{\rm (scl)}$ along the
classical segment $j$ connecting two points of tangency of the path to
the circle is given by
\begin{equation}
  G_j^{\rm (scl)} = f_j\,\frac {1}{4} \left( \frac {2 \i}{\pi k
    |M_{12}|} \right)^\frac{1}{2} \e^{\i kL_j+i\pi\mu_j/2} \,,
\end{equation}
where $L_j$ represents the path length, $\mu_j$ the number of
conjugate points and $f_j$ the product of all the Fresnel factors
along the segment. The monodromy matrix is defined such that
$M_{12}=\partial x'_\bot/\partial \phi''$ with $\phi$ and $x_\bot$
denoting the direction of the path and the coordinate normal to it,
respectively. The Green function along a creeping segment is given by
\begin{equation}
G_j^{\rm (diff)} = \left \{
\begin{array}{lr}
4\pi\sum_n r_n \e^{\i l_n\,\gamma_j}& {\rm deep\; shadow}
\\ & \\
-2c (kR)^{1/3}\e^{\i kR\gamma_j-i\pi/6} & {\rm penumbra}
\end{array}
\right .
\end{equation}
and $r$ in (\ref{nsc}) counts the number of repetitions for orbits which
are multiple traversals of a primitive orbit.

Equation (\ref{nsc}) gives the diffractive contributions in terms of
geometrical information from creeping orbits. However, if one is
interested in the contribution of a purely classical orbit which
includes a very shallow reflection from the circle and must therefore
be treated in the penumbra approximation, it is also possible
to give a correction to the standard semiclassical expression without
explicitly calculating the corresponding creeping orbit. For this
purpose, the length of the creeping orbit is approximated by the
length of the classical orbit since both orbits approach each other
as the reflection becomes closer to tangency. The monodromy matrix along the
classical segment of the creeping orbit is replaced by the monodromy
matrix along the classical orbit {\em excluding} the almost tangent 
reflection which is incorrectly described in the standard theory.
In terms of this reduced monodromy matrix $M$ and the angle of reflection $\theta$ 
the Gutzwiller amplitude for the contribution of the orbit is
\begin{equation}
{1\over \pi} \left |Tr M - 2 + {2 M_{12}\over R \cos\theta}\right |^{-1/2}
\sim {1\over \pi}\left | {R \cos\theta \over 2 M_{12} }\right|^{1/2}\,,
\end{equation}
where we have assumed $\cos\theta\ll 1$.
Comparing this to (\ref{nsc}) we obtain the factor 
\begin{equation}
\label{corrfac}
{c \over \sqrt{\pi \cos\theta}} (kR)^{-1/6}\e^{\i\pi/12}\,,
\end{equation}
by which the standard result for the contribution of the orbit
is enhanced due to diffraction.

Finally we would like to point out the different dependence on $k$ 
of the contributions from various periodic orbits. 
The contribution of an isolated periodic orbit, with all segments in
the lit region, is $\Or(k^0)$, as is expected in the semiclassical
approximation disregarding diffraction. For each segment in the deep
shadow region, the contribution is multiplied by $\Or[\exp(-C
k^{1/3})]$, where $C$ depends on the creeping angle corresponding to
this segment.  
For each glancing segment in the penumbra 
a factor $\Or(k^{-1/6})$ 
is obtained. It is worth noting that,
unless the orbit segment is precisely tangent, it will eventually
fall, for high enough $k$, either in the lit or in the deep shadow
category.

\section {Penumbra diffraction in the spectrum of the annular sector}
\label{sec:annulus}
The simplest billiard with a domain which is exterior to a circle, is
the annulus defined by two concentric circles with radii $R_1$ and
$R_2$. The annulus billiard is integrable, and its periodic orbits
form one parameter families (or manifolds). The ratio $R_2/R_1$ may be
chosen such that two families of primitive periodic orbits, and their
repetitions, will traverse the penumbra of the inner circle (one
with direct segments and one with glancing segments). As a first
numerical test of the expressions for diffractive periodic orbits, we
consider the contribution of these orbits to the spectral density of
the annulus. The annulus has the advantage that since it is
integrable, its periodic orbits are easy to find, and its
eigenvalues may be calculated with a high accuracy and for high
energies, reaching the limit $(kR_1)^{1/3} \gg 1$. As will be
discussed, however, diffractive periodic orbits which bounce more than
once on the exterior circle are not easily described by the theory
presented in this paper because of a problem which is specific to the
annular geometry. For this reason we will consider only the shortest
diffractive periodic orbit of the desymmetrized annulus which bounces
just once on the outer circle. The example of the annulus is thus
limited, and in particular it does not allow to check whether the expressions for
the Green function of the circle give the correct results after a
saddle point integration.

\input{FIG/fig-annulus}
In this section we consider a desymmetrized annulus, the annular
sector of angle $\alpha=\pi/4$, and with $R_2/R_1\approx\sqrt{2}$ (see
figure \ref{fig:annulus}(a)). In the annulus with this radii ratio,
the square orbits of the outer circle are nearly tangent to the inner
one. The orbits of the annular sector are obtained from those of the
full annulus by desymmetrization.  However, as a result of the
symmetry of the square orbits, their desymmetrizations repeat the
corresponding primitive orbits of the annular sector four times. There
are two types of primitive diffractive orbits in the annular sector,
the direct ones, arising from square orbits of the full billiard, and
glancing orbits. They are referred to as T-orbits and bounce once off
the external arc.  For $R_2/R_1 = \sqrt{2}$ they merge into a single
tangent orbit.

The eigenstates of the annular sector are given by
\begin{equation}
  \psi_{l,n}(r,\theta)= \left[ Y_l(k_{l,n}R_1) J_l(k_{l,n}r) -
  J_l(k_{l,n}R_1) Y_l(k_{l,n}r) \right] \sin (l \theta) \,,
  \label{annulusstates}
\end{equation}
for $l=4,8,12,\ldots$ and $n=0,1,\ldots$ where $k_{l,n}$ are solutions
of the secular equation
\begin{equation}
  Y_l(kR_1) J_l(kR_2) - J_l(kR_1) Y_l(kR_2) = 0 \,.
  \label{annulussecular}
\end{equation}
We calculated the eigenvalues for $R_1=1$ and $R_2=\sqrt{2}$ with
$k<2800$ (244\,397 eigenvalues) to an accuracy of $10^{-10}$,
and for several values of $R_2\approx\sqrt{2}$ in the region
$1050<k<1550$ to the same accuracy. 
In order to extract the contribution of the T-orbits to the density of states we
consider the truncated length spectrum (\ref{eq:lengthspec_win})
described in the introduction.
The weight $w(k)$ is taken in all calculations of this section to be a
Gaussian with a center $k_0$ and a variance $\sigma^2=50^2$ (see
figure \ref{fig:annulus}(b)). The peaks in $|D(x)|$ are Gaussians
with a variance $1/\sigma^2$ centered at lengths of periodic orbits.
Our choice of the variance assures that the peak
corresponding to the T-orbits is well separated from all other peaks.
Thus we can study the contribution of T-orbits without considering any
other orbit.

To find the semiclassical contribution of the T-orbits in the annular
sector, we consider a desymmetrized form of the circle Green function,
which satisfies Dirichlet boundary conditions on the straight lines of
the annular sector. It is given by
\begin{eqnarray}\fl
  \tilde G(x,y,x',y') = G(x,y,x',y') -G(x,y,x',-y') -G(x,y,y',x')
  \nonumber \\ +G(x,y,y',-x') +G(x,y,-y',x') -G(x,y,-y',-x')
  \nonumber \\ -G(x,y,-x',y') +G(x,y,-x',-y') \,.
  \label{desymgreen}
\end{eqnarray}
The contribution of orbits which bounce once off the outer arc
$\Gamma$ to the density of states is found using (\ref{nperiodic}) as
\begin{equation}
  {\rm Im} \frac{2}{\pi} \frac {\d}{\d k} \int_{\Gamma} {\rm
    d}s \, \left. \left[ \frac {\partial \tilde G}{\partial \hat n_s}
  ({\bf r}_s,{\bf r}_{s'}) \right] \right|_{s'=s} \,.
  \label{onebounce}
\end{equation}
Different terms of the desymmetrized Green function (\ref{desymgreen})
contribute to different orbits. The fourth and fifth terms are those
which contribute to the T-orbit manifold. Their contribution is the same,
as they correspond to the same orbit traversed in opposite directions.
The integrand is independent of the integration variable and the
contribution of the T-orbit manifold is therefore given by
\begin{equation}
  {\rm Im} \frac{R_2}{2} \frac {\d}{\d k} \left. \left[ \frac
  {\partial}{\partial r} G(r,0;\,0,r) \right] \right|_{r=R_2} \,.
  \label{onebounce2}
\end{equation}
For the Green function we use the expressions given in section
\ref{sec:circgreen}. It is then either of the form $A\e^{\i kL}$ or a sum of
two such terms.  To leading order, which our calculations are limited
to, both derivatives in (\ref{onebounce2}) act on the term
$\e^{\i kL}$ only.

\input{FIG/fig-annlog}
In figure \ref{fig:annlog}(a) we present the results for the exactly
tangent periodic orbit ($R_2 = \sqrt{2}$). The success of the penumbra
approximation in this case is evident. There are two contributions,
which are $\Or(k^{1/2})$ for the direct path, and $\Or(k^{1/3})$ for the
tangent path. The remaining error is very small, and
decreases like $k^{-1/3}$ (the measured slope of the error line in the
graph is -0.317), suggesting that the next order contribution comes
from the glancing part of the Green function. This could be expected since
the penumbra approximation which was used for the scattering
matrix (\ref{stransapp}) is accurate to $\Or(k^{-2/3})$.

In figure \ref{fig:annlog}(b) we present the results for the peak from
the T-orbits as a function of the external radius of the annular sector
while $k_0$ is fixed.  Starting from $R_2>R_1$, the semiclassical
approximation is seen to break down as the orbit approaches tangency.
On the other side the creeping approximation breaks down when the
creeping orbit approaches tangency. The penumbra approximation is best
at exact tangency. On both sides of the penumbra there is a region
where none of the approximations is successful. Some form of uniform
approximation would be needed to cover all regions well. Note that all the
values of $b_{\rm d}-R_1$ presented are inside the penumbra ($b_{\rm r}=
1 - k_0^{-2/3}$ for $b_{\rm d} - R_1 \approx 0.15$, see section \ref{sec:lit}). 
The criterion $\nu=1$ 
yields that the semiclassical expression for the contribution of the direct 
periodic orbit is valid when
$b_{\rm d}-R_1 > 0.035$, and the error of this approximation can be
seen to grow when $b_{\rm d}-R_1$ becomes smaller than this value.

At the beginning of this section it was stated that there is a
problem, unique to the annulus and the desymmetrizations thereof, in
accounting for orbits which bounce more than once on the exterior
circle. The length of an orbit with two consecutive creeping parts in
the annulus is unchanged if the point on the exterior circle between
these parts is allowed to vary (as long as both parts remain
creeping). Thus, an orbit bouncing $n$ times on the exterior circle
and creeping in all its segments is in fact part of a $n$ parameter
family of orbits. The same situation occurs in the penumbra region
for the tangent orbits.  As this problem is unique to the annulus, we
did not pursue it further.

\section{Diffraction effects in the quantized Sinai billiard}
\label{sec:sb}
In this section we study the Sinai billiard, i.e.\,an example for a
chaotic billiard exterior to a circle. We demonstrate the
importance of corrections to the standard Gutzwiller result due to the
diffraction on the concave part of the boundary and check
quantitatively the different approximations for the circle Green
function derived in section \ref{sec:circgreen}. Moreover we discuss
also examples for orbits, where none of the described approximations
accounts for the diffraction corrections in a satisfactory way.

The classical and creeping periodic orbits were calculated using the
minimum and the unique coding principles \cite{Bun94,Sie91}.  The
quantum data on the Sinai billiard have been obtained using the
scattering approach to quantization \cite{DS92b}. The scattering
system we consider here has been described in the introduction (see
figure \ref{fig:sinai}) and a detailed discussion of the numerical
evaluation of the corresponding scattering matrix $S(k)$ can be found
in \cite{SS95}.

However, the scattering approach to quantization does not only provide
a framework for an efficient quantization of the Sinai billiard, it
allows also to formulate the semiclassical theory in a way which is
particularly well suited to study higher order corrections to the
standard results. In order to demonstrate this we will study both
the semiclassical density of states for the Sinai billiard, and the
semiclassical approximation to the corresponding S-matrix.

\subsection{Analysis of the spectral density}
\label{sec:spdens}
We begin with the energy spectrum of the quarter Sinai billiard
with $a=1$ and $R=0.5$. The results are analyzed using the length spectrum
(\ref{eq:lengthspec_win}). Semiclassically, every periodic
orbit (manifold) contributes to $D_w(x)$ in a small vicinity of its
length, and this allows us to pinpoint individual contributions. 
The selected spectral interval $0 \leq k \leq 300$ contains 5667 levels.
The weight function $w(k)$ was taken as a Gaussian centered around
$k_0=150$ whose width is $\sigma=40$. In the figures we show
$|D_w(x)|$ which is sensitive to both amplitude and phase deviations.

\subsubsection{Exactly tangent orbits\label{sec:nr-extan}}
\input{FIG/fig-sb-ls-ab}
In figure \ref{fig:sb-ls-ab} one observes clear deviations between the
quantum (exact) and semiclassical length spectra localized near the
bouncing ball manifold at $x=2$ and its double traversal at $x=4$. The
semiclassical expression contains the leading contributions from the
bouncing ball families \cite{SSCL93}, the unstable isolated periodic
orbits and the edge orbit \cite{SSCL93}. Thus the deviations are
mainly due to penumbra diffraction. As for the annulus, we use 
an appropriately desymmetrized circle Green function 
\begin{eqnarray}\fl
\tilde G[(x, y), (x', y')] =
G[(x, y), (x', y')] + G[(x, y), (-x', y')] \nonumber \\
+G[(x, y), (x', -y')] + G[(x, y), (-x', -y')]\,,
\end{eqnarray}
in order to calculate the penumbra corrections. In the multiple
reflection expansion (\ref{nperiodic}) we concentrate on the terms
that give the bouncing ball contributions. For the shortest family and
a single traversal we consider
\begin{eqnarray}\fl
d_{\rm bb,1}(k) =
- \frac{2}{\pi} \Im \frac{\d}{\d k} \left\{
\int_{0}^{a} {\rm d}x \frac{\partial G 
[(x, y_1), (x, y_2); k]}
{\partial y_1} \right|_{y_1=(-y_2)=a} \nonumber \\ 
+ \left. \left. \int_{0}^{a}
{\rm d}y \frac{\partial G [(x_1, y), (x_2, y); k]} {\partial x_1}
\right|_{x_1=(-x_2)=a} \right\}.
\label{eq:dbb1}
\end{eqnarray}
Due to the $x \leftrightarrow y$ symmetry, the two terms are equal,
and it is enough to consider only one of them. We substitute for
$G$ its leading term approximation in the penumbra, $G\approx G_{\rm d} + G_{\rm g}$
and using equation (\ref{nsc}) and the results of \ref{app:bb-dir}, we get
\begin{eqnarray}
d_{\rm bb,1}(k) &=& d_{\rm bb,1}^{\rm d}(k)+ d_{\rm bb,1}^{\rm g}(k) \\ 
d_{\rm bb,1}^{\rm d}(k) &=&  \frac{2(a-R) a^{1\over2}k^{1\over2}}
{\pi^{\frac{3}{2}}} \cos \left(2ka - \frac{\pi}{4} \right) \\ 
d_{\rm bb,1}^{\rm g}(k) & = & - \frac{2c}{\pi^\frac{3}{2}} \frac{L^{\frac{1}{2}}
R^{\frac{1}{3}}}{k^{\frac{1}{6}}} \cos \left(2ka + \frac{\pi}{12}
\right).
\label{eq:dbb1-g}
\end{eqnarray}
The first term is the semiclassical contribution due to the bouncing
ball family \cite{SSCL93}. The second term is a genuine diffractive
contribution, that can be attributed to the exactly tangent orbit at
the closure of the bouncing ball manifold near the circle. The
contribution of this exactly tangent orbit is $\Or(k^{-1/6})$
which is slightly smaller than $\Or(k^0)$ for unstable periodic orbits.
 
In figure \ref{fig:sb-ls-ab}(a) we present a portion of the length
spectrum near $x=2$. The differences between the quantum and the
semiclassical predictions (including bouncing ball contributions) are
about 10 percent. If we supplement the semiclassical expression with
the tangent contribution in (\ref{eq:dbb1-g}), the deviation reduces
to about 1 percent. This clearly assesses the penumbra theory for
exactly tangent orbits.

A more complicated situation arises for the double repetition of the
above bouncing ball family. The semiclassical contribution is given by
an integral over a multiplication of two circle Green function, each
decomposed into $G_{\rm d} + G_{\rm g}$, which results in three terms:
\begin{eqnarray}
\label{eq:dbb2}
\fl d_{\rm bb,2}(k) =
- \frac{4}{\pi} \Im \frac{\d}{\d k} \int_{0}^{a}
{\rm d}x_1 {\rm d}x_2 \left. \frac{\partial G [(x_1, y_1), (x_2,
y_2)]}{\partial y_1} \frac{\partial G [(x_2, y_2), (x_1,
y_1)]}{\partial y_2} \right|_{y_1=(-y_2)=a} \nonumber \\
\lo = d_{\rm bb,2}^{\rm dd}(k) + d_{\rm bb,2}^{\rm dg}(k) + d_{\rm bb,2}^{\rm gg}(k).
\end{eqnarray}
The terms are interpreted as ``direct-direct'', ``direct-glancing''
and ``glancing-glancing'' contributions, with obvious
notation. Straightforward calculations (see also 
\ref{app:bb-dir})
give
\begin{eqnarray}
\label{eq:dbb2-dd}
d_{\rm bb,2}^{\rm dd}(k) & = &
\frac{\sqrt{2}(a-R)}{\pi^{3/2}} a^{1/2}k^{1/2}\cos
\left(4ka - \frac{\pi}{4} \right) - \frac{a}{\pi^2} 
\cos (4ka) \\
\label{eq:dbb2-dg}
d_{\rm bb,2}^{\rm dg}(k) & = & 
- \frac{\sqrt{2}c}{\pi^{3/2}} a^{1/2} R^{1/3} 
k^{-{1/6}} \cos \left(4ka + \frac{\pi}{12} \right) \\
\label{eq:dbb2-gg}
d_{\rm bb,2}^{\rm gg}(k) & = &
\frac{c^2}{\pi^2} R^{2/3} k^{-{1/3}} \cos \left(4ka +
\frac{\pi}{6} \right)\,.
\end{eqnarray}
The results for $d_{\rm bb,1}^{\rm g}$, $d_{\rm bb,2}^{\rm dg}$ and $d_{\rm bb,2}^{\rm gg}$
agree with the general expression (\ref{nsc}) and can be interpreted as
the contributions from isolated orbits which are exactly tangent to
the circle.  The term $d_{\rm bb,2}^{\rm dd}$ contains the semiclassical
contribution of the bouncing ball family and an interesting correction
which is of the same order as an unstable periodic orbit. This
correction comes from the non-zero average of the squared Fresnel
factor and thus is of diffractive origin. There is no unstable
periodic orbit that gives this contribution. These predictions are
fully verified against the numerical data shown in
figure \ref{fig:sb-ls-ab}(b). Indeed, the penumbra corrections reduce
the deviations very significantly near $x=4$.

\subsubsection{Almost tangent orbits}
\label{sec:ato}
\input{FIG/fig-sb-ls-cd}
We now turn to the investigation of almost tangencies, that occur in
generic isolated and unstable periodic orbits. In figure
\ref{fig:sb-ls-cd} we indicated a few periodic orbits for which there
are significant deviations due to penumbra effects. We choose to
concentrate on the pair of periodic orbits of lengths $5.10845$
(direct) and $5.10908$ (glancing) which are plotted in figure
\ref{fig:sb-ls-cd}(a). The orbits are geometrically similar, except
that the direct orbit has a segment that just misses tangency with the
circle, while for the glancing orbit the corresponding segment
reflects from the circle in a very shallow angle. This angle of
reflection is about $1^\circ$, which is well inside the penumbra and
fully justifies the implementation of the corrections. To calculate
the corrected contributions, we replaced the semiclassical Green
functions with their penumbra counterparts $G_{\rm d}$ and $G_{\rm g}$
(see (\ref{greentransd}), (\ref{greentranst})). For the direct orbit,
the only change was a multiplication of the standard semiclassical
contribution by by a Fresnel factor, whose value was
\[
\frac{F(\infty) - F(\nu \approx -0.35)}{\sqrt{2 \i}} 
\approx 0.71 \e^{-0.23 \i}
\]
for $k=k_0=150$. Including this correction reduces the deviations
significantly, as can be clearly seen from
figure \ref{fig:sb-ls-cd}(a). To account for the glancing corrections
we use the approximation (\ref{corrfac}).
While the difference in the lengths of the classical orbit and the 
corresponding creeping orbit (with negative creeping angle) is very small  
($\approx 10^{-6}$), the prefactor
significantly grows by a factor of $\approx 4.7$. The
contribution of the glancing orbits further reduces the deviation by a
factor of $2$, as seen in the figure.

\subsubsection{Ghost orbits}
\label{sec:ghost}
One of the most interesting applications of penumbra corrections is
for ghost and creeping orbits, which are classically forbidden. 
Ghost orbits which are almost tangent in the shadowed
part of the penumbra, are expected to give appreciable contributions,
comparable to standard semiclassical contributions of real periodic
orbits with similar lengths. To find such ghost orbits in the Sinai
billiard, one needs therefore to look at periodic orbits that are
pruned at a radius slightly smaller than $R=0.5$ which we use for the
quantum results. Indeed, we observe a pair of geometrically similar
periodic orbits that coalesce and prune at $R \approx 0.48$. After
enlarging the radius back to $R=0.5$, we get a pair of
``direct-shadowed'' and ``glancing-shadowed'' penumbra orbits (see
figure \ref{fig:sb-ls-cd}(b)). The lengths of the orbits are $\approx
5.2409$ and $5.2413$, respectively. The creeping angle of the glancing
orbits is $\approx 1.4^{\circ}$, which is small enough to justify the
penumbra approximation. The Fresnel parameter $\nu$ for the direct
orbit is $\approx 0.25$, that gives a multiplicative Fresnel factor
$\approx 0.39 \exp (0.31\,\i)$ which indeed indicates almost tangency.
The direct contribution is by a factor $\approx 3$ larger (and with opposite sign)
than that of the glancing orbit, and thus we should expect to see a
noticeable peak in the length spectrum. Our expectations are
fulfilled, as can be seen in figure \ref{fig:sb-ls-cd}(b). We can
identify a peak in the quantum length spectrum near $x=5.24$ with
large deviations between the quantum and the standard semiclassical
results.  They correspond to the ghost orbits, and if we include
ghost contributions, the deviations significantly decrease, which
indicates the success of the theory. We tried to na\"{\ii}vly implement
the geometrical theory of diffraction as in \cite{VWR94}, which takes
into account only the creeping orbit. Summing over many creeping modes
to get a convergent answer, we obtained large deviations from the
quantum results in the length spectrum, as expected due to the small
creeping angle. 

\subsection{The semiclassical S-matrix}
\label{sec:stod}
In this section we consider the semiclassical approximation to the
S-matrix involved in the scattering quantization. As discussed in the introduction,
all the spectral information on the billiard is contained in
the total phase and all the traces of the S-matrix. We would like to
show how this information may be extracted and used to study very fine
details of the spectrum which would otherwise not be accessible.  The
fact that the S-matrix is a continuous function of $k$ rather than a sum
of delta peaks necessitates an analysis which is slightly different
from that presented in the sections \ref{sec:annulus} and \ref{sec:spdens}.  The central
spectral quantity which we consider here is the number counting
function $N(k)=\int^k {\rm d}k'\,d(k')$ rather than the density of
states. According to (\ref{eq:density}) it can be decomposed as 
\begin{equation}\label{ns}
N(k) = N^{(R)}(k)+\sum_{n=1}^\infty\,N^{(n)}(k)
\end{equation}
with
\begin{equation}\eqalign{
N^{(R)}(k)&\equiv{1\over 2\pi}\Theta(k)\,,\\\bs
N^{(n)}(k)&\equiv{(-1)^n\over n\pi}{\rm Im}\, {\rm Tr}\, S(k)^n\,.
\label{ntns}}
\end{equation}
Each of the terms in this decomposition can be analyzed separately. We
will concentrate here on the first two terms $N^{(R)}$ and $N^{(1)}$.
$N^{(R)}$ will be referred to as resonance counting function for
reasons explained in \cite{DS92a}.  Beside the oscillating
contributions from trapped periodic orbits it contains also a smooth
part which is identical to the smooth part of the billiard spectrum up
to the third term in the expansion of Weyl's law . This is not
completely obvious, since it is known for the scattering from the
outside of a billiard \cite{SU96} that in general the above statement
holds only for the area term. For the Sinai billiard we have clarified
this point in the introduction by considering the limiting case of an
empty square billiard where an analytic expression for the total phase
is available.

\input{FIG/fig-ls-cmp}
In order to pinpoint the contributions from individual periodic
orbits we consider again length spectra which are now obtained by a
fast Fourier transform using the Welch window \cite{numrec} and based
on discrete points with a spacing $\Delta k$.  Note that due to the
discreteness of the transformation the contributions from long orbits
appear at $x=L^{(\rm P)}\,{\rm mod}\,2\pi/\Delta_k$ and may interfere with the
contributions from the shorter orbits of interest.  An example is
provided in figure \ref{ls-cmp}, which displays the length spectra
obtained from the oscillatory parts of $N(k)$ (top), $N^{(1)}(k)$
(middle) and $N^{(R)}$ (bottom) using a logarithmic scale.  From the
number counting function we obtain a few well pronounced peaks at
short lengths which are on top of a very large background containing
the contributions from all long orbits. Although these long orbits are
extremely unstable, the exponential proliferation of orbits ensures
that the amplitude of their combined effect diminishes only slightly
with the orbit length.

\input{FIG/fig-orb-tp}
\input{FIG/table}
The situation is different for the trace and the total phase of S,
where the number of contributing orbits does not grow exponentially
with the length. Nevertheless the long orbits become exponentially
unstable and this is why we see almost no background in the lower two
parts of figure \ref{ls-cmp}. So the observation of very fine
structures in the length spectrum becomes possible, e.g.\,the tiny peak at
$x\approx 3$ in the spectrum of the resonance counting function, which is
due to diffraction effects as we shall explain in the sequel. It would
hardly be possible to discover and study such peaks among all the
leading order contributions from the unstable periodic orbits with a
similar length. The resonance counting function is particularly well
suited for a semiclassical analysis, since it contains contributions
only from the very sparse set of trapped orbits, i.e.\,those orbits
which never hit the section $\Gamma_{\rm S}$ of figure \ref{fig:sinai}. Some of
these orbits -- classical and diffractive -- are displayed in
figure \ref{fig:orb-tp}.
 
The most important of the oscillatory contributions is of order
$k^{1/2}$ and comes from the family of neutral bouncing ball orbits {\bf bb}
with length $2a$ and its multiple traversals. It is
responsible for the large peaks at $x=2$ and $x=4$ in the length
spectrum of the resonance counting function which is displayed in fig.\
\ref{fig:tp-d05} for $a=1$ and $R=0.5$.
The dashed line is obtained by subtracting Weyl's law from
the exact quantum result and contains therefore all oscillatory
contributions to $N^{(R)}(k)$.

\input{FIG/fig-tp-d05}
Disregarding diffraction but including the edge orbit {\bf e} running along the
right billiard wall, the contribution of the bouncing ball family has
been derived in \cite{SSCL93} and is also contained in
equation (\ref{eq:finphase}). Beside this, the only standard semiclassical
contribution to the total phase comes from the isolated unstable orbit
{\bf u} running along the left billiard boundary.  The semiclassical
amplitude of such an edge orbit has been derived e.g.\,in
\cite{Sie91}. For the thin solid lines in figure \ref{fig:tp-d05} all standard
semiclassical contributions have been subtracted.  The remaining peaks
in the solid line are now exclusively due to higher order corrections
to Gutzwiller's result.  At the length of the unstable orbit {\bf u} ($x=1$)
this is sufficient to reduce the amplitude of the peak by more than
two orders of magnitude. The subtraction of the leading order
semiclassical expression is less successful at the bouncing ball
lengths, since there we have very large diffraction corrections, which
are explicitly given in (\ref{eq:dbb1-g}) and
(\ref{eq:dbb2-dd})-(\ref{eq:dbb2-gg}). All these terms have been
subtracted from the quantum mechanical data to obtain finally the
thick solid line. Indeed, the magnitude of the peaks at the bouncing
ball lengths is now also reduced considerably.

\input{FIG/fig-kd}
The analysis can be further supported by considering the dependence of
the magnitude of the peaks on $k$ as displayed in figure \ref{fig:kd}
using a double logarithmic scale. The curves are obtained by
restricting the Fourier transformation of the data to small intervals
centered around a mean $k$ given in units of $\pi/a$ at the abscissa
of the plots.  The width of the intervals is chosen large enough to
guarantee a sufficient resolution in the length spectrum. The
approximate exponents $a$ for the $k$-dependence of the peaks in the
semiclassical domain have been determined by a linear fit to the
curves at high values of $k$ and are given next to the curves.  In
figure \ref{fig:kd}(a) we demonstrate in this way that the contribution
from a standard periodic orbit is correctly described by the
Gutzwiller formula up to corrections of the order $k^{-1}$.  At $x =
2$ (b) and $x = 4$ (c) the dashed line corresponding to the
oscillating part of the resonance counting function is close to a
$k^{1/2}$-behaviour which indicates that the peak is dominated by the
bouncing ball contribution. The exponent for the peak at $x=2$ in the
solid line is close to $-1/6$ which agrees with the penumbra
contribution of the tangent orbit (\ref{eq:dbb1-g}). When this
expression is also subtracted from the data (fat line) we are left
with a peak whose magnitude is much smaller and moreover falls off
faster than $k^{-1/2}$, i.e.\,the description of the tangent orbit
is quantitatively correct and further corrections are of lower
order.  The situation is similar at $x = 4$.  However the leading
order correction to the bouncing ball result is now $\sim k^0$ and
does not come from an isolated tangent orbit but from the diffraction
correction in (\ref{eq:dbb2-dd}) to the family itself. When all
diffraction effects have been subtracted (fat line) the height of the
peak is reduced by an order of magnitude, but the peak does not fall
off as fast as for $x=2$. This apparent discrepancy is at least
partially due to the finite available $k$-range since it disappears
gradually as the interval of computation is further enlarged (not
displayed). We also recall that another error may be due to the
explained ``folding'' of the peaks from long orbits on top of the
length spectrum.

Beside the discussed corrections to the standard semiclassical
contributions, the solid line in figure \ref{fig:tp-d05} displays a
number of additional peaks which are exclusively due to diffractive
orbits. The open Sinai billiard supports four families of primitive
creeping periodic orbits. The two shortest members from each of these
families are displayed in figure \ref{fig:orb-tp}. These are orbits
which are not due to a bifurcation of the type described in the
introduction or in the last section. Rather they remain creeping even
for very small radius.  The creeping length of the orbit $c_1^{(0)}$
is exactly zero, and therefore this orbit has always to be treated
within the penumbra approximation and results in (\ref{eq:dbb1-g}).
The other orbits are in the penumbra or deep shadow region depending
on the value of $k$. Table \ref{tab:orb-tp} contains the necessary
geometrical information to evaluate the contributions from the
shortest members of the four families.  In figure \ref{fig:tp-d05} the
lengths of the creeping orbits are indicated with vertical lines and
the most important of them are denoted on top of the plots.  It is
clearly seen that each peak in the length spectrum corresponds to a
particular periodic orbit.  The parameters are such that the creeping
orbits are well described by the deep shadow approximation throughout
the whole $k$ range.  Indeed the magnitude of the peaks in the thin
solid line is reduced when the contribution of these orbits is
subtracted according to (\ref{nsc}) (fat curve), although the
reduction is not as striking as for the exactly tangent orbits.

\input{FIG/fig-tp-n05}
\input{FIG/fig-tp-n08}
From the expressions derived in the 
\ref{sec:mixed} we expect
that in the case of Neumann boundary conditions the creeping orbits
have a larger amplitude than in the case of Dirichlet boundary
conditions. Indeed the peaks at the corresponding lengths in
figure \ref{fig:tp-n05} (Neumann b.\,c.) are more pronounced than in
figure \ref{fig:tp-d05} (Dirichlet b.\,c.) and the success of the deep
shadow approximation for these orbits is even more evident (note the
different scaling of the ordinate according to the different magnitude
of the peaks).  It is also possible to observe and correctly account for a
peak at $x=\pi/2$ due to the circumference orbit {\bf C} which never
leaves the circle at all.  As the radius of the circle grows, the
lengths of the creeping orbits become closer to each other and the
length spectrum is more complex (figure \ref{fig:tp-n08}).
Nevertheless we see that the magnitude of the peaks can be reduced
considerably, when the semiclassical contributions are subtracted
according to (\ref{nsc}).  Unlike classical orbits, an arbitrary
combination of primitive creeping orbits can be joined to form a new
periodic orbit which also gives a contribution to the spectrum. Of
particular importance are the combinations including the tangent orbit
$\bf c_1^{(0)}$, since then the creeping angle is relatively small. An
example for a creeping orbit of this type is the peak denoted with
$\bf c_1^{(0)}+c_1^{(2)}$ in figure \ref{fig:tp-n08}.

\input{FIG/fig-trs}
\input{FIG/fig-trs5}
\input{FIG/fig-orb-tr}
Now we would like to discuss the semiclassical approximation to ${\rm
Tr}\, S$ in more detail. Again we will show that important corrections
to the leading order semiclassical result are due to diffraction
effects. In figure \ref{trs} we display the length spectrum of ${\rm
Tr}\, S$, which can be expressed semiclassically in terms of all
periodic orbits of the Sinai billiard hitting the section $\Gamma_{\rm S}$
exactly one time.  This includes the contributions from the two
bouncing ball families present in the Sinai billiard with $R=0.5$
which result in the two most prominent peaks in the spectrum at
$x=2$ and $x=2\sqrt{2}$. Most of the other
large peaks can be related to isolated unstable periodic orbits as it
can be seen from the vertical bars in the upper part of the figures
which denote the lengths of all the classical orbits contributing to
$S$. The remaining difference between the semiclassical approximation
and the quantum data after the leading order contributions from all
classical orbits have been included according to the standard
Gutzwiller result is displayed using the grey shaded areas. We see
that the semiclassical result accounts very well for some of the peaks
at small lengths: all the peaks up to $x=4$ are reduced by the
subtraction of the semiclassical result by at least a factor of 10
(note the logarithmic scale). The result becomes increasingly worse as
the length grows and from $x\approx 7$ the leading order semiclassical
theory fails completely.

In order to explain where the deviations come from we have further
enlarged the region $4\le x\le 5$ in figure \ref{trs5}. There are 12
isolated classical orbits in this interval which are displayed in
figure \ref{orb-tr}. We observe large deviations between the
semiclassical result and the quantum data for the orbits with $L=4.12,
4.22, 4.61$ and $4.62$. These orbits have in common that one of their
reflections from the circle is very shallow, i.e.\,the classical
description excluding diffraction effects approaches the limits of its
validity. However, unlike the orbits discussed in the context of the
spectral density, where a considerable improvement of the
semiclassical result could be achieved when the penumbra corrections
are taken into account, the orbits of figure \ref{orb-tr} are not really
well inside penumbra but rather at the border between the illuminated
region and the penumbra. Consequently the results of section
\ref{sec:tod} do not apply. As already mentioned at the end of
section \ref{sec:annulus}, a uniform geometric theory of diffraction would be
needed in order to correct the contributions from such orbits.

The peaks at $4.47$ and $4.75$ do not correspond to classical orbits
but represent the contribution from creeping and ``ghost'' orbits.
$L=2\sqrt{5}a=4.472$ is the length of the first completely
shaded bouncing ball family, which does not give a contribution to
leading order. However, an important diffraction correction resulting
in the observed peak in the length spectrum is due to those orbits in
the family which traverse the penumbra. 

\input{FIG/fig-lorb-trs}
Finally we turn to the conspicuous peaks in the length spectrum of
${\rm Tr}\, S$ which are located in the vicinity of $x=8, 10,
12,\dots$ and can be explained by the orbits displayed in
figure \ref{lorb-trs}. Below the orbits the length and the dominant
eigenvalue of the monodromy matrix are given.  The orbits become very
unstable as the length increases and the standard Gutzwiller
expression fails completely to predict the amplitude of their
contribution, which is due to the almost tangent reflection from the
circle. Although the penumbra approximation is much better and
predicts at least the order of magnitude of the contributions, it is
not capable of giving a satisfactory quantitative description and
therefore not displayed. While for the shorter orbits the reflection
from the circle is not yet inside the penumbra, the problem with the
longer orbits is that more and more of the straight segments are close
to the circle and need an additional diffraction correction which
leads to an increasing error. Moreover additional corrections due to
intermittency which were recently derived in
\cite{Tan96} may be necessary to predict the amplitudes correctly
since the long orbits are very close in phase space to the family {\bf
  bb} of bouncing ball orbits transversal to the channel.

\section{Frequency of penumbra traversals}
\label{sec:penstat}
The numerical examples of section \ref{sec:spdens} illustrated the
success of the theory derived in section \ref{sec:tod} to account for the
significant penumbra corrections for particular periodic orbits. A
natural question would be: how many periodic orbits should be
corrected in this way? To answer this question we should consider two
factors. First, the borderlines of the penumbra are $k$ dependent and
the fraction of phase space occupied by the penumbra is of order
$(kL)^{-2/3}$, where $L$ is a typical length of the billiard. This
means that the penumbra shrinks to $0$ as $k
\rightarrow \infty$. In particular, we can also conclude that any  
given periodic orbit (that has no exactly tangent segments) will be
outside the penumbra for $k$ large enough. The second factor to
consider is that in order to quantize the billiard up to a wavenumber
$k$ with a resolution of the mean level spacing one needs to consider periodic
orbits up to the Heisenberg length $L_{\rm H} \approx kL^2$, or, equivalently up
to number of bounces $n_{\rm H} \approx kL$. This enhances the chance to
visit a given area of phase space as $k$ grows. To obtain the overall
effect of these two contradicting trends, let us consider for each
periodic orbit $\Delta \equiv \min_j |l_j-kR|$, where $l_j$ is the
angular momentum of the $j$th segment of the orbit with respect to the
circles center. The orbit traverses the penumbra at least once if
$\Delta \lesssim (kR)^{1/3}$. If we assume ergodicity, then each
segment of the orbit has an a-priori probability
\begin{equation}
p \approx (kL)^{-2/3}
\end{equation}
to traverse the penumbra. Assuming statistical independence of the
segments, and homogeneous coverage of phase space by long periodic
orbits, the probability that an orbit with $n$ bounces avoids the penumbra is
\begin{equation}
(1-p)^n \approx \exp \left[ -n (kL)^{-2/3} \right] = \exp \left(
-\frac{n}{n_{\rm H}^{2/3}} \right).
\label{pavoid}
\end{equation}
Because of the exponential proliferation of periodic orbits, the
overwhelming majority of periodic orbits satisfy $n\approx n_{\rm H}$ and
thus $n_{\rm H}^{2/3}\lesssim n$.  This means according to (\ref{pavoid})
that in the semiclassical limit {\em most} of the periodic orbits
traverse the penumbra at least once, and for them the semiclassical
approximation fails and should be corrected. To emphasize this point
we rephrase our findings about the semiclassical limit in the
following way:
\begin{itemize}
\item Given a periodic orbit (which is not exactly tangent), its
standard semiclassical contribution is recovered for $k$ large enough.
\item For a given $k$, most periodic orbits which are shorter than
$\approx n_{\rm H}^{2/3}$ are described by the standard
semiclassical approximation.
\item Although their number grows with $k$, their fraction out of the 
relevant periodic orbits becomes smaller, and the great majority of
periodic orbits are affected by penumbra corrections.
\end{itemize}
\input{FIG/fig-pdelta}
To verify these ideas we calculated $\Delta/(kR)$ for all periodic
orbits of length up to $10$ of the quarter Sinai billiard. There were
$20\,150$ primitive orbits, with total of $320\,002$ segments. The
coarse--grained distribution is shown in figure \ref{fig:pdelta}. It
is sharply peaked near the minimal value $\Delta = 0$, which indicates
that almost all periodic orbits include a nearly tangent chord, as
predicted. To get a quantitative estimate, let us take the Heisenberg
length to be the maximal periodic orbit length: $L_{\rm H} = 10$. We need to
estimate the relevant $k$. We use the definition of the Heisenberg
time
\begin{equation}
T_{\rm H}(E) = h \bar{d}(E)\,,
\end{equation}
which for billiards can be written as
\begin{equation}
2 \pi \bar{d}(k) = L_{\rm H}(k).
\end{equation}
If we use the leading order expression for $\bar{d}(k)$ for billiards
\begin{equation}
\bar{d}(k) = \frac{A k}{2 \pi}\,,
\end{equation}
where $A$ is the area of the billiard we get
\begin{equation}
k = \frac{L_{\rm H}}{A} \approx 12.4
\end{equation}
in our case. The relative phase space area occupied by the penumbra is
estimated as
\begin{equation}
p \approx \frac{2 (k R)^{\frac{1}{3}}}{k b_{max}} \approx 0.21\,,
\end{equation}
where $b_{max} = \sqrt{2}$ is the maximal impact parameter in the
billiard. The factor $2$ is due to taking into account impact
parameters which are both larger and smaller than $R$. The number of
bounces is estimated by
\begin{equation}
n_{\rm H} = \frac{L_{\rm H}}{c} = \frac{L_{\rm H} \gamma}{\pi A} \approx 15\,,
\end{equation}
where $c$ is the mean chord and $\gamma$ is the billiards
perimeter. Due to the exponential proliferation of orbits, their
majority will have a length close to $L_{\rm H}$ and thus $n_{\rm H}$
chords. The probability of such orbits to avoid the penumbra is
\begin{equation}
q = (1-p)^{n_{\rm H}} \approx 0.03
\end{equation}
and consequently about $97\%$ of the orbits are expected to
traverse the penumbra. The penumbra borders in terms of $\Delta/(k R)$
are estimated as
\begin{equation}
\frac{\Delta}{k R} \leq (k R)^{-\frac{2}{3}} \approx 0.3
\end{equation}
which includes according to the numerical data $95.7\%$ of the orbits,
in good agreement with the theory.

\section{Conclusions}
\label{sec:conc}
In this paper we have considered the semiclassical quantization of
billiard systems. Using asymptotic approximations to the circle Green
function we have derived corrections to the standard Gutzwiller
formula which account for the quantum diffraction from concave parts
of the billiard boundary in the penumbra. This is the nearly tangent
parameter region where neither the leading order semiclassical result
nor the deep shadow approximation \cite{LK59,VWR94} are valid. There
are two types of corrections: the first can be expressed in terms of
nearly tangent classical periodic orbits (including ``ghost'' orbits
which cut straight through the concave billiard boundary). The
contributions from these orbits differ from the Gutzwiller result by a
prefactor of the order one, i.e.\,the correction can be as large as
the standard semiclassical amplitude itself. The other type of
correction can be expressed in terms of creeping orbits including
those with negative creeping angles.  Although these orbits are also
obtained from the deep shadow approximation, their contribution is
different in the penumbra: the amplitude does not decay exponentially
in $k$ but only as $k^{-1/6}$ for each creeping segment of the orbit.
The appearance of this new type of orbits in the semiclassical
quantization formulas raises questions about the structure of the set
of all creeping orbits, e.g.\,how they can be computed using an
extremum principle together with a code and if there is a one to one
correspondence between the nearly tangent classical orbits and the
creeping counterparts by which they have to be replaced in the case of
penumbra diffraction. Although we have some preliminary results for
the Sinai billiard, the answer to these questions is beyond the scope
of the present paper.

The derived corrections to the standard Gutzwiller result have been
tested in both, the integrable annular billiard and the chaotic Sinai
billiard. The success of the theory was obvious as long as the orbits
under consideration were indeed well inside the penumbra region. 

In the case of the Sinai billiard we have used beside the length
spectrum obtained from the set of billiard eigenvalues also an
alternative method of spectral analysis which is based on the
scattering approach to quantization. Here discrete length spectra
are directly obtained from the total phase or the traces of the
involved S-matrix which has the advantage that only limited subsets of
periodic orbits contribute. The resulting sparse spectrum is
particularly well suited to observe deviations from the standard
Gutzwiller result and we could in this way check the quality of the 
penumbra and the deep shadow approximation to a high precision.

The importance of penumbra diffraction corrections becomes obvious if
one estimates the number of orbits which are affected by them. It
turns out that the amplitudes of most of the orbits contributing to
the spectrum up to some fixed value $k$ must be corrected. In view of
this fact it would certainly be desirable to further extend the
results which we presented in this paper to the parameter regions,
where none of the so far known expressions is sufficiently accurate.

\ack

The research reported in this work was supported by grants from the US
Israel Binational Science Foundation and the Minerva Center for
Nonlinear Physics. HS acknowledges support from the DFG and the
Minerva Foundation.

\appendix

\section{The circle Green function with Neumann and mixed
  boundary conditions}
\label{sec:mixed}
The different approximations to the circle Green function are
found in a manner similar to section \ref{sec:circgreen} when
homogeneous boundary conditions are imposed on the circle.  In this
appendix we give the modifications to the expressions derived in
section \ref{sec:circgreen} for this case. Equations
(\ref{greensum}-\ref{greenint}) are still the starting point for all
calculations, and the boundary conditions affect only the scattering matrix
$S_l(kR)$.

For Neumann boundary conditions ($\partial_{\bf \hat n}\psi({\bf r})=0$)
the scattering matrix is given by
\begin{equation}
  S_l(kR) = -\frac{H_l^{-'}(kR)}{H_l^{+'}(kR)} \,.
  \label{neusmatrix}
\end{equation}
In the lit region, $G_{\rm d} ({\bf r},{\bf r'})$ (\ref{greenlitd}),
the contribution of the direct path, is unaffected, while the
contribution of the reflected path $G_{\rm r} ({\bf r},{\bf r'})$
(\ref{greenlitr}) is multiplied by $-1$. In the deep shadow region,
the contribution of the creeping path $G_{\rm c}^{(0)} ({\bf r},{\bf
r'})$ is still given by (\ref{greencreep}-\ref{creepcoeff}), but with
different poles $l_n$ and residues $r_n$. The poles are given by
(\ref{spoles}), where in this case $x_n$ are the zeros of ${\rm
Ai}'(x)$. The residues are given by
\begin{equation}
  r_n = \frac {\e^{-\i \pi / 6}}{2\pi x_n {\rm Ai}(-x_n)^2} \left(
  \frac{kR}{2} \right)^\frac {1}{3} \,.
  \label{neuresidues}
\end{equation}
In the penumbra the contribution of the direct path $G_{\rm d} ({\bf
r},{\bf r'})$ (\ref{greentransd}) is unaffected by the boundary
conditions. The constant $c$ in the contribution from the tangent path
$G_{\rm g} ({\bf r},{\bf r'})$ (\ref{greentranst}) is now given by
\begin{equation}
  c = - \frac {1}{2^{1/3}} \int_{-\infty}^0 {\rm d}x \, \frac {{\rm
      Ai}' (x\e^{-2\pi \i/3})}{{\rm Ai}'(x\e^{2\pi \i /3})} + \frac
  {\e^{-2\pi \i/3}}{2^{1/3}} \int_0^\infty {\rm d}x \, \frac {{\rm Ai}'
    (x)}{{\rm Ai}'(x\e^{2\pi \i/3})} \,.
  \label{neucint}
\end{equation}
The first term is again the complex conjugate of the second
\cite{RW56}, and by a numerical integration over the second term we
have
\begin{equation}
  c \approx -0.864251443481 \,.
  \label{neuc}
\end{equation}

We now consider the general case of mixed boundary conditions
($\kappa\psi({\bf r})+\partial\hat n\psi({\bf r})= 0$). Neumann
boundary conditions are the case of $\kappa=0$, and Dirichlet boundary
conditions are the limit $\kappa\rightarrow\infty$. The semiclassical
quantization for mixed boundary conditions is considered in
\cite{SPS+95}. The scattering matrix is given by
\begin{equation}
  S_l(kR)=-\frac{\kappa H_l^-(kR)-kH_l^{-'}(kR)}{\kappa
    H_l^+(kR)-kH_l^{+'}(kR)} \,.
  \label{mixedsmatrix}
\end{equation}
In the lit region $G_{\rm d} ({\bf r},{\bf r'})$ (\ref{greenlitd}) is again
unaffected. The contribution of the reflected path, $G_{\rm r} ({\bf r},{\bf
  r'})$ (\ref{greenlitr}) is multiplied by a phase $\e^{\i\phi}$, where
$\phi$ is given \cite{SPS+95} by
\begin{equation}
  \phi = 2 \arctan \left( \frac {k}{\kappa} \cos \theta \right) \,.
  \label{mixedphase}
\end{equation}
Note that for $\kappa\rightarrow\infty$ (Dirichlet) $\phi=0$, and for
$\kappa=0$ (Neumann) $\phi=\pi$. The poles $l_n$ for the contribution
of the creeping path $G_{\rm c}^{(0)} ({\bf r},{\bf r'})$
(\ref{greencreep}-\ref{creepcoeff}) in the deep shadow region are
still given by (\ref{spoles}), where in this case $x_n$ are the
solutions of
\begin{equation}
  \frac{{\rm Ai}'(-x)}{{\rm Ai}(-x)} = \e^{\i\pi/3} \frac{\kappa}{k}
  \left( \frac{kR}{2} \right)^\frac{1}{3} \,.
  \label{mixedzeq}
\end{equation}
The residues are given by
\begin{equation}
  r_n = \frac {\e^{-i\pi/6}}{2\pi \left[ {\rm Ai}'(-x_n)^2 + x_n {\rm
      Ai}(-x_n)^2 \right] } \left( \frac{kR}{2} \right)^\frac {1}{3}
  \,.
  \label{mixedresidues}
\end{equation}
In the penumbra $G_{\rm d} ({\bf r},{\bf r'})$ (\ref{greentransd})
is again unaffected. The constant $c$ in $G_{\rm g} ({\bf r},{\bf r'})$
(\ref{greentranst}) must be replaced by the $k$-dependent expression
\begin{equation}
  c \left(\frac{\kappa}{k} \left[ \frac{kR}{2} \right]^\frac{1}{3}
    \right) \,.
  \label{mixedc}
\end{equation}
After some manipulations $c(z)$ is given in a form convenient for
numerical integration by
\begin{eqnarray}
\fl  c(z) = \frac{1}{2^{1/3}} \int_0^\infty {\rm d}x \, \frac {z {\rm
      Ai} (x) + \e^{-2\pi \i/3} {\rm Ai}' (x)}{z {\rm Ai} (\e^{-2\pi \i/3}x)
    + \e^{2\pi \i/3} {\rm Ai}' (\e^{-2\pi \i/3}x)} \nonumber \\ +
  \frac{1}{2^{1/3}} \int_0^\infty {\rm d}x \, \frac {z {\rm Ai} (x) +
    {\rm Ai}' (x)}{z {\rm Ai} (\e^{2\pi \i/3}x) + \e^{2\pi \i/3} {\rm Ai}'
    (\e^{2\pi \i/3}x)}\,.
  \label{mixedcz}
\end{eqnarray}

For some applications it is convenient to express mixed boundary
conditions as $b\cos\alpha\psi({\bf r}) + \sin\alpha\partial_{\hat
n}\psi({\bf r}) =0$. The parameter $\alpha$ interpolates between
Dirichlet ($\alpha=0$) and Neumann ($\alpha=\pi/2$) boundary
conditions. In \cite{SPS+95} the derivative
\begin{equation}
  \left. \frac{\partial}{\partial\alpha} d(k; b,\alpha)
  \right|_{\alpha=0}
  \label{densityderiv}
\end{equation}
is introduced as a tool for analyzing the spectrum of the Sinai
billiard. It was conjectured there that the derivative of tangent
contributions should semiclassically vanish.
Using (\ref{mixedcz}) we find that
\begin{equation}
  \left. \frac{\d}{\d\alpha} c \left(\frac{b}{k} \left[
    \frac{kR}{2} \right]^\frac{1}{3} \cot\alpha \right)
  \right|_{\alpha=0} = \e^{2\pi \i/3} \frac{k}{b} \frac{1}{(kR)^{1/3}} \,,
  \label{cderivative}
\end{equation}
which together with (\ref{eq:dbb1-g}), (\ref{eq:dbb2-dg}) and
(\ref{eq:dbb2-gg}) indeed gives a contribution $\Or(k^{-1/2})$ smaller
than for a standard unstable periodic orbit.

\section{Calculation of direct contributions}
\label{app:bb-dir}

In this appendix we calculate the purely ``direct'' contributions to
$d_{\rm bb, 1}(k)$ and $d_{\rm bb, 2}(k)$ that appear in section
\ref{sec:nr-extan}. The purpose is to find genuine diffractive
(penumbra) contributions beyond the semiclassical ones
\cite{SSCL93}. There is an inherent difficulty in this problem, since
diffractive effects are localized near the circle, while the bouncing
ball family is more global and covers a considerable volume of
configuration space. In the lack on a uniform approximation for the
direct contribution $G_{\rm d}$, it is natural to consider a splitting
of the integration region into ``near'' and ``far'' regions, using the
appropriate expressions in each region. However, in our case it is not
completely clear what is the correct splitting, and what is the
correct transition between the penumbra and the illuminated and
shadowed regions. Thus we choose to use the penumbra approximation of
$G_{\rm d}$ for the whole integration region, and after performing the
calculation to consider more closely the origin of corrections, if
any.

To get $d_{\rm bb, 1}^{\rm d}(k)$ we substitute in equation (\ref{eq:dbb1}) the
explicit form of $G_{\rm d}$ (equation (\ref{greentransd})), and to leading
order in $k$ we have
\begin{equation}
\label{eq:dbb1d}
d_{\rm bb, 1}^{\rm d}(k) = \frac{\sqrt{2 k L}}{\pi^\frac{3}{2}} 
\Im \left\{ \e^{\i k L + \i \frac{\pi}{4}} \int_{0}^{\frac{L}{2}}
\tilde{F} \left( 2 \sqrt{\frac{k}{\pi L}} (R-x) \right)
{\rm d}x \right\}\,,
\end{equation}
where we denoted $\tilde{F}(x) \equiv [F(\infty)-F(x)] / \sqrt{2 \i}$.
The integral is evaluated by using the indefinite integral of the
Fresnel function
\begin{equation}
\label{eq:frei1}
I_1(t) \equiv \int^t \tilde{F}(t') {\rm d}t' = t \tilde{F}(t) -
\frac{1}{\pi} \sqrt{\frac{\i}{2}} \e^{\i \frac{\pi}{2} t^2}.
\end{equation}
It can be simplified, if we consider the asymptotic approximation of
the Fresnel function for $|t| \gg 1$:
\begin{equation}
\label{eq:freasy}
F(t) = F(\infty) \ {\rm sign}(t) - \frac{\i}{\pi t} \e^{\i \frac{\pi}{2}
t^2} + \Or(|t|^{-3}).
\end{equation}
Combining (\ref{eq:freasy}) with (\ref{eq:frei1}) we get
\begin{equation}
I_1(t) \approx \frac{1-{\rm sign}(t)}{2} t = \left\{
\begin{array}{cc} t & t < 0 \\ 0 & t > 0 \end{array}
\right.\,.
\end{equation}
Thus finally
\begin{equation}
d_{\rm bb, 1}^{\rm d}(k) = \frac{(L-2R) \sqrt{k L}}{\sqrt{2}
\pi^{\frac{3}{2}}} \cos \left( k L - \frac{\pi}{4} \right) + 
\Or(k^{-\frac{1}{2}})\,.
\end{equation}
This recovers the semiclassical contribution of the bouncing balls
\cite{SSCL93}. There are no diffraction corrections to $d_{\rm bb,
1}^{\rm d}(k)$ that are larger than the semiclassical error of the
bouncing ball contributions. We emphasize that the asymptotic
approximation of $F(t)$ was invoked {\em after} the integration was
performed. This is completely justified, because the argument of both
limits is $\Or(k^\frac{1}{2}) \gg 1$. Replacing the {\em integrand} with
the asymptotic form would be unjustified, since the interesting region
near the circle is not asymptotic.

For the double repetition the situation is more interesting. To
leading order in $k$ we have
\begin{equation}
\label{eq:dbb2dd}
d_{\rm bb, 2}^{\rm dd}(k) = \frac{\sqrt{k L}}{\pi^\frac{3}{2}} \Im \left\{
\e^{2 \i k L + \i \frac{\pi}{4}} \int_{0}^{\frac{L}{2}} \tilde{F}^2
\left( 2 \sqrt{\frac{k}{\pi L}}(R-x) \right) {\rm d}x \right\}.
\end{equation}
The relevant integral is
\begin{eqnarray}
\fl I_2(t) \equiv \int^t \tilde{F}^2 (t') {\rm d}t' = t \tilde{F}^2 (t)
- \frac{\sqrt{2 \i}}{\pi} \tilde{F}(t) \e^{\i \frac{\pi}{2} t^2} -
\frac{1}{\sqrt{2} \pi} F(\sqrt{2}t) \\
\lo \approx t \left[ \frac{1-{\rm sign}(t)}{2} \right] -
\frac{\sqrt{\i}}{2 \pi} {\rm sign}(t) = \left\{
\begin{array}{rc} 
t+\sqrt{\i}/(2 \pi) & t < 0 \\
- \sqrt{\i}/(2 \pi) & t > 0
\end{array} \right.\,,
\label{eq:i2asy}
\end{eqnarray}
where the last line was obtained using the asymptotic approximation of
$F(t)$. Equation (\ref{eq:i2asy}) indicates, that in the lit region ($t <
0, |t| \gg 1$) we obtain a linear contribution, together with a
constant that comes from the integration near $t \approx 0$. In the
shadowed region we have only a constant contribution from the $t
\approx 0$ region. Inserting (\ref{eq:i2asy}) into 
(\ref{eq:dbb2dd}) we get
\begin{equation}
\fl d_{\rm bb, 2}^{\rm dd}(k) = \frac{\sqrt{k L}}{\pi^\frac{3}{2}} \left(
\frac{L}{2} - R \right) \cos \left( 2 k L - \frac{\pi}{4} \right) -
\frac{L}{2 \pi^2} \cos (2 k L) + \Or(k^{-\frac{1}{2}}).
\end{equation}
Thus, in addition to the semiclassical contribution due to the
bouncing balls, we obtain a genuine diffractive contribution. It is
$\Or(k^0)$ which is the same as for an unstable periodic orbit and
therefore must be retained. The explicit form of $I_2(t)$ in
(\ref{eq:i2asy}) indicates that indeed this diffraction contribution
is obtained from the region near the circle and is not a result of a
more global effect.

\section*{References}


\end{document}

%% file: FIG/fig-sinai.tex
\begin{figure}[tbp]
  \centerline{\psfig{figure=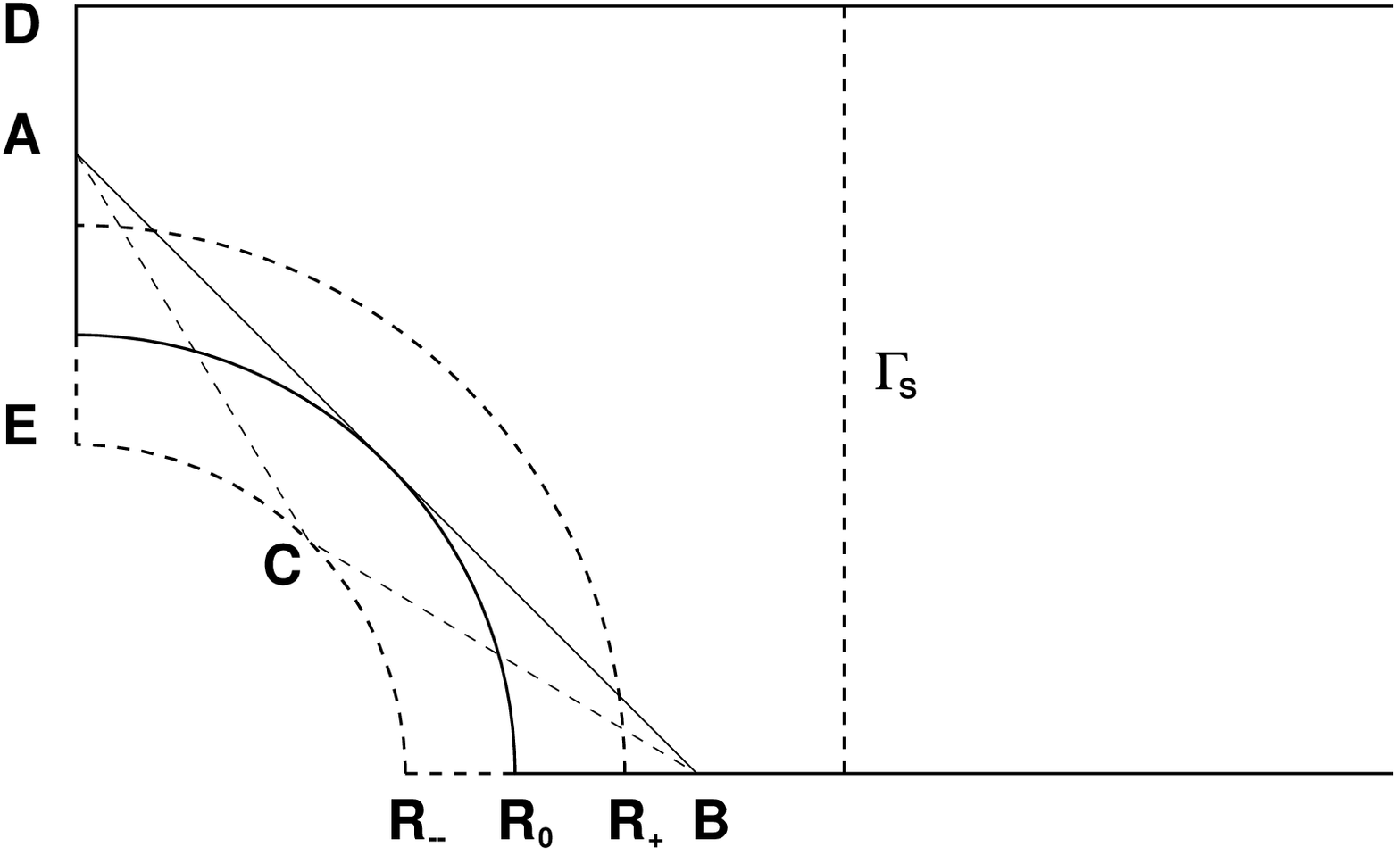,angle=-90,width=12cm}}
  \caption{  \label{fig:sinai}
The quarter Sinai billiard and the attached waveguide.}
\end{figure}

%% file: FIG/fig-regions.tex
\begin{figure}[tbp]
  \centerline{\psfig{figure=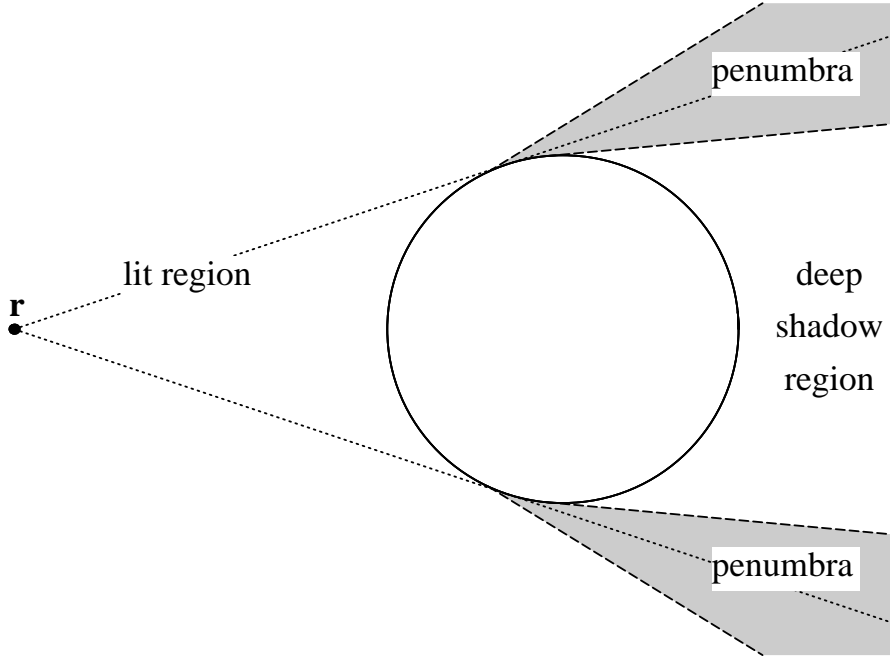,width=12cm}}
  \caption{The different regions of $\bf r'$ for a given $\bf r$.
    The penumbra occupies a small region on both sides of the
    geometrical shadow line (dotted).
  \label{fig:regions}}
\end{figure}

%% file: FIG/fig-lit.tex
\begin{figure}[tbp]
  \begin{center}
    \setlength{\unitlength}{0.3cm}
    \begin{picture}(40,28)(-25.18,-8)
      \put(-25.18,-8){\psfig {figure=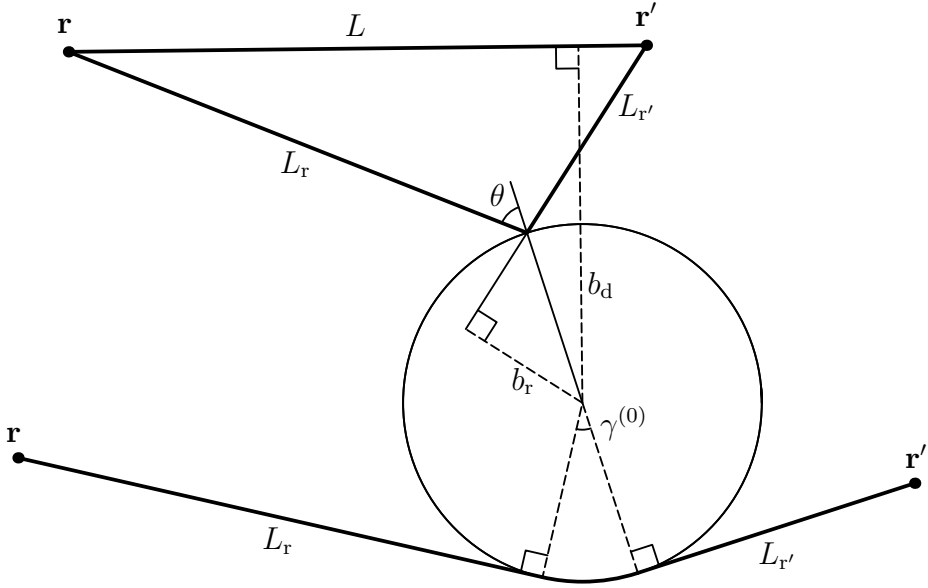,width=40\unitlength}}
      \put(-22.93,16.51){\makebox(0,0)[b]{$\bf r$}}
      \put(2.85,16.8){\makebox(0,0)[b]{$\bf r'$}}
      \put(-25.18,-1.66){\makebox(0,0)[b]{$\bf r$}}
      \put(14.82,-2.8){\makebox(0,0)[b]{$\bf r'$}}
      \put(-10.04,16.35){\makebox(0,0)[b]{$L$}}
      \put(-12.7,11.26){\makebox(0,0)[t]{$L_{\rm r}$}}
      \put(1.58,13.20){\makebox(0,0)[l]{$L_{\rm r'}$}}
      \put(-13.48,-5.43){\makebox(0,0)[t]{$L_{\rm r}$}}
      \put(8.65,-6.1){\makebox(0,0)[t]{$L_{\rm r'}$}}
      \put(0.8,-0.8){\makebox(0,0)[l]{$\gamma^{(0)}$}}
      \put(-3.4,8.81){\makebox(0,0)[br]{$\theta$}}
      \put(0.24,5.32){\makebox(0,0)[l]{$b_{\rm d}$}}
      \put(-2.6,1.65){\makebox(0,0)[t]{$b_{\rm r}$}}
    \end{picture}
  \end{center}
  \caption{The geometry of trajectories in the lit region
    (in the upper part) and of creeping trajectories in the deep
    shadow region (in the lower part).
  \label{fig:lit}}
\end{figure}

%% file: FIG/fig-lplane.tex
\begin{figure}[tbp]
  \begin{center}
    \setlength{\unitlength}{2.1cm}
    \begin{picture}(6,3.5)(-3,-1.5)
      \put(-3,-1.5){\psfig{figure=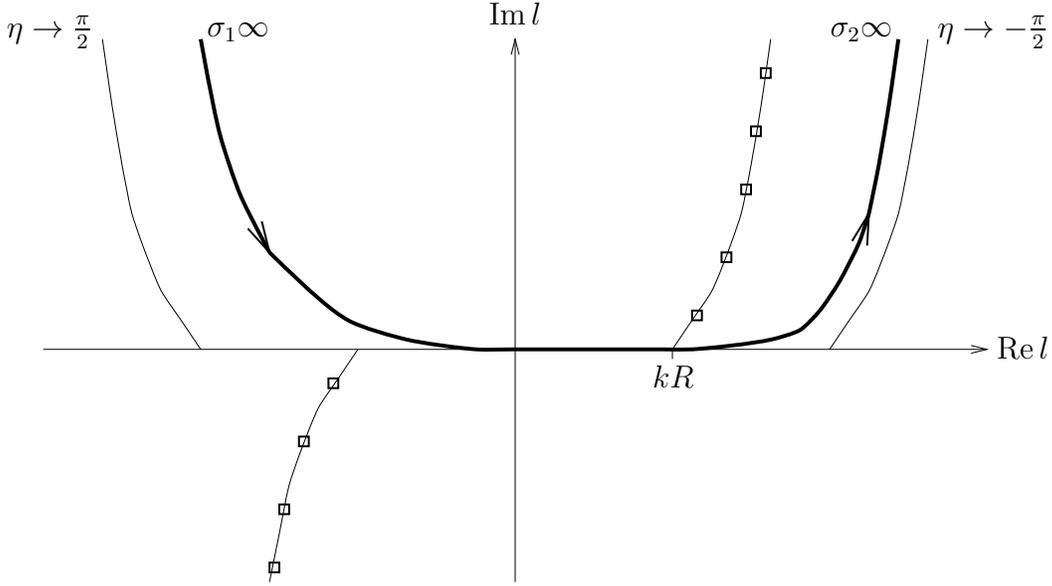,width=6\unitlength}}
      \put(3.05,0){\makebox(0,0)[l]{${\rm Re}\, l$}}
      \put(0,2.05){\makebox(0,0)[b]{${\rm Im}\, l$}}
      \put(2.3875,2){\makebox(0,0)[r]{$\sigma_2 \infty$}}
      \put(-1.95,2){\makebox(0,0)[l]{$\sigma_1 \infty$}}
      \put(2.675,2){\makebox(0,0)[l]{$\eta \rightarrow
            -\frac{\pi}{2}$}}
      \put(-2.675,2){\makebox(0,0)[r]{$\eta \rightarrow
            \frac{\pi}{2}$}}
      \put(1,-0.1125){\makebox(0,0)[t]{$kR$}}
    \end{picture}
  \end{center}
  \caption{The complex $l$ plane.
    The squares represent the poles of the $S$ matrix. The bold line
    shows the contour from $\sigma_1 \infty$, going through $kR$ and
    around the poles of the $S$ matrix to $\sigma_2 \infty$. The
    contour does not cross the two lines defined by
    (\protect{\ref{defsigma1}})-(\protect{\ref{defsigma2}}), with
    $\eta \rightarrow \pm \pi / 2$.
  \label{fig:lplane}}
\end{figure}

%% file: FIG/fig-glance.tex
\begin{figure}[tbp]
  \begin{center}
    \setlength{\unitlength}{0.3cm}
    \begin{picture}(40,21.94)(-26.71,-8)
      \put(-26.71,-8){\psfig {figure=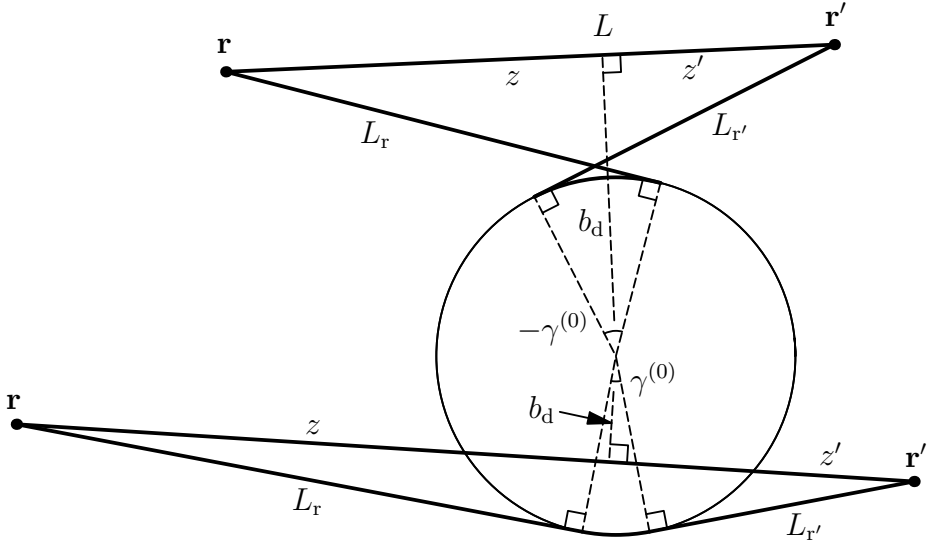,width=40\unitlength}}
      \put(-17.38,13.52){\makebox(0,0)[b]{$\bf r$}}
      \put(9.73,14.74){\makebox(0,0)[b]{$\bf r'$}}
      \put(-26.71,-2.25){\makebox(0,0)[b]{$\bf r$}}
      \put(13.29,-4.81){\makebox(0,0)[b]{$\bf r'$}}
      \put(-4.60,12.70){\makebox(0,0)[t]{$z$}}
      \put(3.40,13.40){\makebox(0,0)[t]{$z'$}}
      \put(-0.60,14.27){\makebox(0,0)[b]{$L$}}
      \put(-13.44,-3.35){\makebox(0,0)[b]{$z$}}
      \put(9.5,-4.75){\makebox(0,0)[b]{$z'$}}
      \put(-10.00,10.50){\makebox(0,0)[tr]{$L_{\rm r}$}}
      \put(4.20,10.85){\makebox(0,0)[tl]{$L_{\rm r'}$}}
      \put(-13.01,-5.8){\makebox(0,0)[tr]{$L_{\rm r}$}}
      \put(7.49,-6.9){\makebox(0,0)[tl]{$L_{\rm r'}$}}
      \put(-1.2,1.33){\makebox(0,0)[r]{$-\gamma^{(0)}$}}
      \put(0.6,-0.93){\makebox(0,0)[l]{$\gamma^{(0)}$}}
      \put(-0.5,6){\makebox(0,0)[r]{$b_{\rm d}$}}
      \put(-2.7,-2.3){\makebox(0,0)[r]{$b_{\rm d}$}}
    \end{picture}
  \end{center}
  \caption{The geometrical setup of trajectories in the penumbra.
    In the upper part $\bf r$ and $\bf r'$ are in the classically
    illuminated region ($b_{\rm d} > R$), and in the lower part they are in
    the classically shadowed region.
  \label{fig:glance}}
\end{figure}

%% file: FIG/fig-annulus.tex
\begin{figure}[tbp]
  \begin{center}
    \setlength{\unitlength}{0.15cm}
    \begin{picture}(28,28.28)(0,0)
      \put(0,0){\psfig{figure=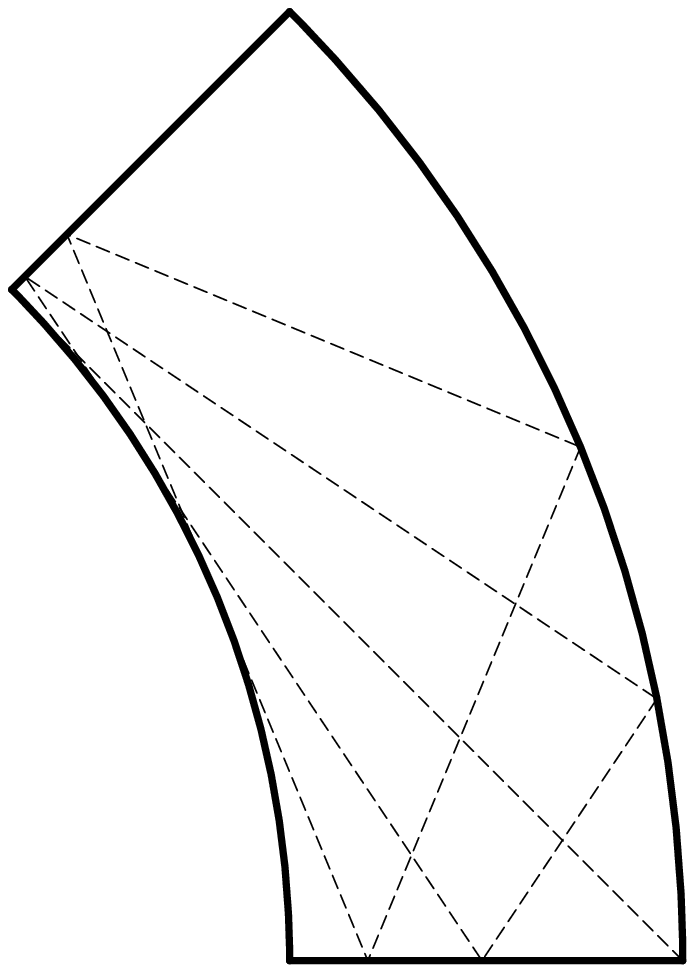,width=20\unitlength}}
      \put(0,6){\makebox(0,0)[l]{(a)}}
      \put(28,6){\makebox(0,0)[r]{(b)}}
    \end{picture}
    \setlength{\unitlength}{2.475cm}
    \begin{picture}(3,2)
      \put(0,0){\psfig{figure=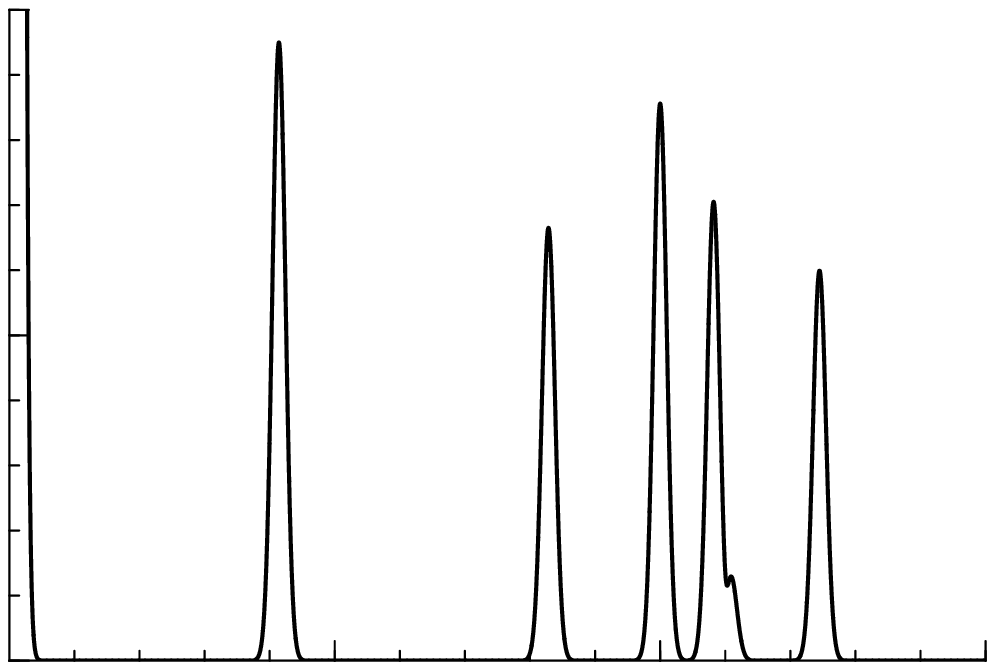,width=3\unitlength}}
      \put(3.1,0){\makebox(0,0)[l]{$x$}}
      \put(0,2.1){\makebox(0,0)[b]{$|D_w(x)|$}}
      \put(0.828,2){\makebox(0,0){${\rm D}_1$}}
      \put(1.657,2){\makebox(0,0){${\rm D}_2$}}
      \put(2.485,2){\makebox(0,0){${\rm D}_3$}}
      \put(2,2){\makebox(0,0){T}}
      \put(2.165,2){\makebox(0,0){P}}
      \put(1,-0.03){\makebox(0,0)[t]{1}}
      \put(2,-0.03){\makebox(0,0)[t]{2}}
      \put(-0.03,0){\makebox(0,0)[r]{0}}
      \put(-0.03,1){\makebox(0,0)[r]{1}}
    \end{picture}
  \end{center}
  \caption{(a) The annular sector, with
    $R_2=\protect{\sqrt 2} R_1$ and $\alpha=\pi/4$. Three T-orbits
    (which in this case are exactly tangent) are shown in dashed
    lines.  (b) The absolute value of the length spectrum, with
    $R_2=\protect{\sqrt 2}$ and $k_0=1300$. Peak T corresponds to the
    T-orbits. Peaks ${\rm D}_n$ correspond to $n$ repetitions of the
    orbits along the diameter. To peak P contribute orbits which do
    not reach the inner arc, and bounce $n\geq2$ times off the outer
    one.
  \label{fig:annulus}}
\end{figure}

%% file: FIG/fig-annlog.tex
\begin{figure}[tbp]
  \begin{center}
    \setlength{\unitlength}{2.8cm}
    \begin{picture}(2.2,2.2)(7.8,-1.9)
      \put(7.425,-1.7){\psfig{figure=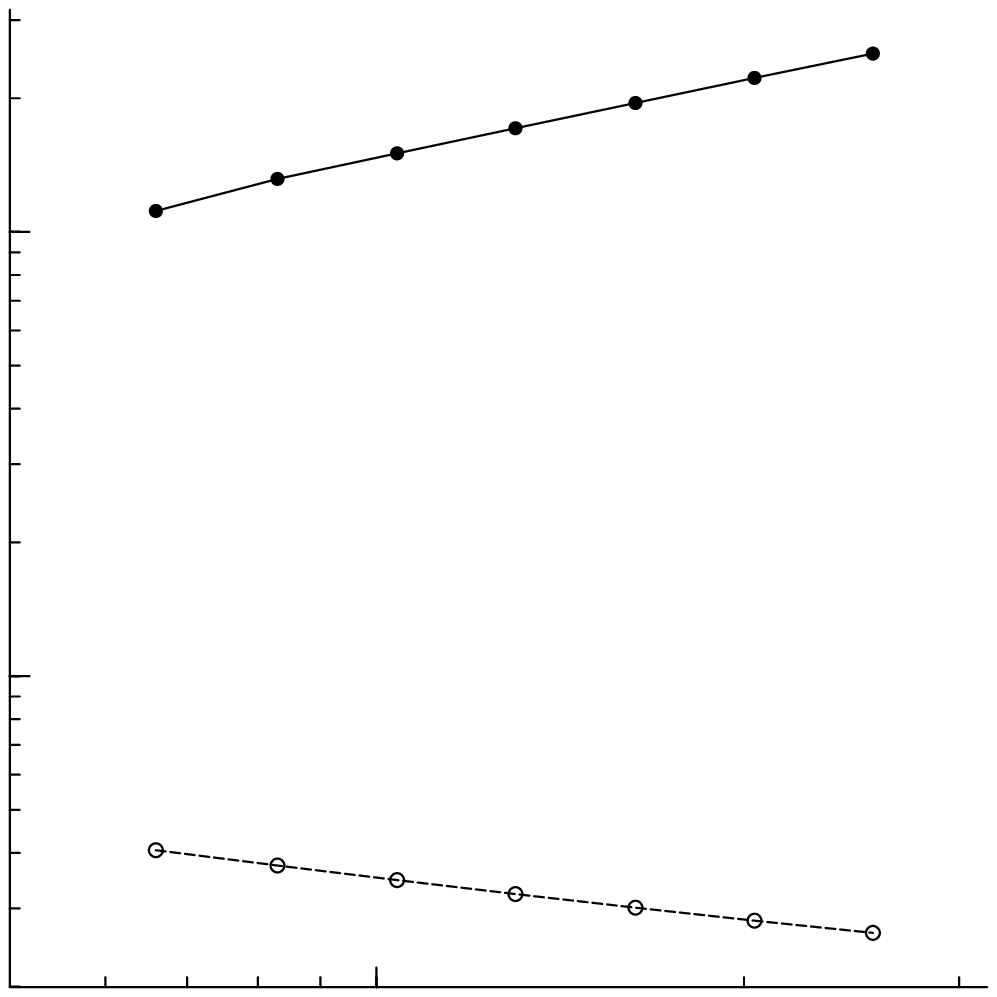,width=2.2\unitlength}}
      \put(9.667,-1.7){\makebox(0,0)[l]{$k_0$}}
      \put(8.25,-1.725){\makebox(0,0)[t]{$1000$}}
      \put(9.078,-1.725){\makebox(0,0)[t]{$2000$}}
      \put(7.4,-1){\makebox(0,0)[r]{$0.1$}}
      \put(7.4,0){\makebox(0,0)[r]{$1$}}
      \put(7.645,0.5){\makebox(0,0)[t]{(a)}}
    \end{picture}
    \setlength{\unitlength}{2.4cm}
    \begin{picture}(2.6,2.6)(-1.2,-1.8)
      \put(-1.2,-1.6){\psfig{figure=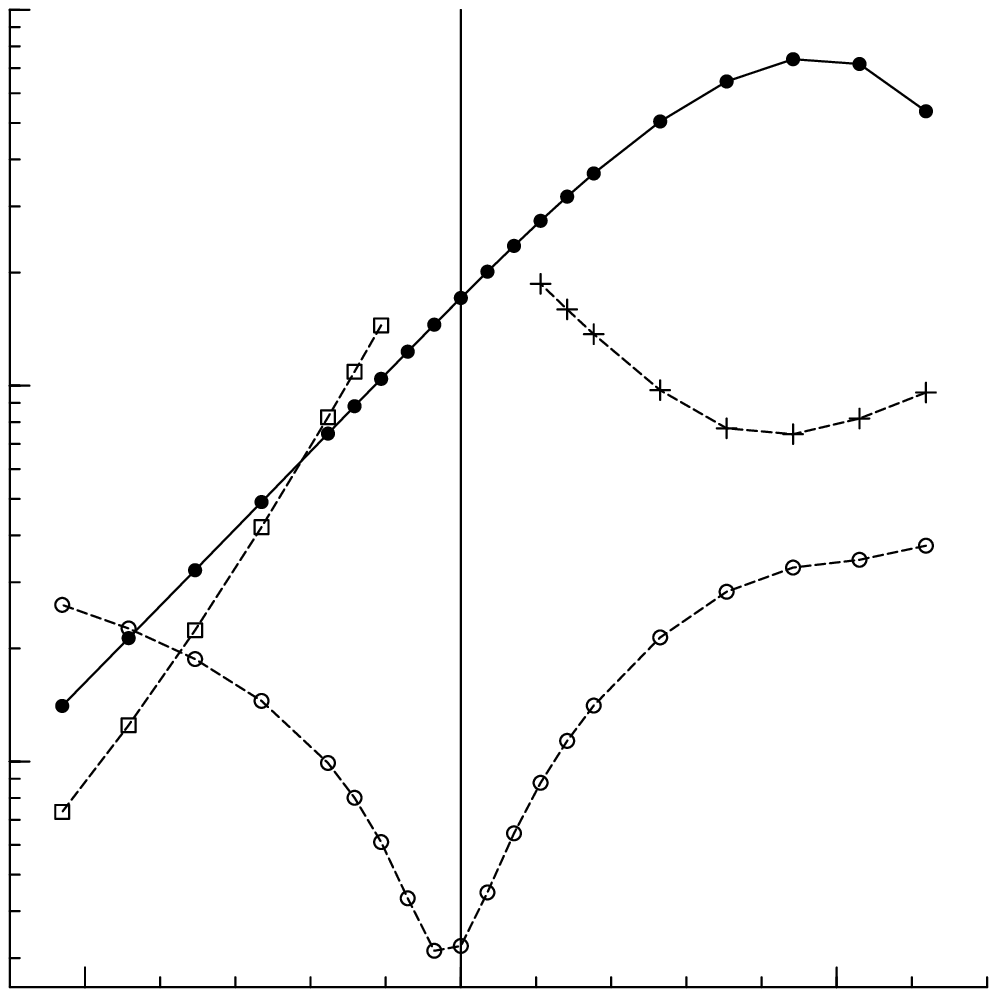,width=2.6\unitlength}}
      \put(1.45,-1.6){\makebox(0,0)[l]{$b_{\rm d}-R_1$}}
      \put(-1,-1.63){\makebox(0,0)[t]{$-0.05$}}
      \put(0,-1.63){\makebox(0,0)[t]{$0$}}
      \put(1,-1.63){\makebox(0,0)[t]{$0.05$}}
      \put(-1.23,-1){\makebox(0,0)[r]{$0.1$}}
      \put(-1.23,0){\makebox(0,0)[r]{$1$}}
      \put(-1.23,1){\makebox(0,0)[r]{$10$}}
      \put(-0.94,1){\makebox(0,0)[t]{(b)}}
    \end{picture}
  \end{center}
  \caption {(a) The maximum of the T-orbits peak in $|D_{\rm w}(x)|$
    (full circles) as a function of $k_0$ for $R_2 =
    \protect{\sqrt{2}}$ and $R_1 = 1$ (exact tangency). The maximal
    error of the penumbra expression for this peak, including both
    direct and glancing contributions is given by the empty circles.
    Note the logarithmic scale. (b) The maximum of the T-orbits peak
    in $|D_{\rm w}(x)|$ (full dots) for different values of $R_2$, with
    $k_0 = 1300$ and $R_1 = 1$. The maximal error of the semiclassical
    approximation for this peak is given by the crosses, for the
    creeping approximation by the squares and for the penumbra
    approximation by the empty circles.  The whole region of $b_{\rm d}-R_1$
    is well inside the penumbra. In both graphs the lines are for
    guiding the eyes only.
  \label {fig:annlog}}
\end{figure}

%% file: FIG/fig-sb-ls-ab.tex
\begin{figure}[tbp]
\samepage
\begin{center}
\leavevmode
\psfig{figure=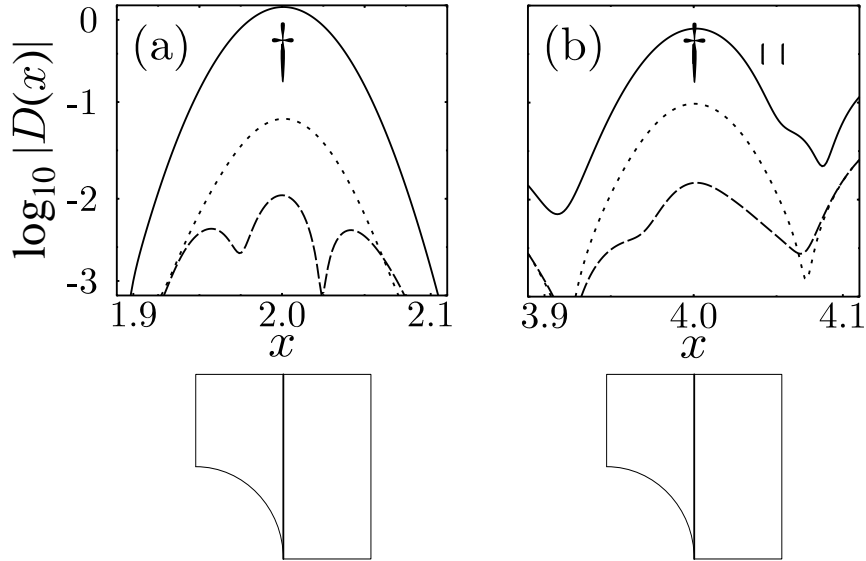,width=12cm}
\end{center}
\nopagebreak
\caption{
\label{fig:sb-ls-ab}
Penumbra corrections of the length spectrum for the case of 
    exactly tangent orbits. The orbits considered are drawn below the
    frames. Solid lines show the quantum (exact) length spectrum, and
    dotted lines show the {\em difference} between the quantum length
    spectrum and the semiclassical approximation supplemented by the
    results of \protect\cite{SSCL93} for the bouncing ball and edge
    contributions. Vertical bars indicate locations of unstable
    periodic orbits, daggers indicate bouncing ball families. Note the
    logarithmic scale. (a) Shortest tangent orbit $(x=2)$. Dashed line
    - glancing contribution also included. (b) Double traversal of the
    orbit considered in (a), $x=4$. Dashed line includes 3 penumbra
    contributions (see text).}
\end{figure}

%% file: FIG/fig-sb-ls-cd.tex
\begin{figure}[tbp]
\begin{center}
\leavevmode
\psfig{figure=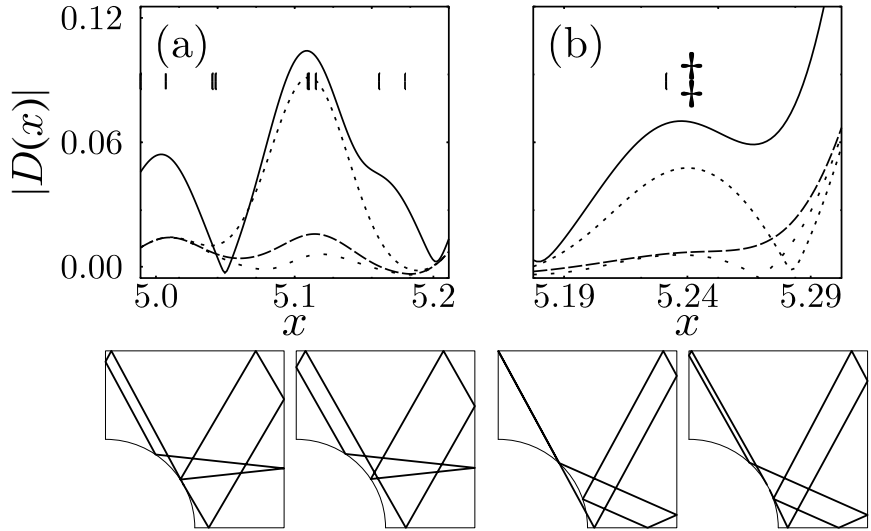,width=12cm}
\end{center}
\caption{
\label{fig:sb-ls-cd}
Penumbra corrections of the length spectrum for the case of 
    almost tangent and ghost orbits. The orbits considered are drawn
    below the frames. Solid lines show the quantum (exact) length
    spectrum, and dotted lines show the {\em difference} between the
    quantum length spectrum and the standard semiclassical
    approximation. Vertical bars indicate locations of unstable
    periodic orbits. (a) Pair of almost tangent periodic orbits at $x
    \approx 5.10 $. Dashed line - direct term included, sparse dotes -
    glancing contribution also included. (b) Pair of classically
    forbidden periodic orbits at $x \approx 5.24$. Their location is
    indicated by a double dagger. Notation is as in (c).}
\end{figure}

%% file: FIG/fig-ls-cmp.tex
\begin{figure}[p]
\centerline{\hspace*{20mm}\psfig{figure=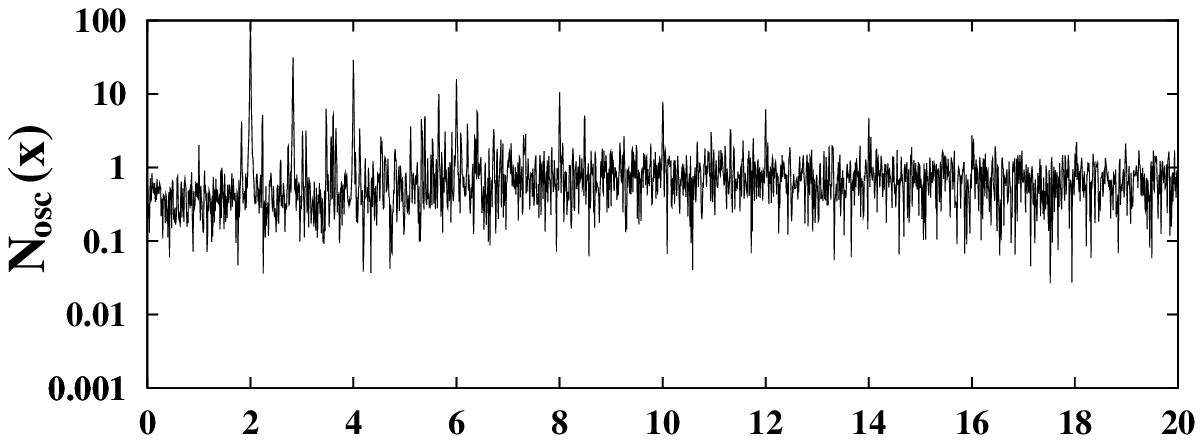,height=5cm}}
\centerline{\hspace*{20mm}\psfig{figure=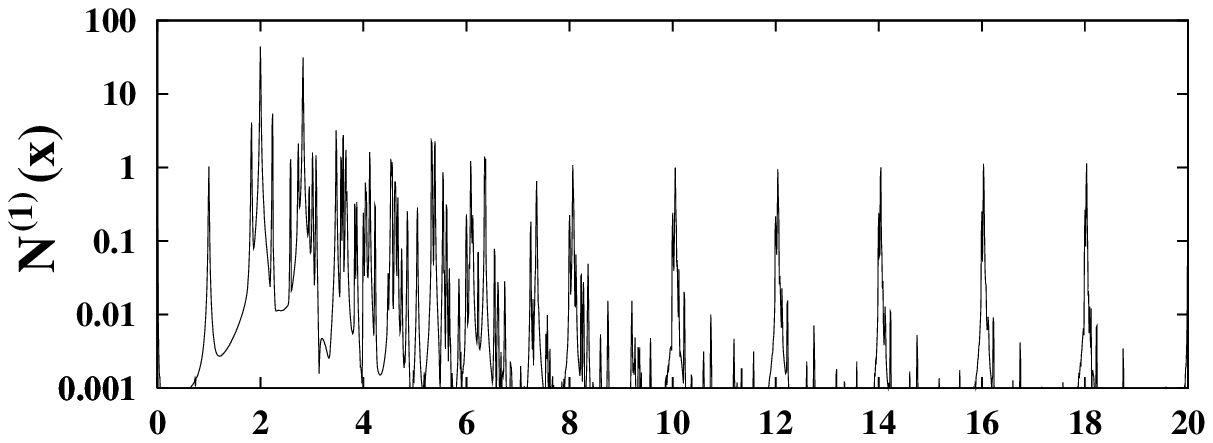,height=5cm}}
\centerline{\hspace*{20mm}\psfig{figure=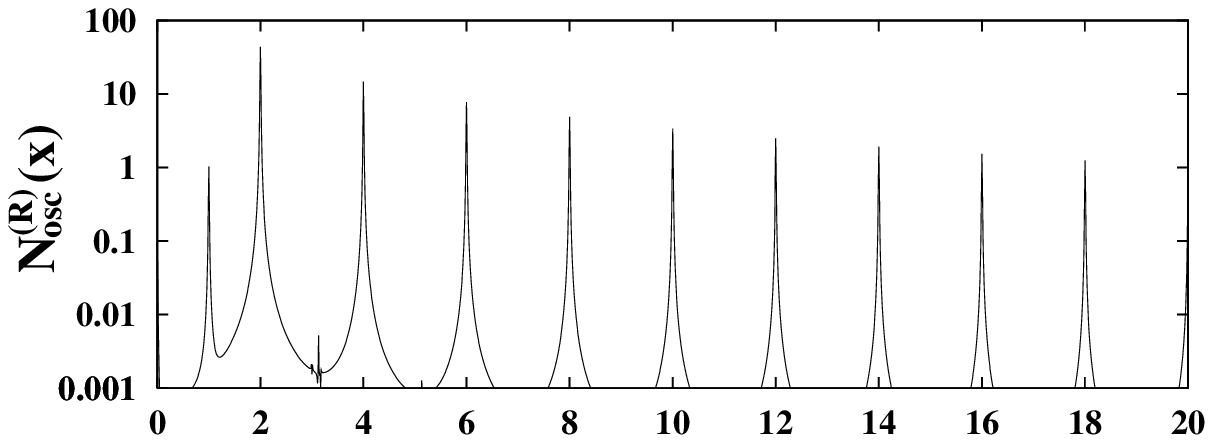,height=5cm}}
\centerline{\Large\bf \hspace*{12mm}x}
\vspace*{1mm}
\caption{\label{ls-cmp} Comparison between the 
length spectrum of the number counting function $N(k)$ (top), the term
$N^{(1)}(k)$ related to the trace of the S-matrix (middle) and the
resonance counting function $N^{(\rm R)}(k)$ (bottom). The length
spectra are obtained from the S-matrix for $a=1$ and $R=0.5$ in the
interval ${k/\pi}=1,\dots,281$ (containing more than 
49\,200 eigenvalues) and with $\Delta_k=\pi/64$. The smooth
parts of $N(k)$ and $N^{(\rm R)}(k)$ have been subtracted according
to Weyl's law.}
\end{figure}

%% file: FIG/fig-orb-tp.tex
\begin{figure}[tbp]
\centerline{
\psfig{figure=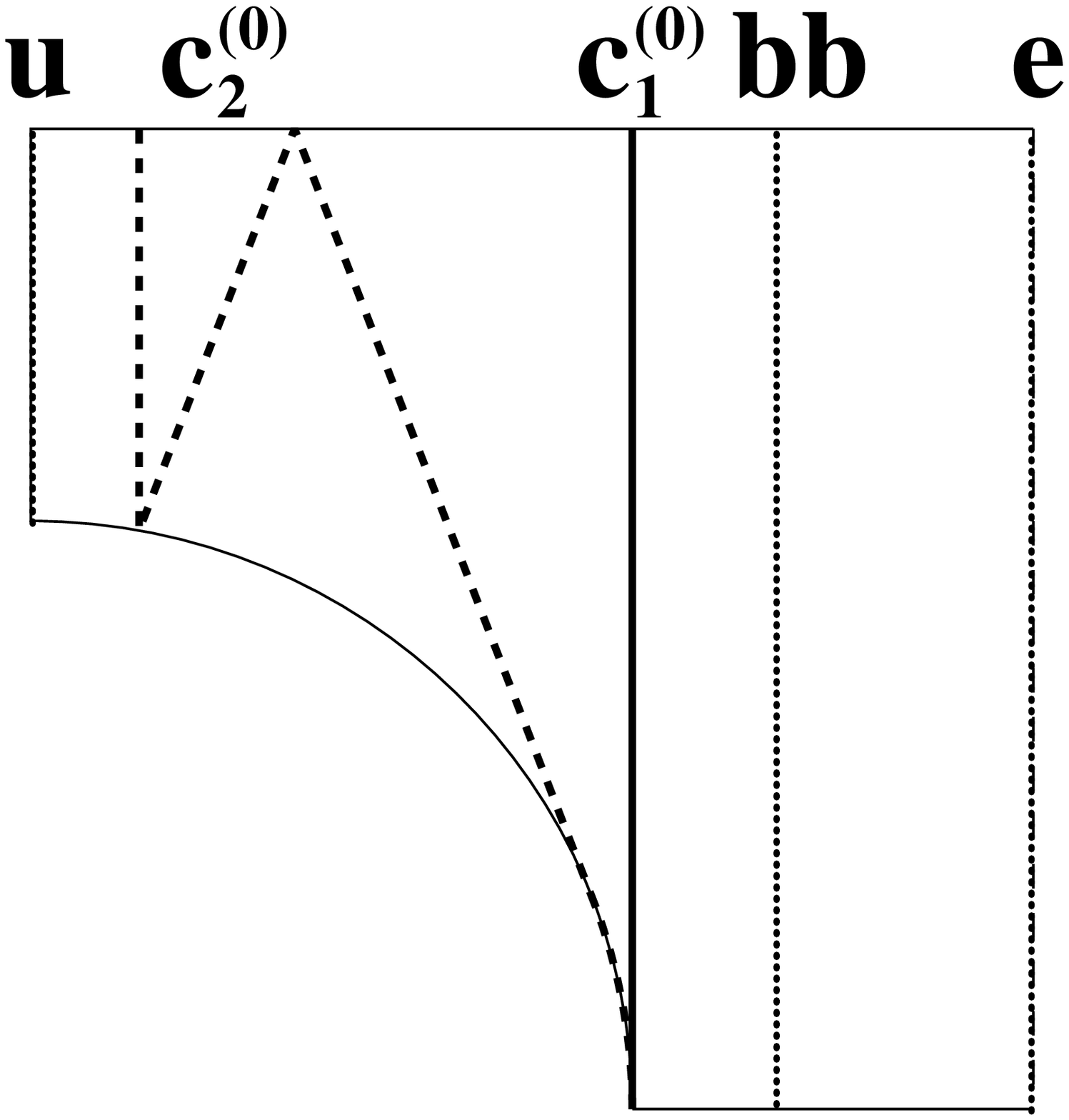,height=37mm,angle=-90}
\psfig{figure=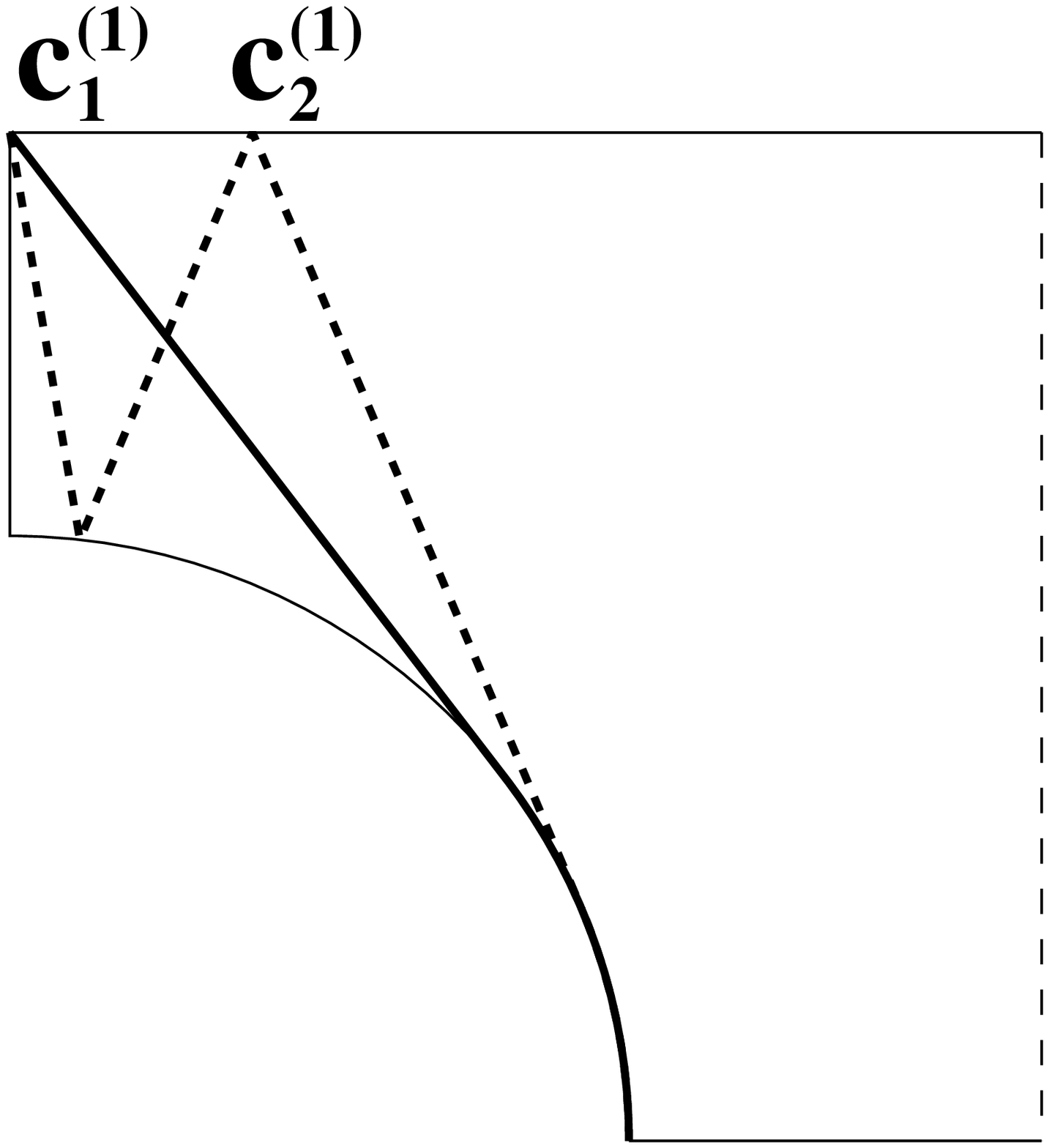,height=37mm,angle=-90}
\psfig{figure=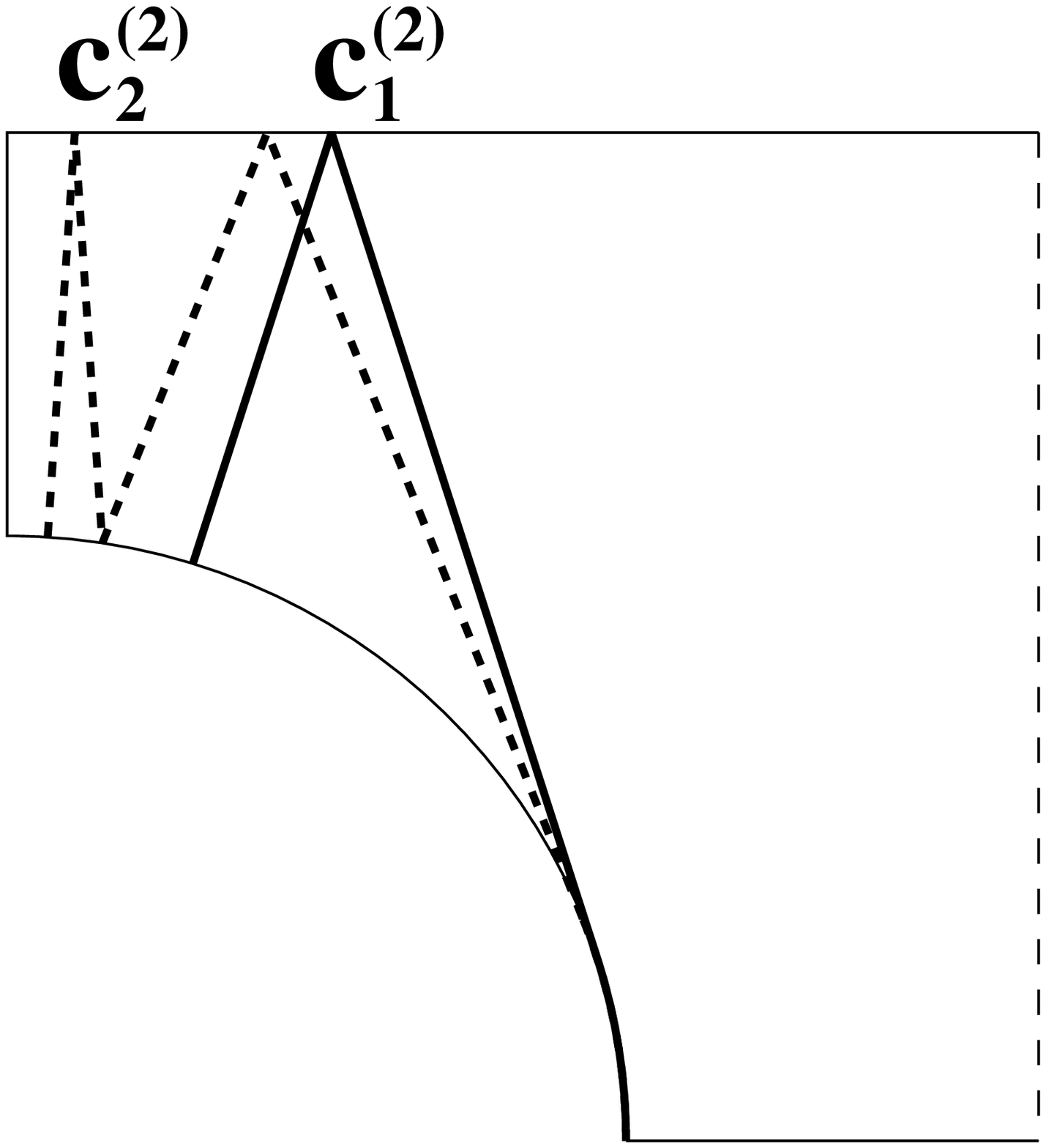,height=37mm,angle=-90}
\psfig{figure=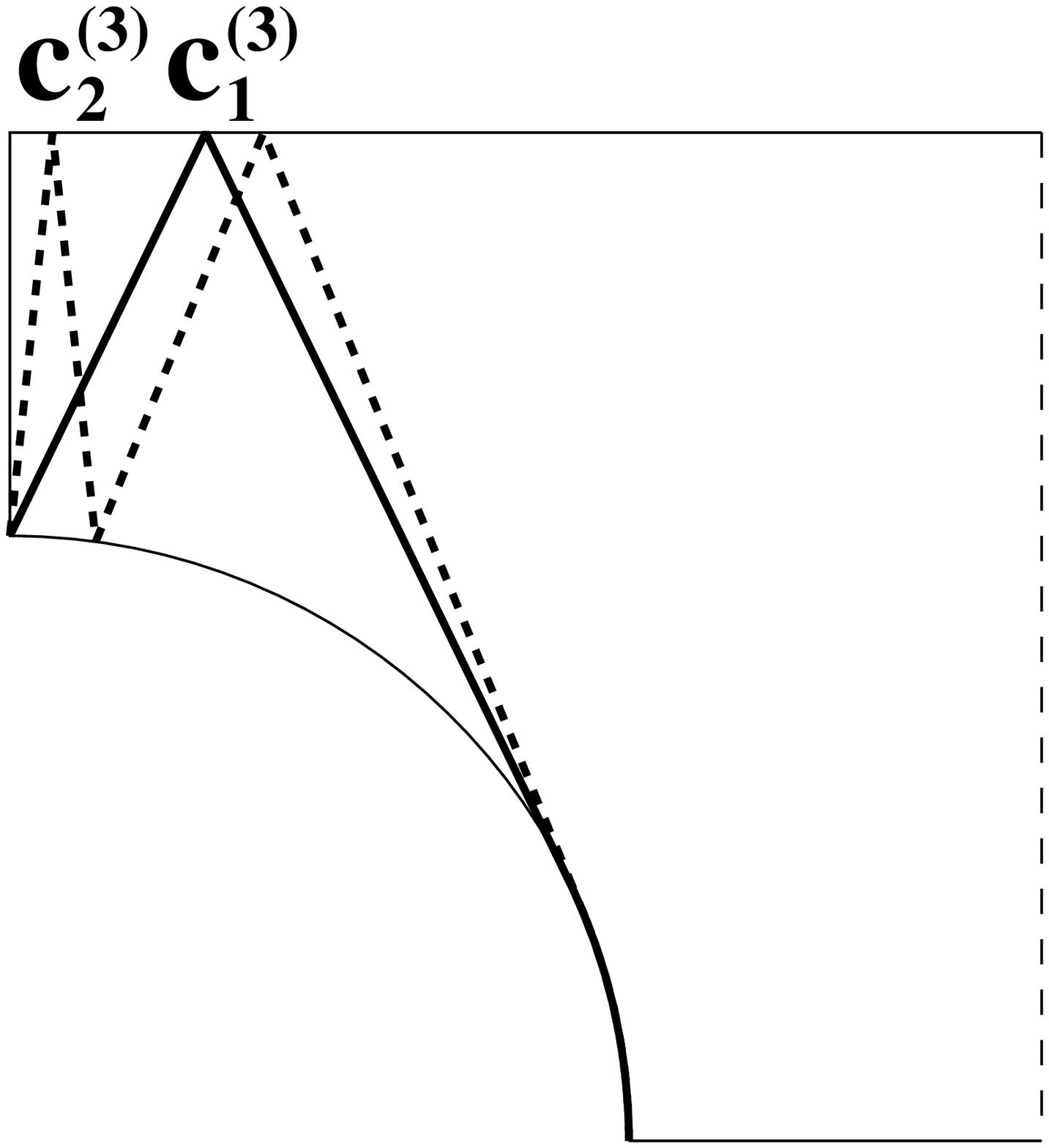,height=37mm,angle=-90}
}
\caption{\label{fig:orb-tp} The figures show some trapped orbits 
  of the opened Sinai billiard. {\bf bb} denotes an example for a bouncing
  ball orbit and {\bf e} is the edge orbit limiting the family. The other
  edge orbit {\bf u} is the only isolated classical orbit contributing to
  the resonance counting function. The orbits denoted with {\bf c} are
  diffractive. }
\end{figure}

%% file: FIG/table.tex
\begin{table}[tbp]
\begin{center}
\begin{tabular}{|c|c|c|c|}
\hline
$ $&$ L^{\rm {(cl)}} $&$ M^{\rm {(cl)}}_{12} $& $ \sin{\gamma/ 2} $\\
\hline
\hline 
$ c_1^{(0)} $&$ 2a $&$ L^{\rm {(cl)}} $&$ 0 $\\ \hline 
$ c_1^{(1)} $&$ 2\sqrt{a^2 - R^2} $&$ L^{\rm {(cl)}} $&$ {R/a} $\\ \hline 
$ c_1^{(2)} $&$  2\sqrt{4a^2 - R^2}-2R $&$ L^{\rm {(cl)}} + 
{(L^{\rm {(cl)}})^2 / 2R} $&$ {R/2a} $\\ \hline 
$ c_1^{(3)} $&$ 2\sqrt{(2a-R)^2 - R^2} $&$ 
L^{\rm {(cl)}} + {(L^{\rm {(cl)}})^2
/ 2R\cos{\gamma/2}}$&$ {R/(2a - R)} $\\ 
\hline
\end{tabular}\vspace*{3mm} 
\caption{\label{tab:orb-tp}
The table gives the necessary geometrical data for the computation
of the contributions from some diffractive orbits to $N^{(R)}(k)$.
$ L^{\rm {(cl)}} $ and $ M^{\rm {(cl)}}_{12} $ are the path length
and monodromy matrix of the classical segment of the orbit, $\gamma$
is the creeping angle.
}
\end{center}
\end{table}

%% file: FIG/fig-tp-d05.tex
\begin{figure}[tbp]
\centerline{\hspace*{20mm}\psfig{figure=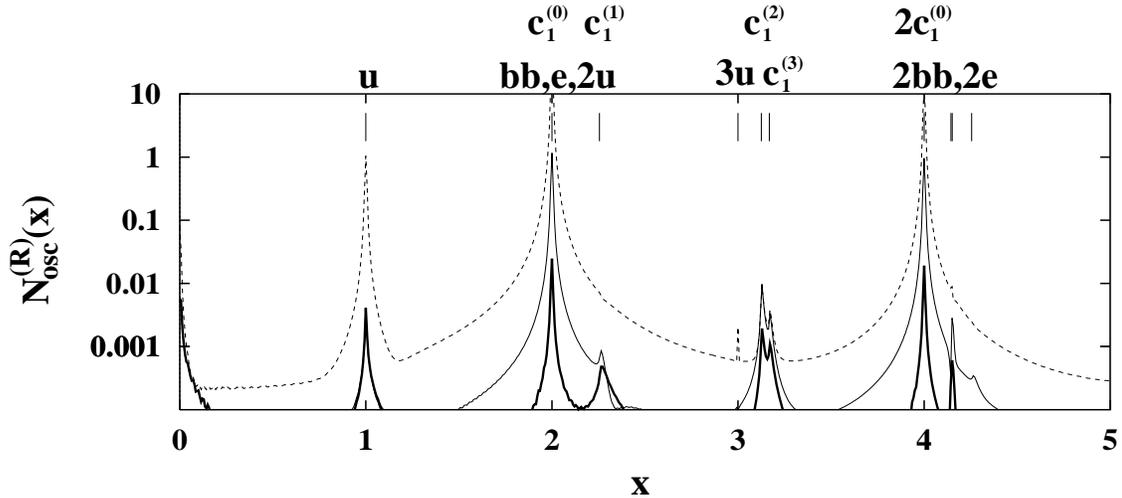,width=16cm,angle=-90}}
\caption{\label{fig:tp-d05} The length spectrum of 
  $N^{\rm (\rm R)}(k)$ obtained from the interval $1 < {k a/\pi} < 281$
  for a billiard with $a=1$, $R=0.5$ and Dirichlet boundary conditions
  on the circle.  The dashed line corresponds to the full quantum
  result with the smooth part subtracted according to the generalized
  Weyl law, i.~e.\ it comprises all oscillatory contributions to the
  resonance counting function. The thin solid line shows the
  deviations of the semiclassical approximation based on all classical
  orbits from the quantum data and for the thick solid line the
  diffractive orbits have also been included into the semiclassical
  approximation. The lengths of contributing orbits are marked with
  vertical bars.}
\end{figure}

%% file: FIG/fig-kd.tex
\begin{figure}[tbp]
\centerline{\hspace*{0mm}
\psfig{figure=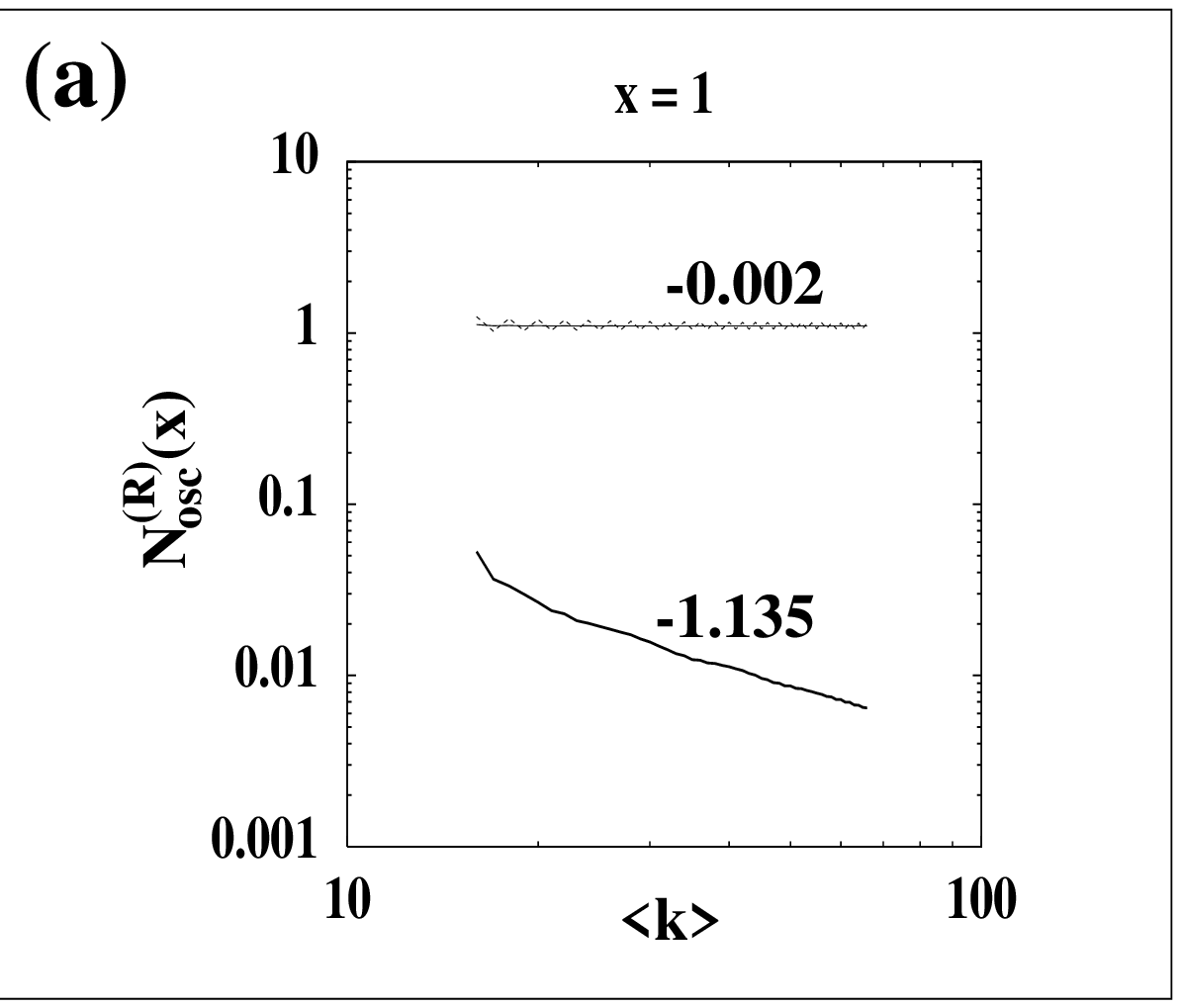,width=5cm,angle=-90}
\psfig{figure=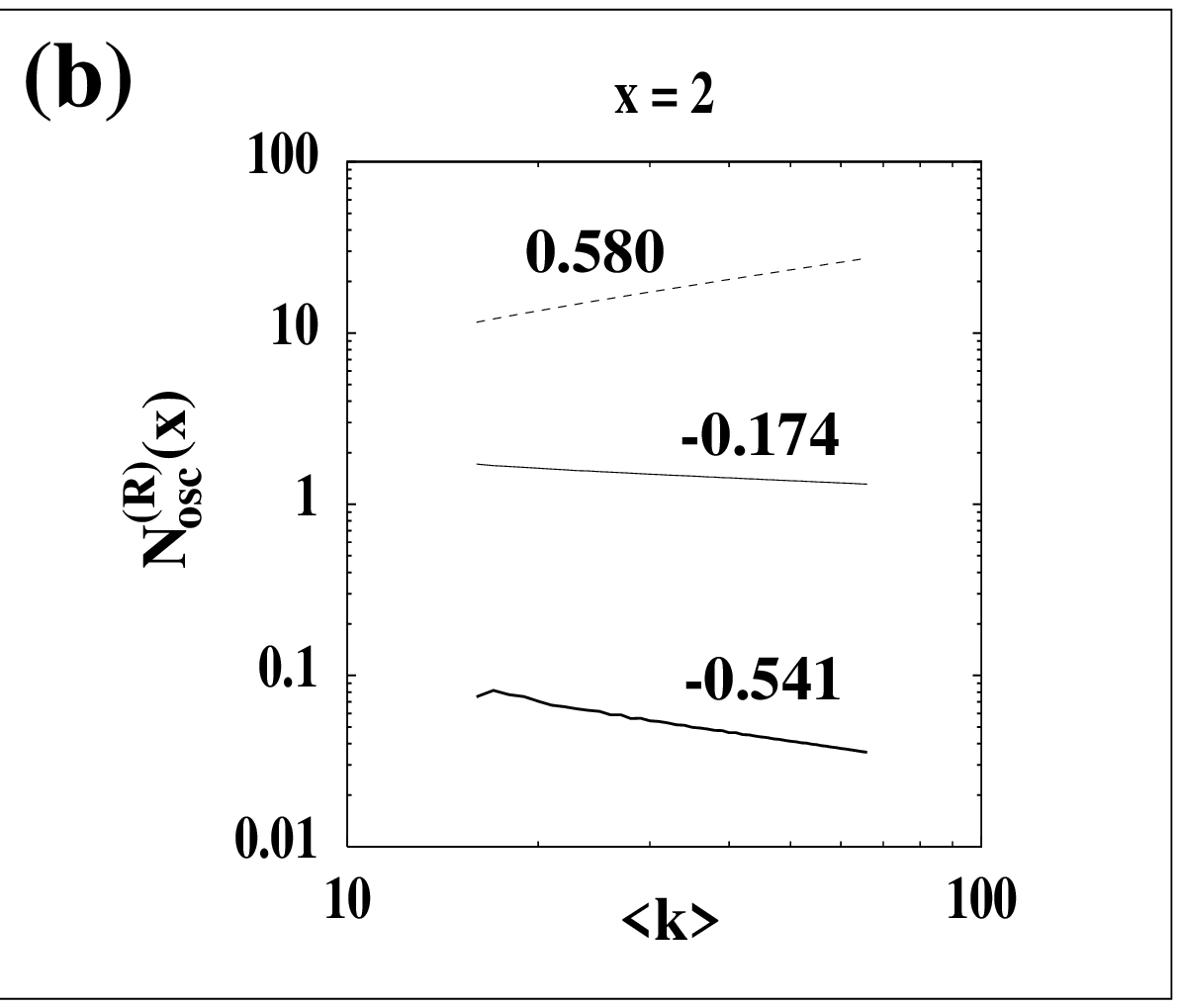,width=5cm,angle=-90}
\psfig{figure=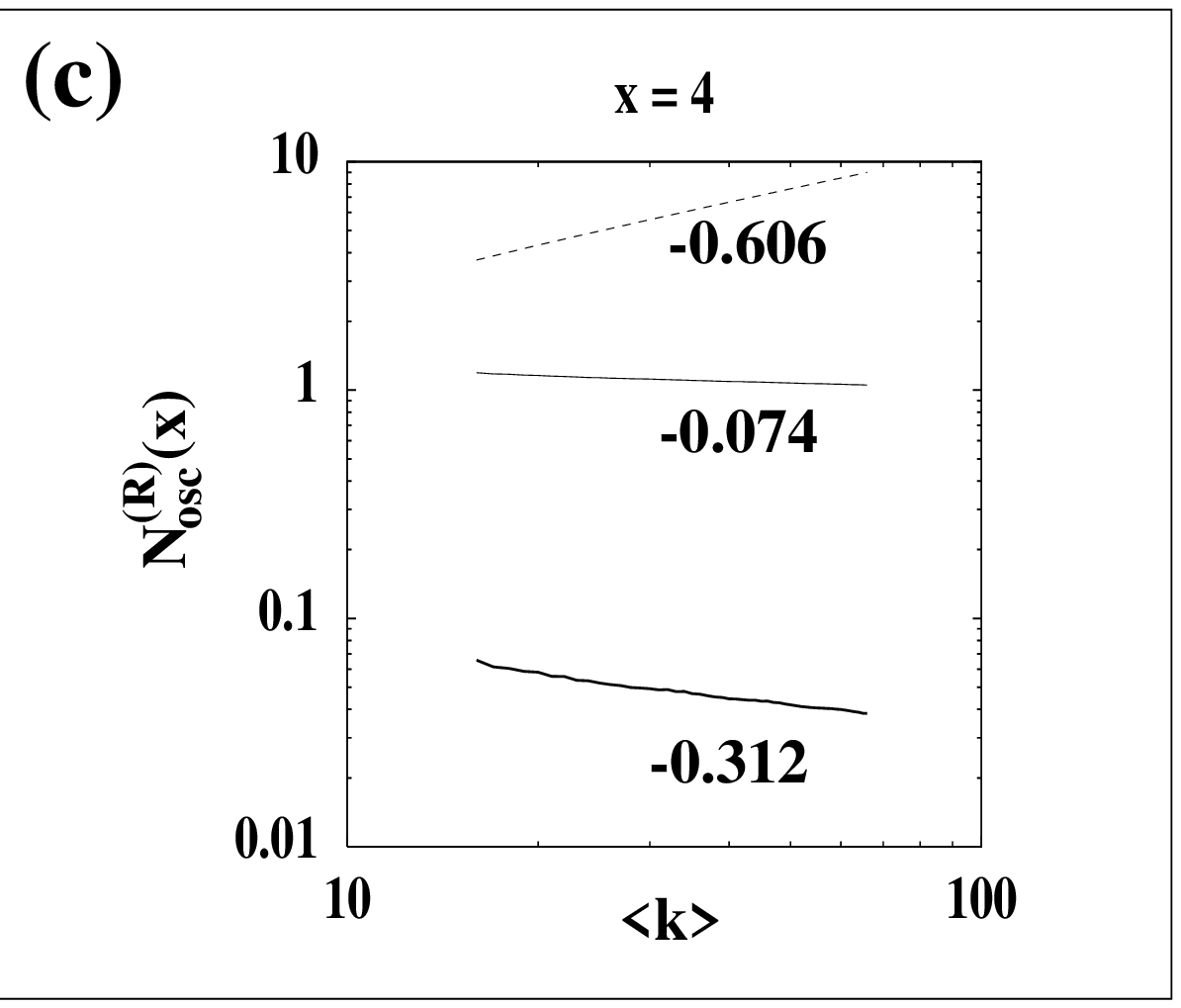,width=5cm,angle=-90}}
\vspace*{3mm}
\caption{\label{fig:kd} 
  The dependence of some peaks in the length spectrum of $N^{\rm
    (R)}(k)$ on the wave number $k$ in a double logarithmic plot.  The
  length spectrum was calculated in $k$-intervals of width ${30
    \pi/a}$ centered at the value which is given on the abscissa in
  units of $\pi/a$. As in figure \protect\ref{fig:tp-d05} the upper curve
  corresponds to the quantum result while the middle and the bottom
  curve show the deviations of the semiclassical approximation
  excluding and including diffraction, respectively.  (a) peak at the
  length of the unstable isolated orbit {\bf u}. (b) and (c) one and
  two traversals of a bouncing ball orbit. The $k$-dependence can be
  fitted by a power law with an exponent given next to the curves.}
\end{figure}

%% file: FIG/fig-tp-n05.tex
\begin{figure}[tbp]
\centerline{\hspace*{20mm}\psfig{figure=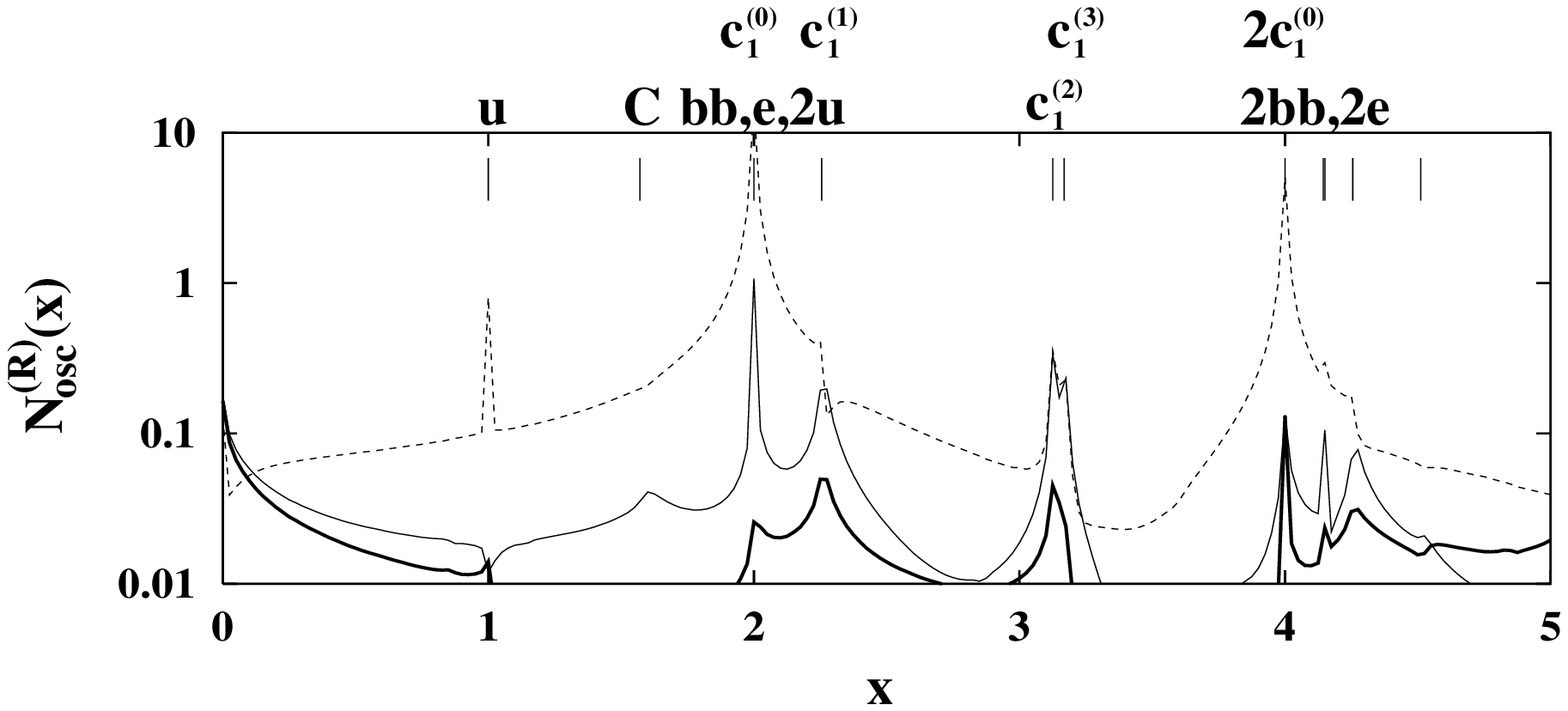,width=16cm,angle=-90}}
\caption{\label{fig:tp-n05} 
  The length spectrum of $N^{\rm (R)}(k)$ computed from the interval
  $1 < {ka/\pi} < 81$ for $a=1$, $R=0.5$ and Neumann boundary
  conditions on the circle. The line types are chosen as in figure 
  \protect\ref{fig:tp-d05}. }
\end{figure}

%% file: FIG/fig-tp-n08.tex
\begin{figure}[tbp]
\centerline{\hspace*{20mm}\psfig{figure=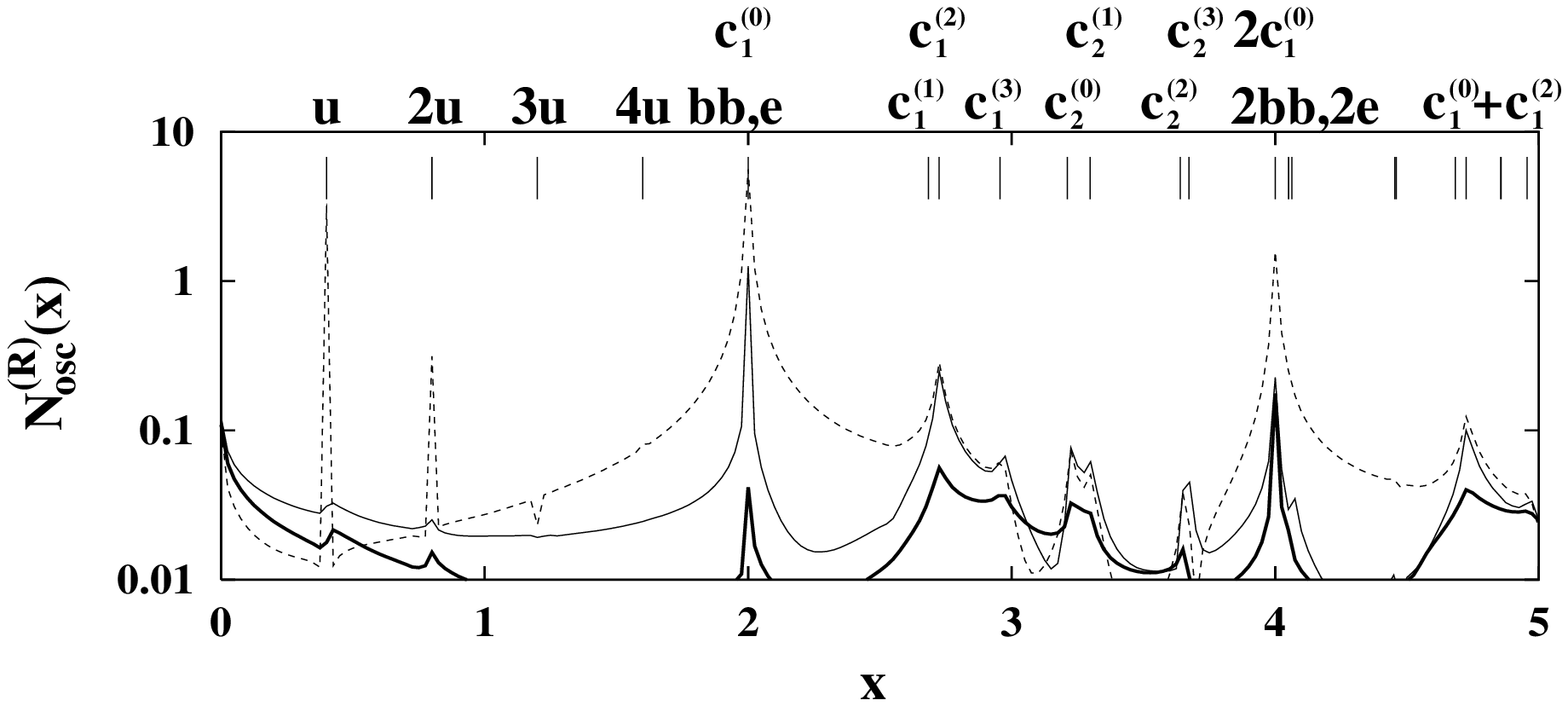,width=16cm,angle=-90}}
\caption{\label{fig:tp-n08}
  The length spectrum of $N^{\rm (R)}(k)$ computed from the interval
  $1 < {ka/\pi} < 81$ for $a=1$, $R=0.8$ and Neumann boundary
  conditions on the circle. The line types are chosen as in figure 
  \protect\ref{fig:tp-d05}. 
}
\end{figure}

%% file: FIG/fig-trs.tex
\begin{figure}[tbp]
\centerline{\psfig{figure=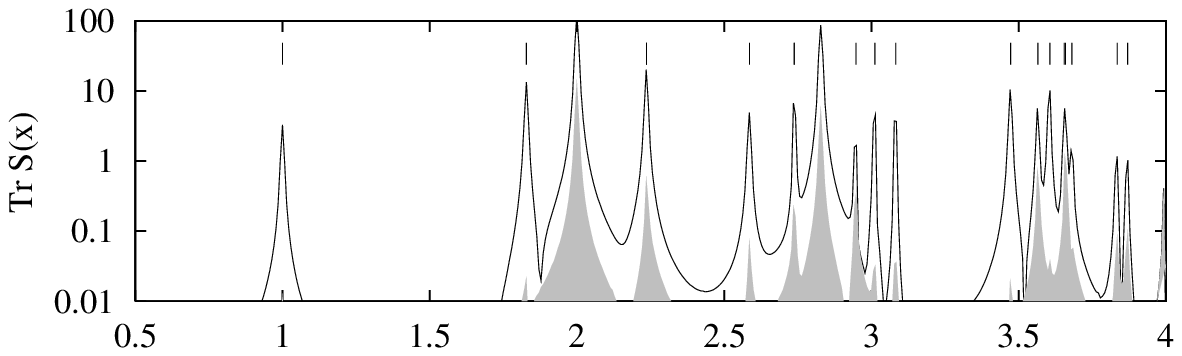,height=4cm}}
\centerline{\psfig{figure=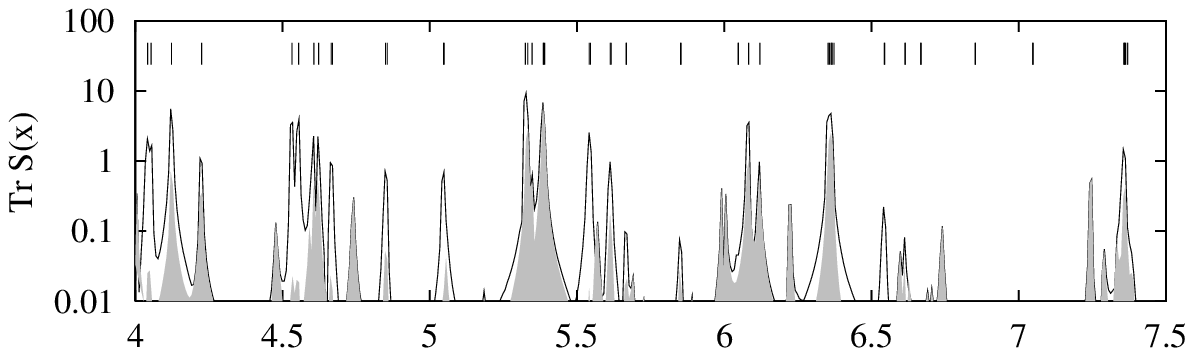,height=4cm}}
\centerline{\psfig{figure=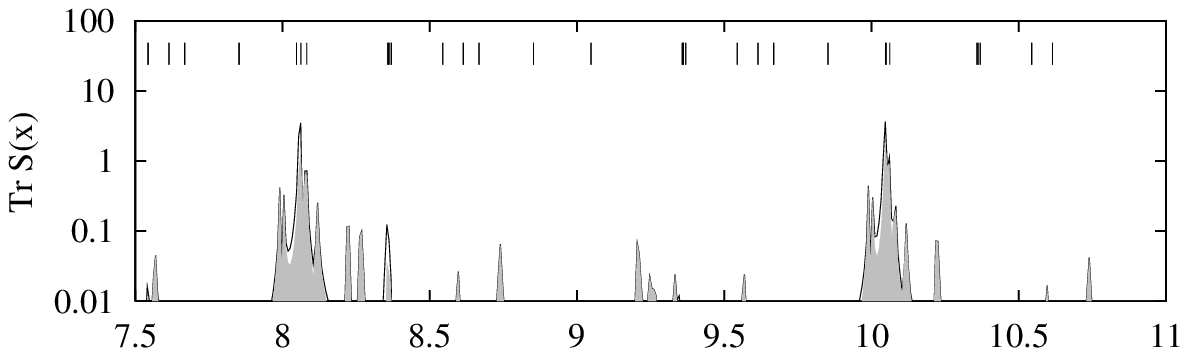,height=4cm}}
\centerline{\large \hspace*{10mm}x}
\caption{\label{trs}
The length spectrum of ${\rm Tr}\, S(k)$ for $a=1$ and $R=0.5$ in the
interval ${ka/\pi}=1,\dots,281$ with $\Delta_k=\pi/64$.  The full line
shows the quantum data and the shaded areas represent the error of the
leading order semiclassical approximation. The location of the
periodic orbits contributing to ${\rm Tr}\, S$ is marked with vertical
bars.}
\end{figure}

%% file: FIG/fig-trs5.tex
\begin{figure}[p]
\centerline{\psfig{figure=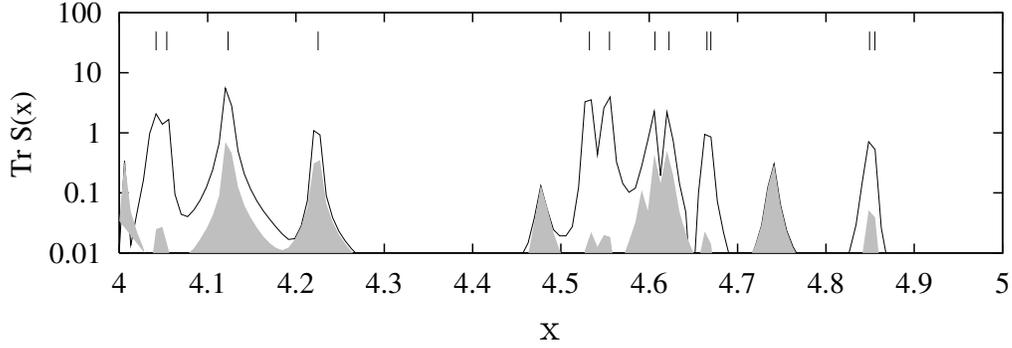,height=4cm}}
\centerline{\large \hspace*{10mm}x}
\caption{\label{trs5}
The length spectrum of ${\rm Tr}\, S(k)$ for $a=1$ and $R=0.5$ in the interval
${ka/\pi}=1,\dots,281$.  The full line shows the quantum data and the
shaded areas represent the error of the standard semiclassical
approximation. The location of the unstable periodic orbits contributing 
to ${\rm Tr}\, S$ is marked with the dotted vertical lines.
}
\end{figure}

%% file: FIG/fig-orb-tr.tex
\begin{figure}[p]
\begin{center}
{
\begin{tabular}{cccc}

\begin{tabular}{c}
\psfig{figure=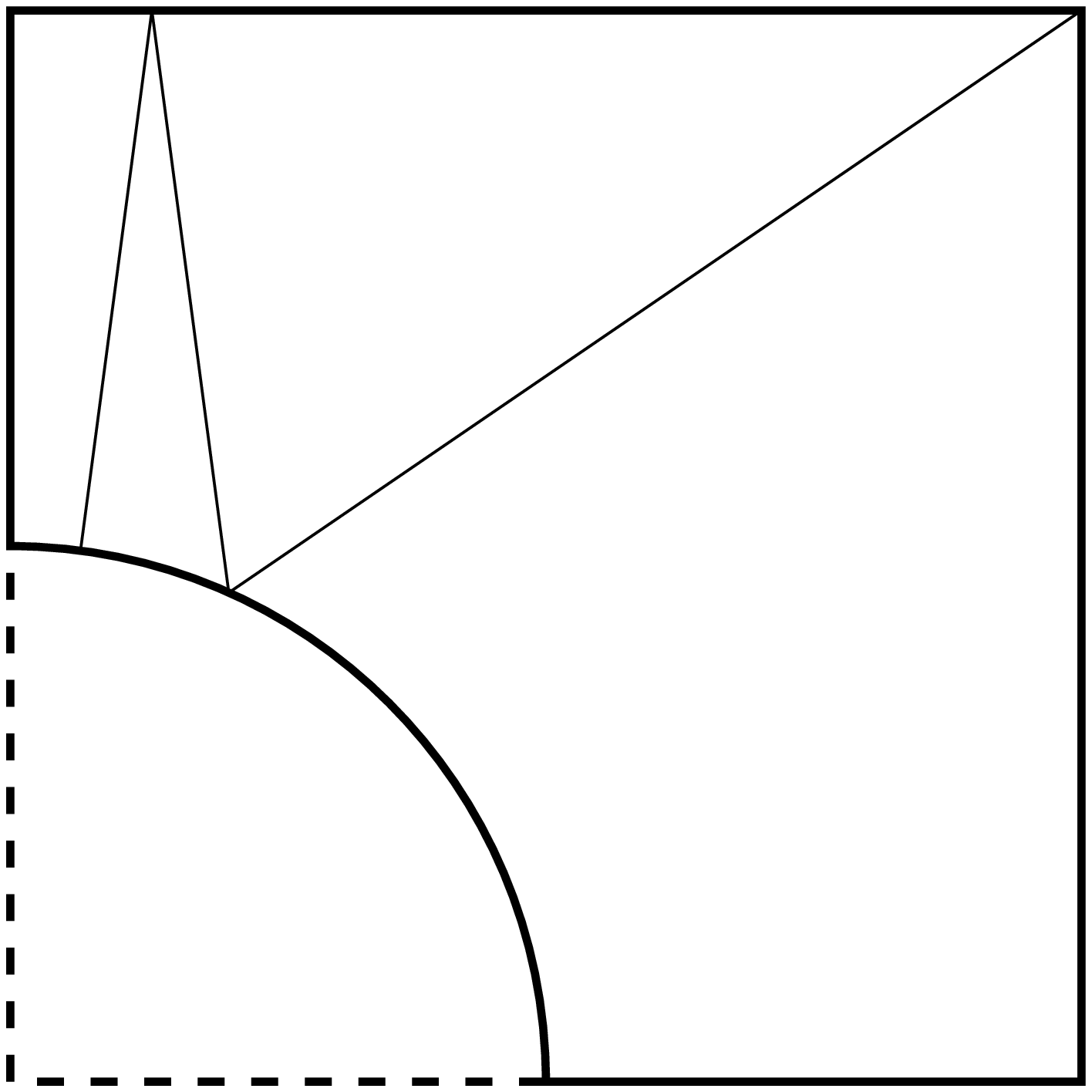,height=25mm} \\
4.04157\\-0.45E+03
\end{tabular}
&

\begin{tabular}{c}
\psfig{figure=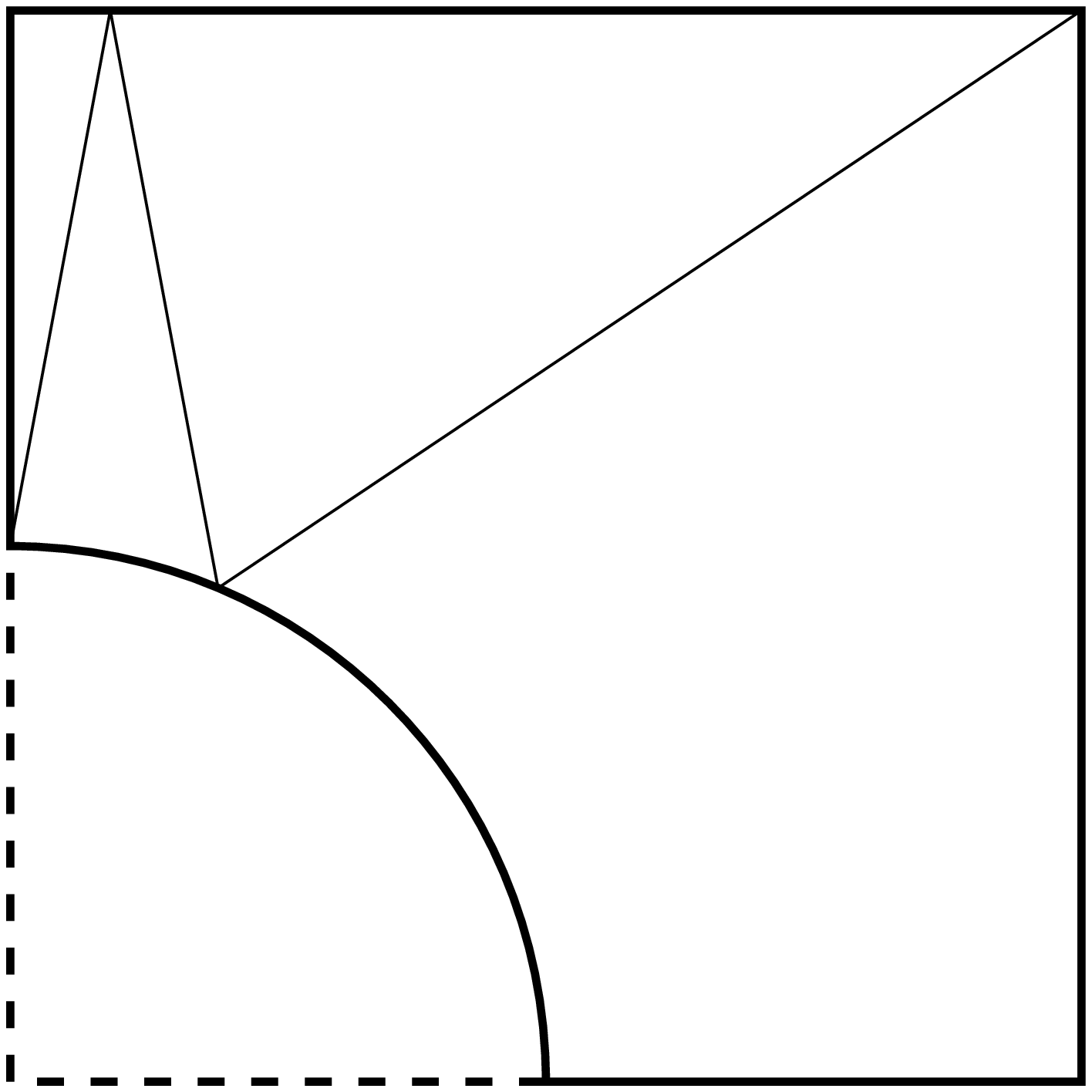,height=25mm} \\
4.05367\\0.48E+03
\end{tabular}
&

\begin{tabular}{c}
\psfig{figure=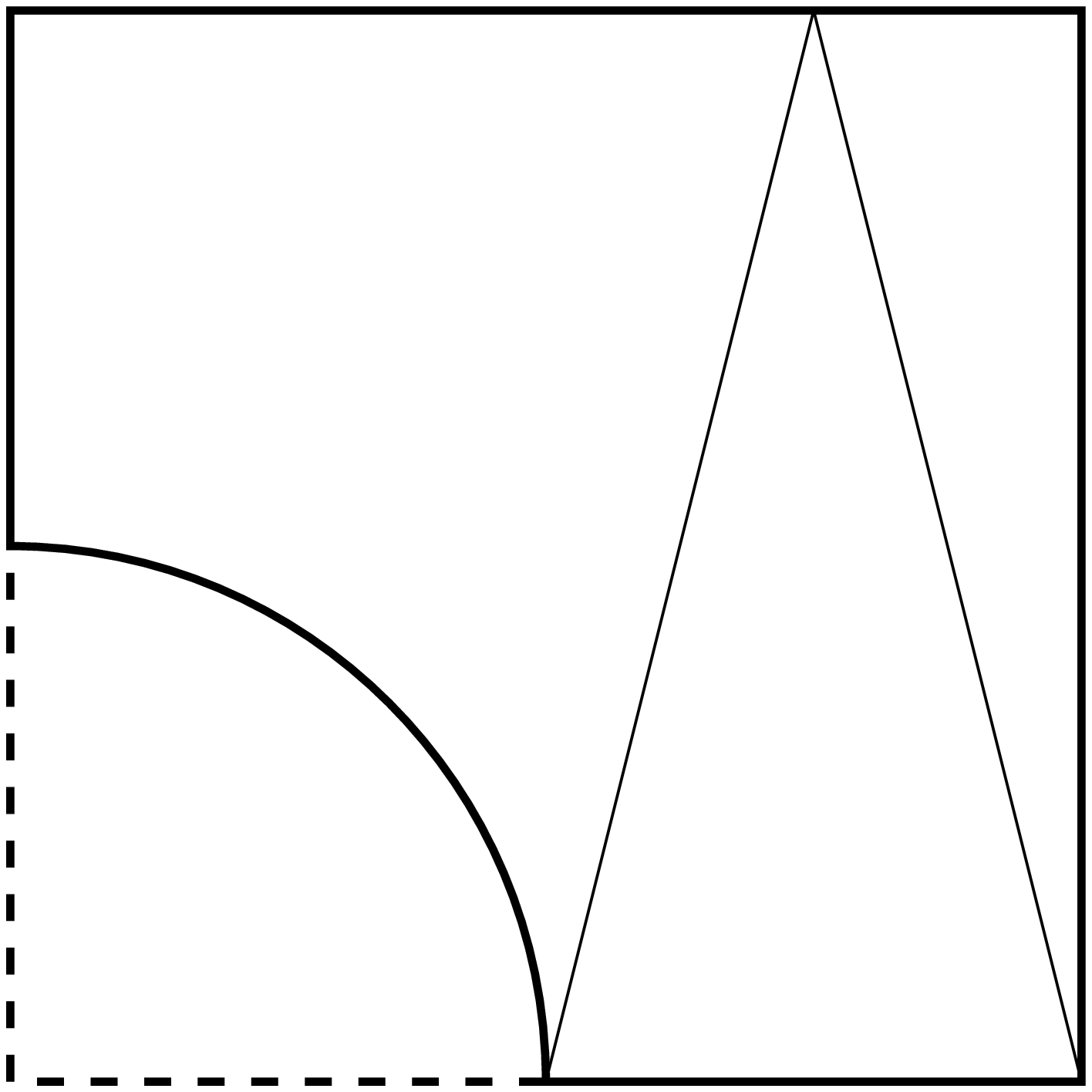,height=25mm} \\
4.12311\\0.70E+02
\end{tabular}
&

\begin{tabular}{c}
\psfig{figure=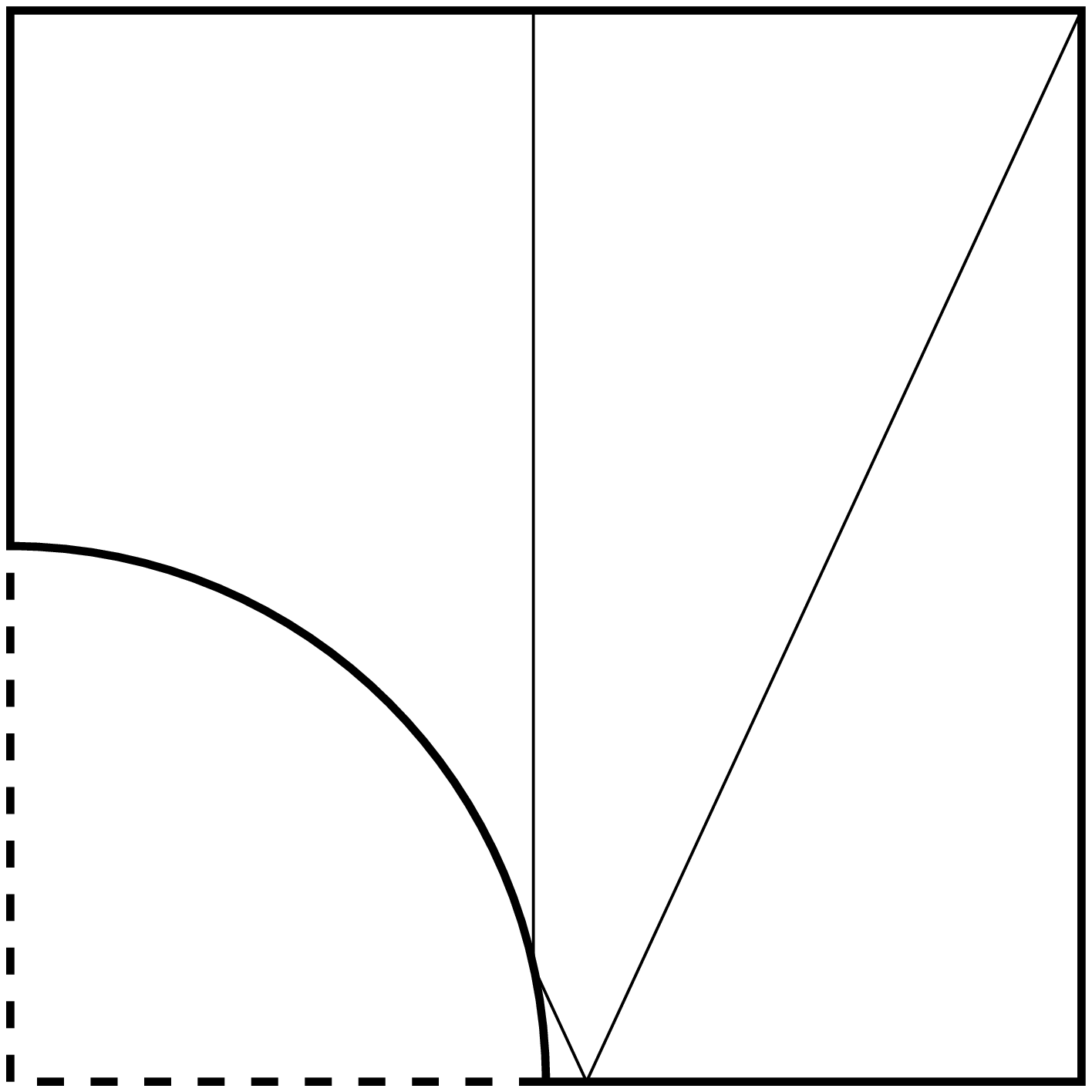,height=25mm} \\
4.22499\\0.17E+04
\end{tabular}
\\

\begin{tabular}{c}
\psfig{figure=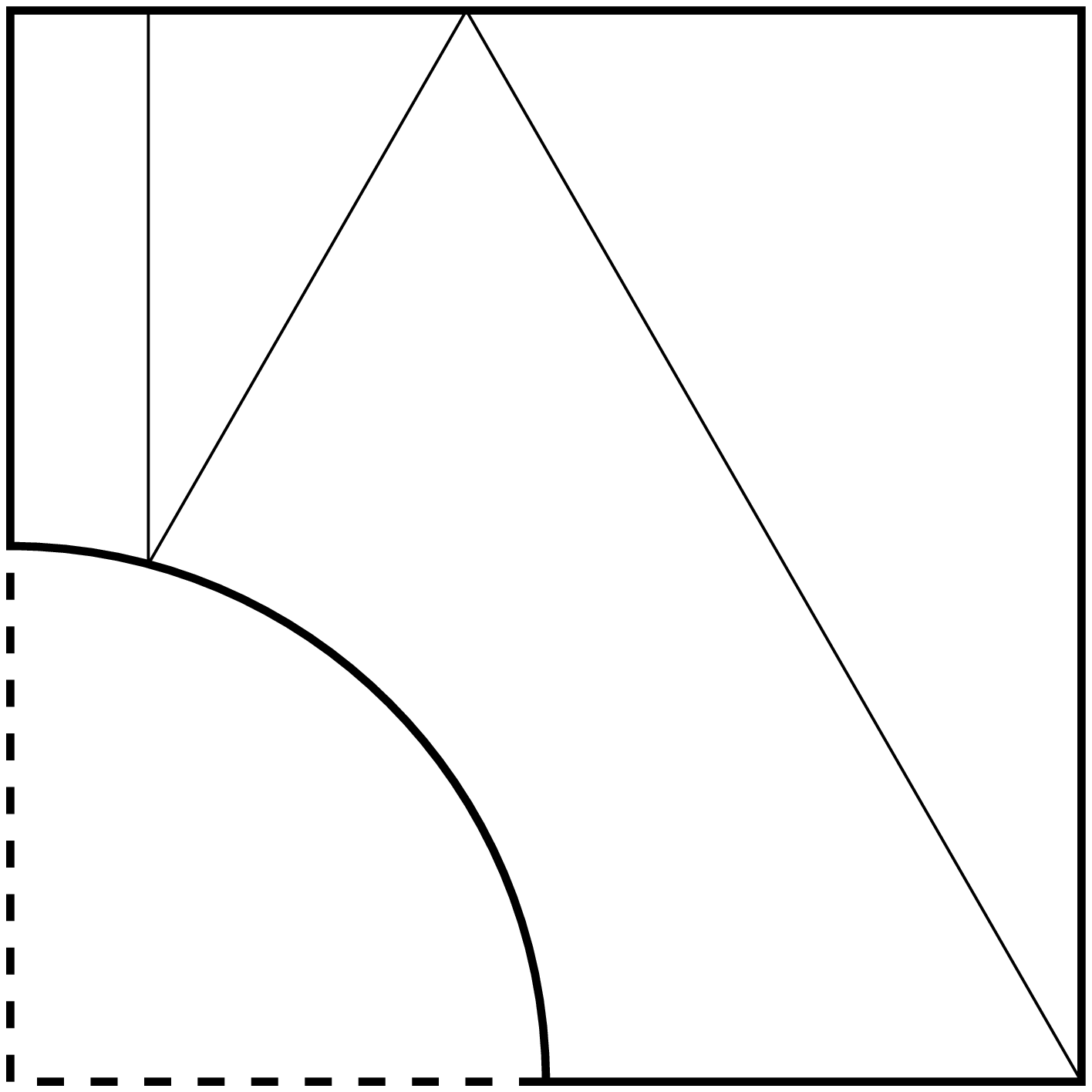,height=25mm} \\
4.53226\\0.10E+03
\end{tabular}
&

\begin{tabular}{c}
\psfig{figure=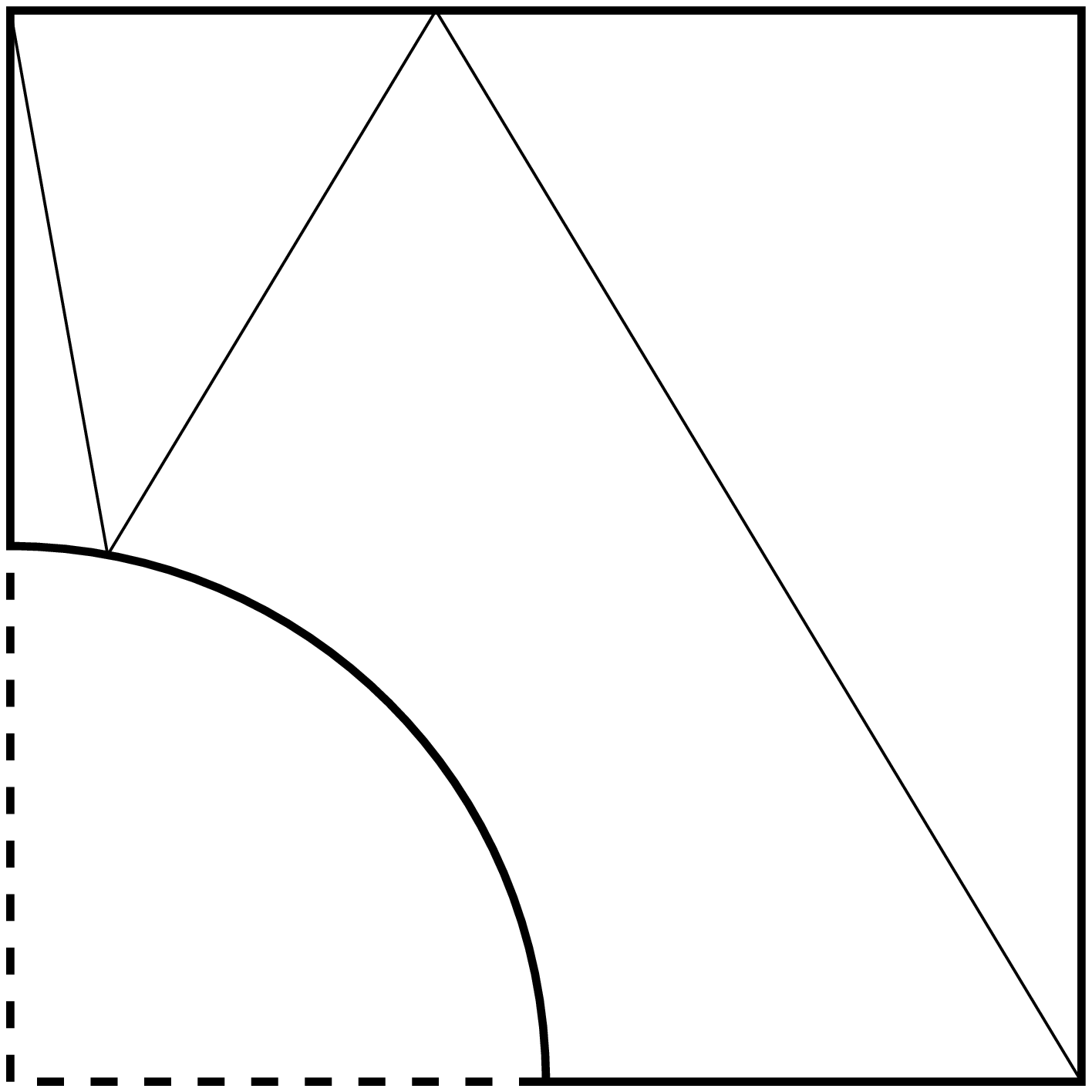,height=25mm} \\
4.55502\\0.11E+03
\end{tabular}
&

\begin{tabular}{c}
\psfig{figure=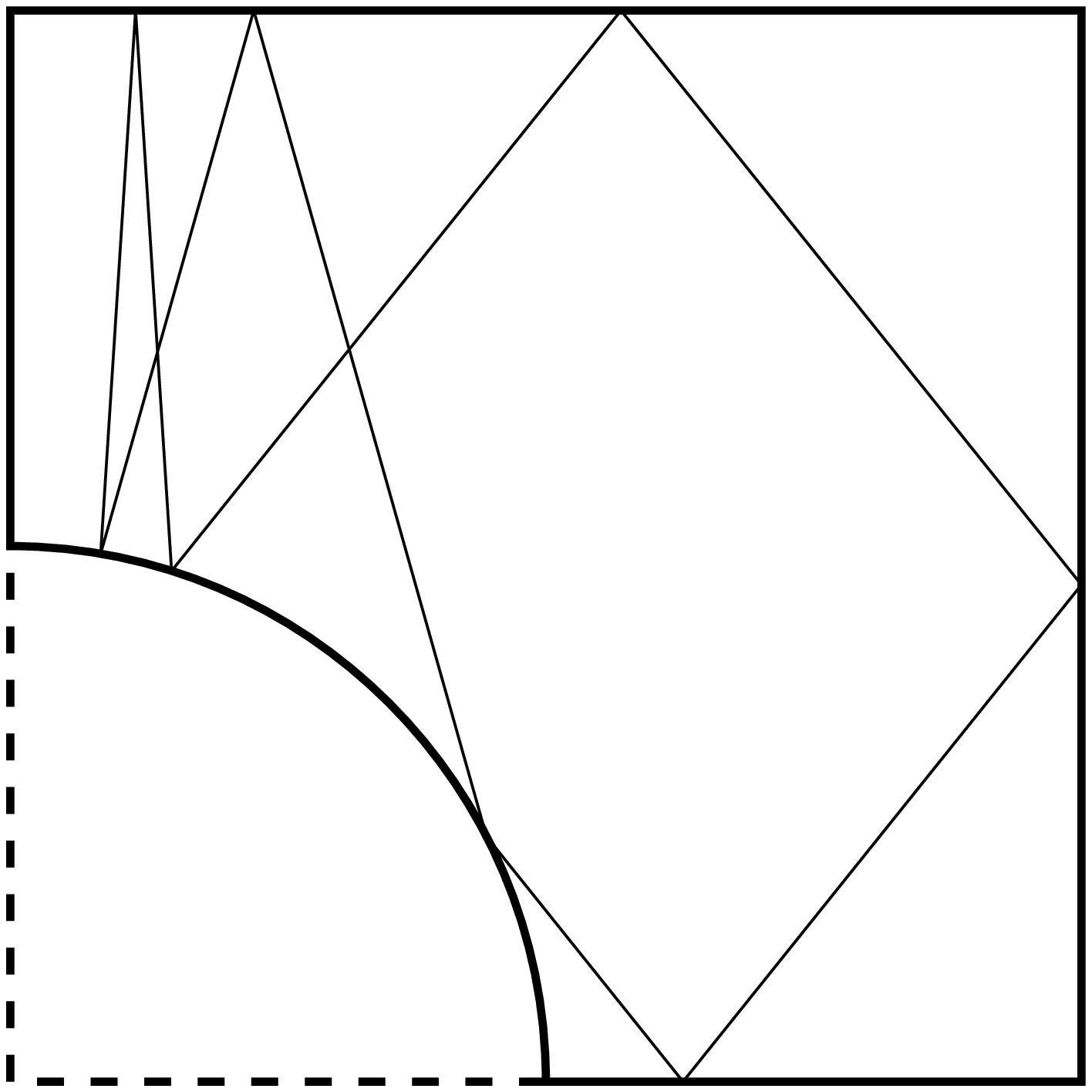,height=25mm} \\
4.60635\\0.21E+04
\end{tabular}
&

\begin{tabular}{c}
\psfig{figure=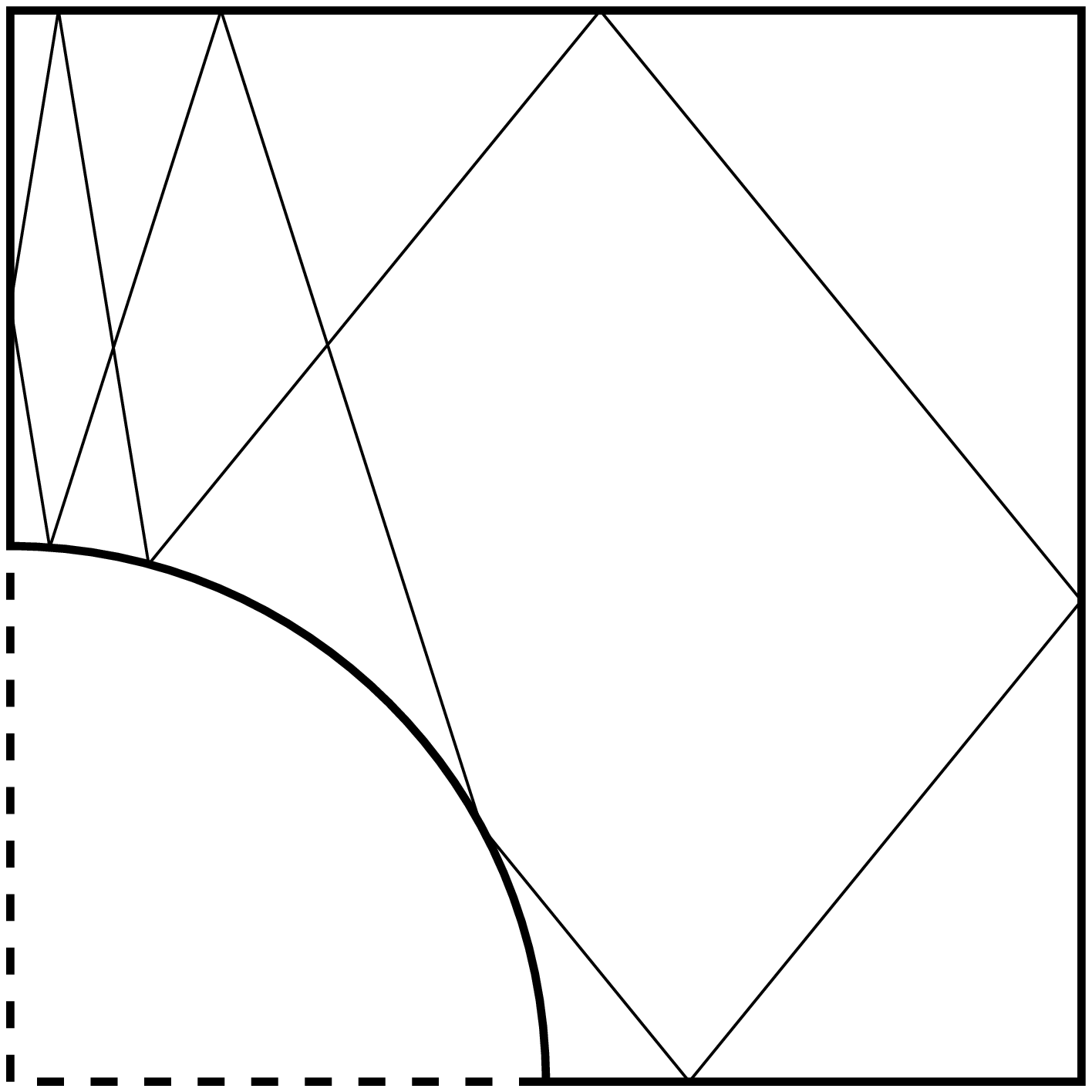,height=25mm} \\
4.62233\\0.23E+04
\end{tabular}
\\

\begin{tabular}{c}
\psfig{figure=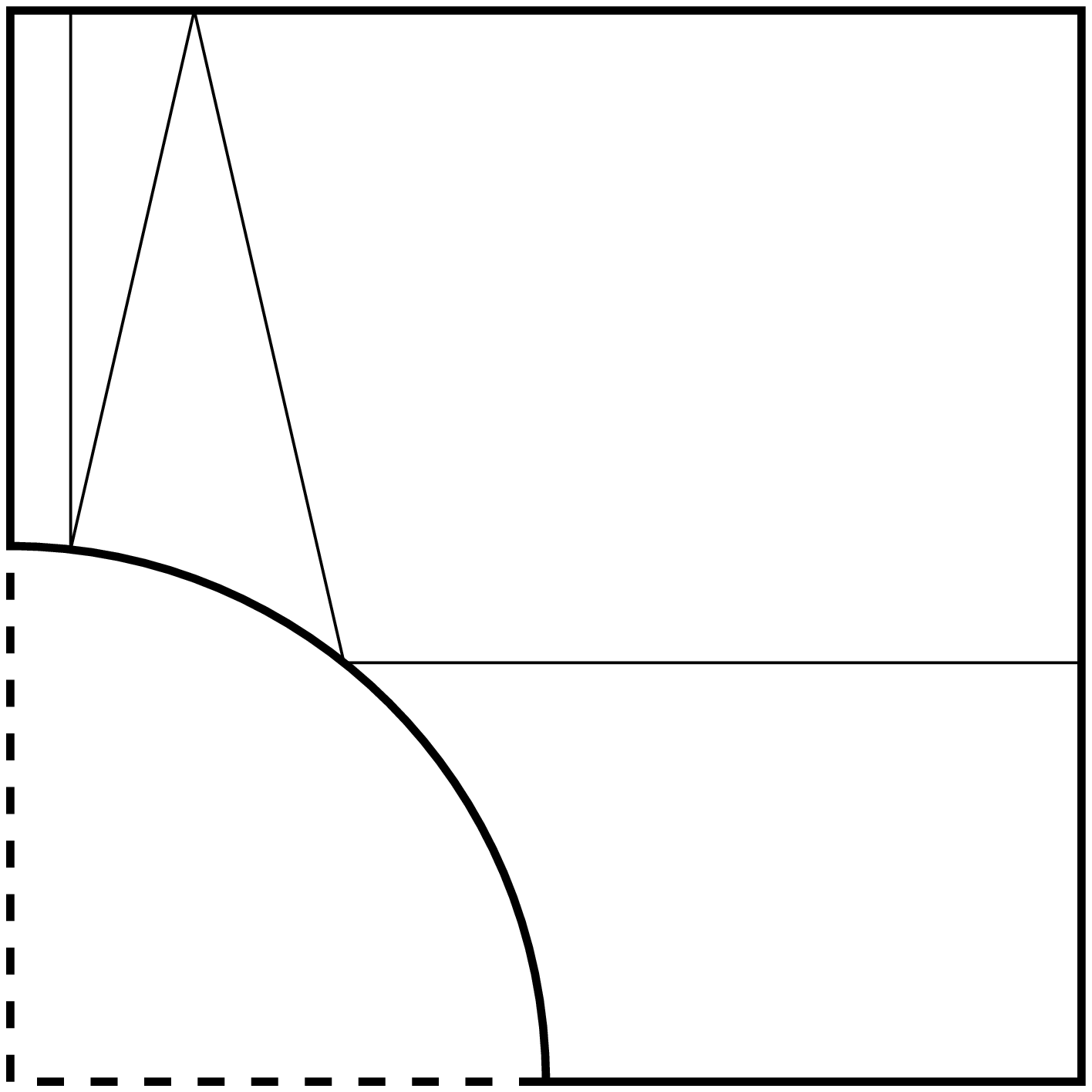,height=25mm} \\
4.66531\\0.38E+04
\end{tabular}
&

\begin{tabular}{c}
\psfig{figure=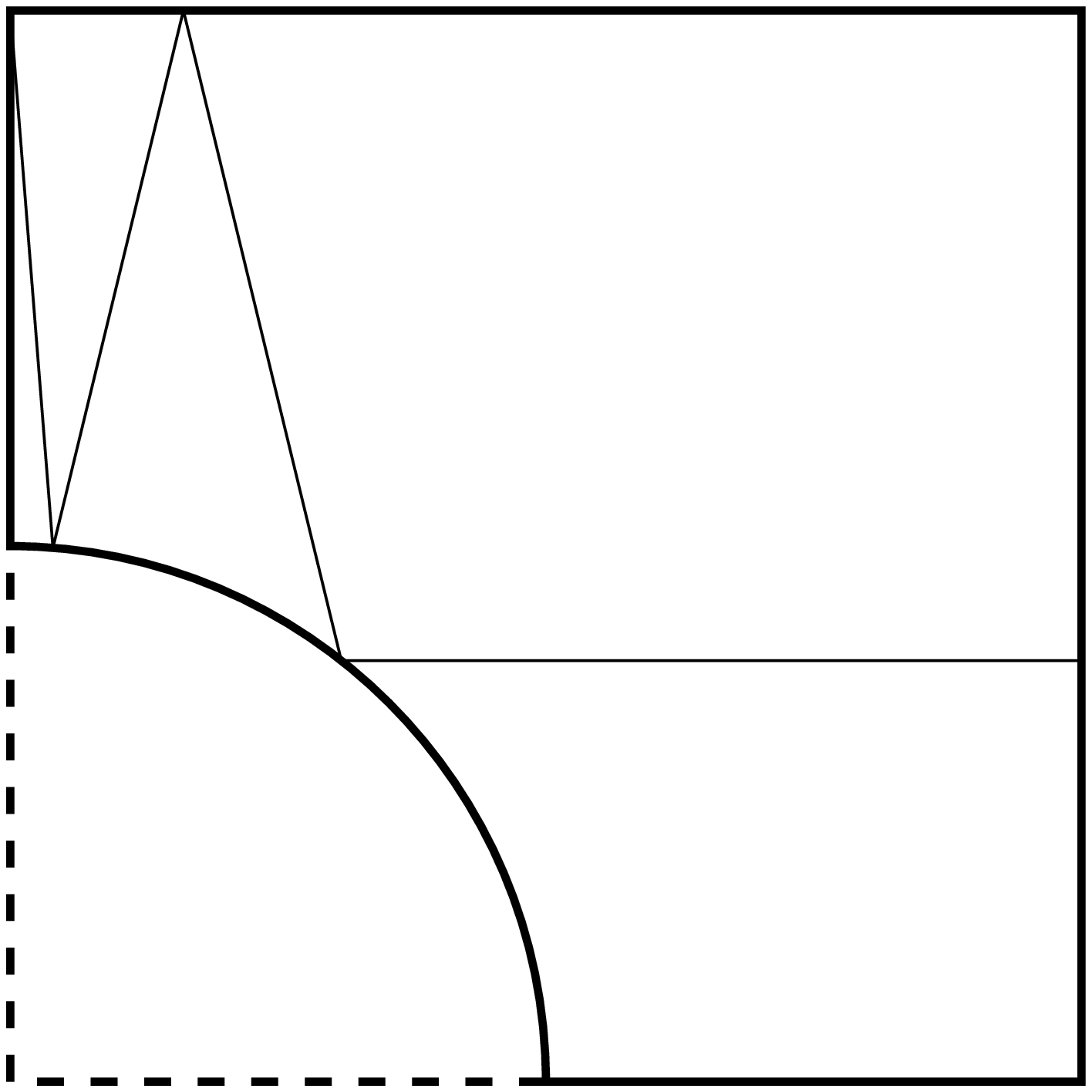,height=25mm} \\
4.66976\\0.39E+04
\end{tabular}
&

\begin{tabular}{c}
\psfig{figure=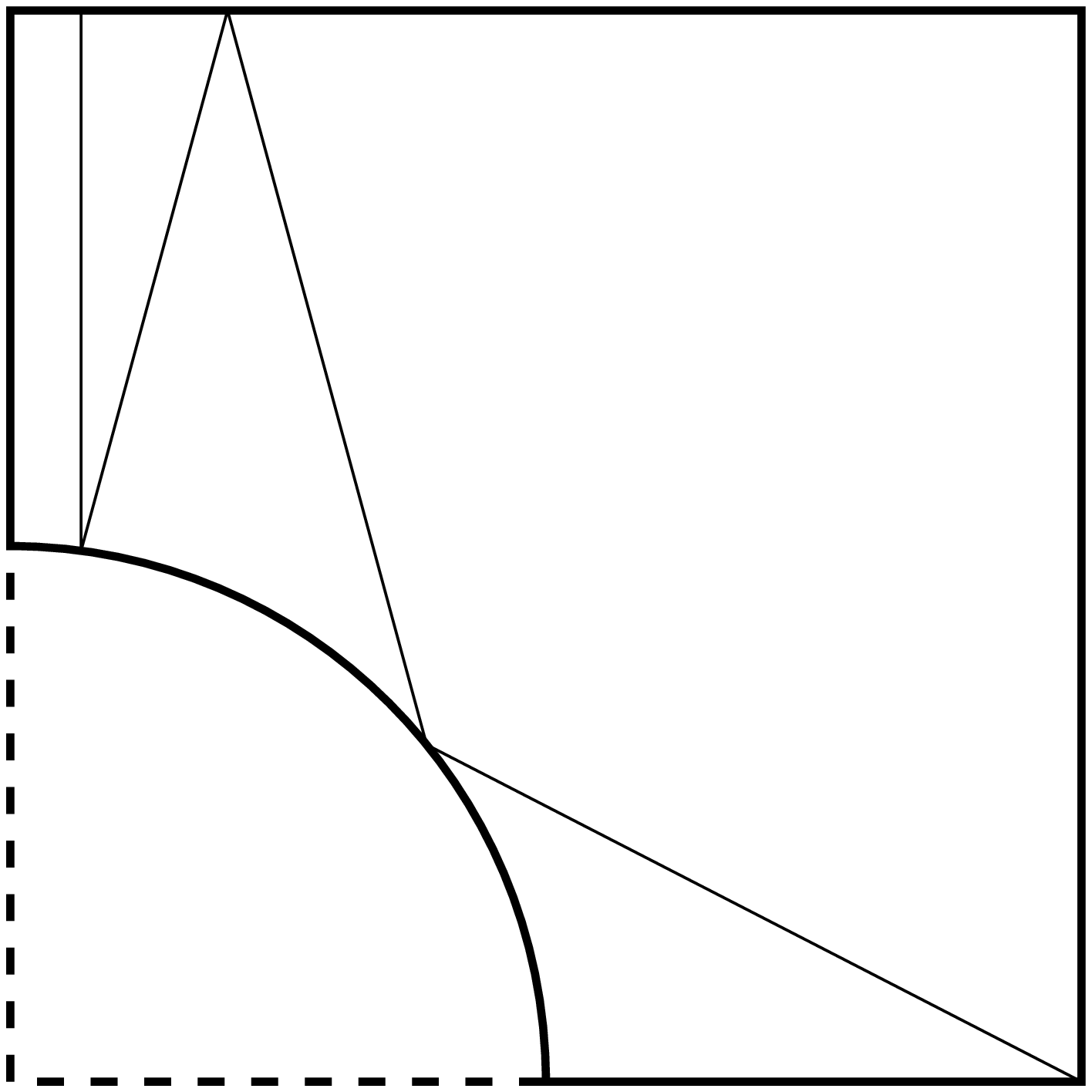,height=25mm} \\
4.84938\\0.89E+04
\end{tabular}
&

\begin{tabular}{c}
\psfig{figure=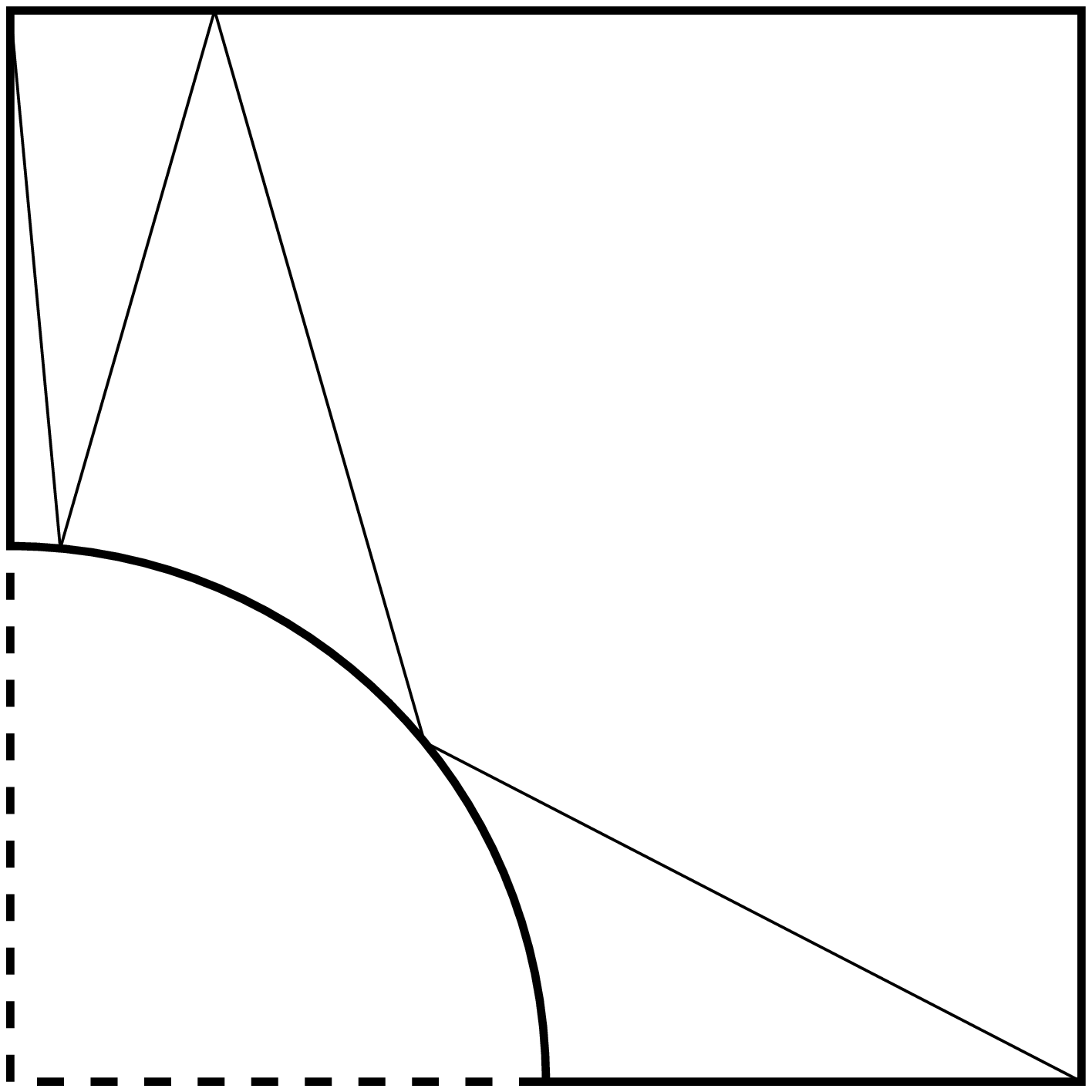,height=25mm} \\
4.85549\\0.94E+04
\end{tabular}

\end{tabular}
}
\end{center}
\caption{\label{orb-tr}
  The unstable periodic orbits contributing to ${\rm Tr}\, S(k)$ with
  lengths in the interval $4\le L \le 5$ corresponding to figure
  \protect\ref{trs5}. Below the figures the length and the
  stability prefactor of the individual orbits are given.  }
\end{figure}

%% file: FIG/fig-lorb-trs.tex
\begin{figure}[tbp]

\begin{center}
\begin{tabular}{cccc}
\begin{tabular}{c}
\psfig{figure=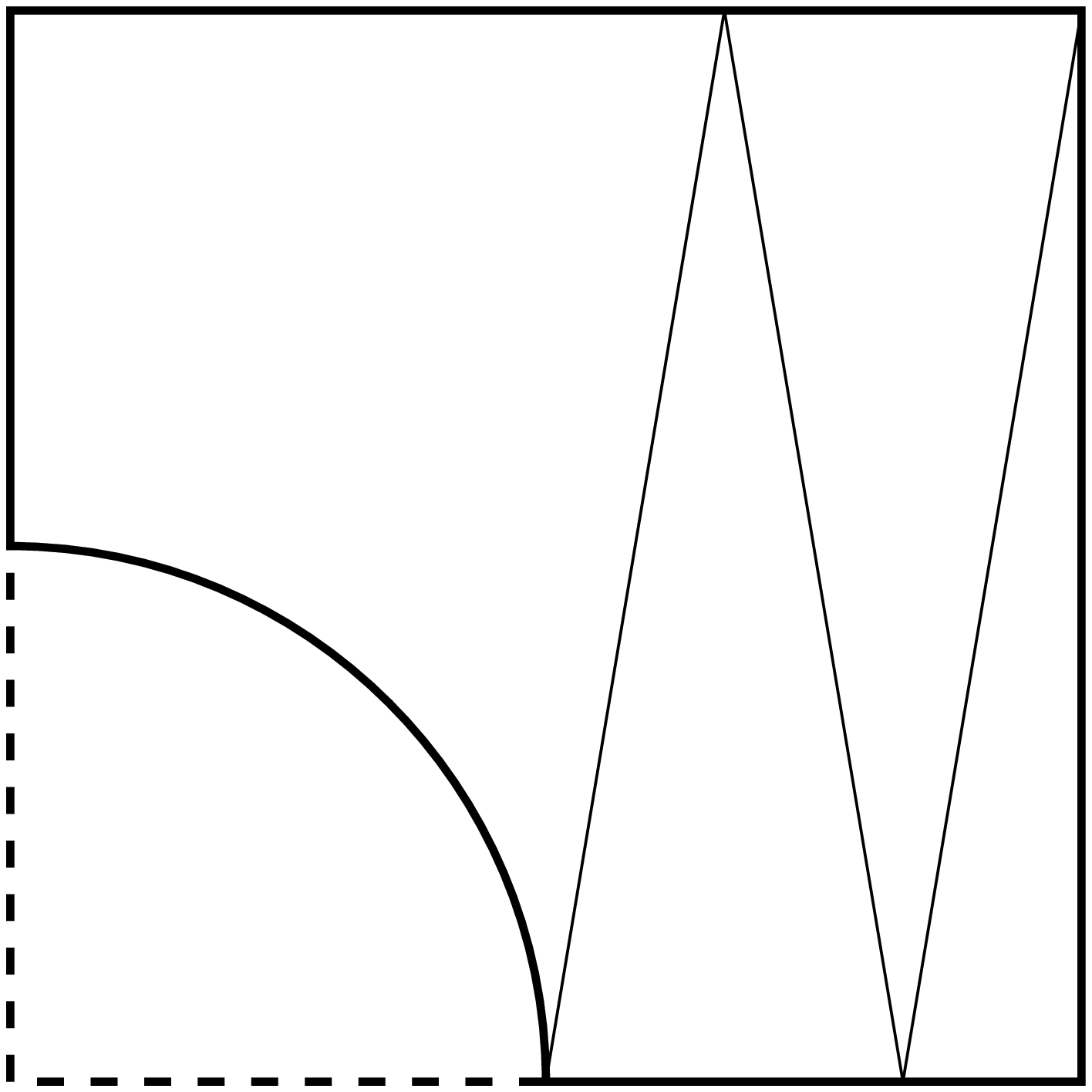,height=25mm} \\
 6.08276 \\ 0.15E+03
\end{tabular}
&

\begin{tabular}{c}
\psfig{figure=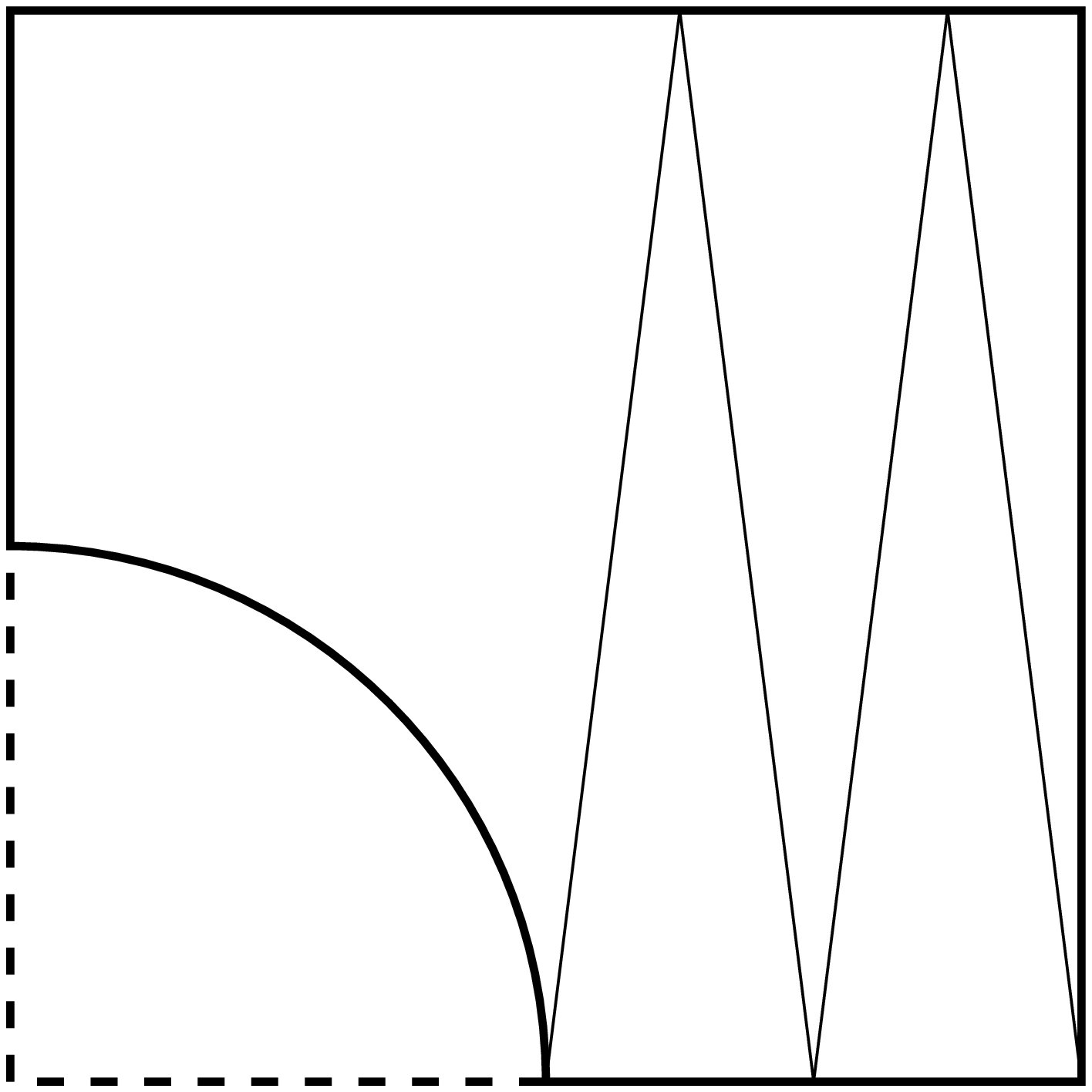,height=25mm} \\
 8.06226 \\ 0.26E+03
\end{tabular}
&

\begin{tabular}{c}
\psfig{figure=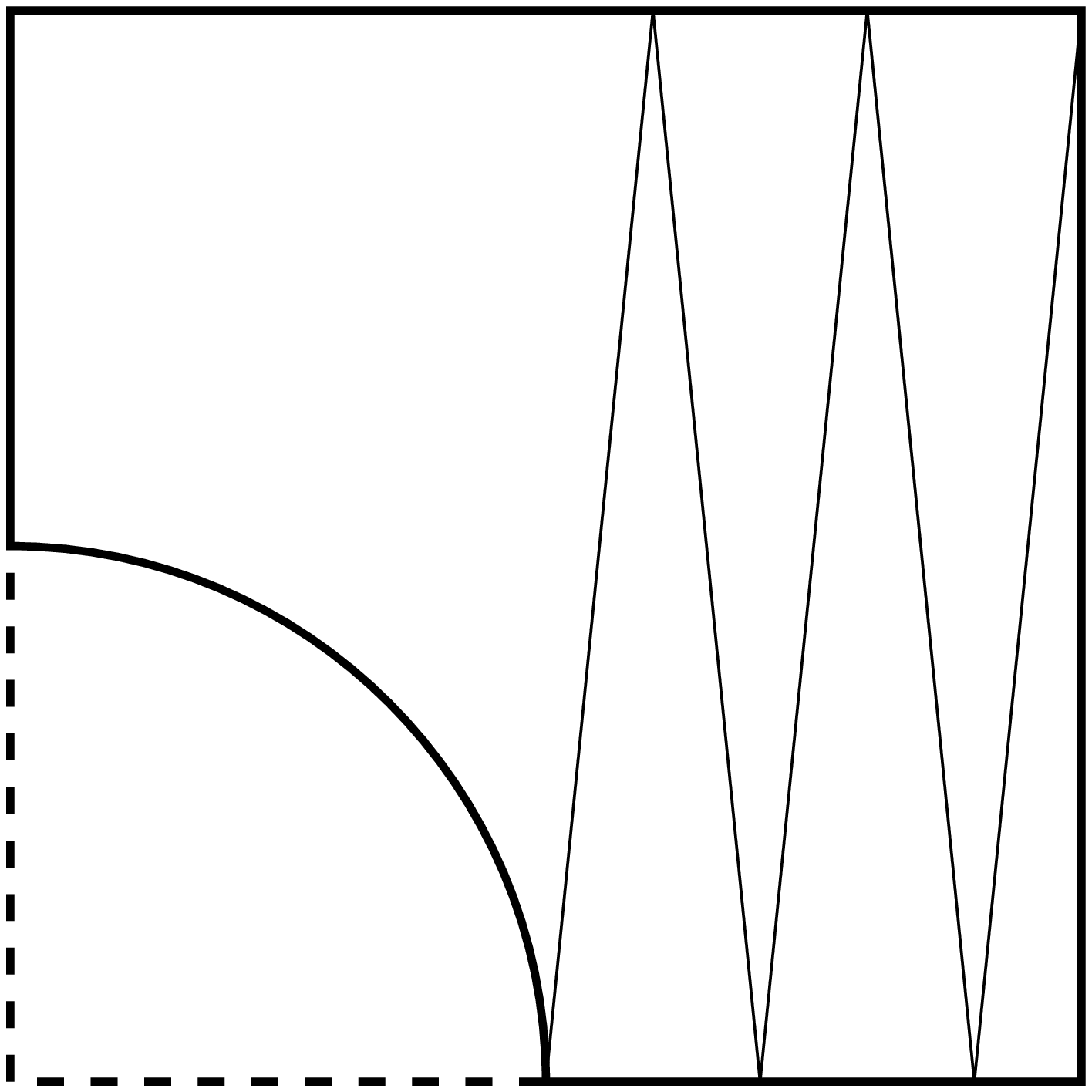,height=25mm} \\
10.04988 \\ 0.41E+03
\end{tabular}
&

\begin{tabular}{c}
\psfig{figure=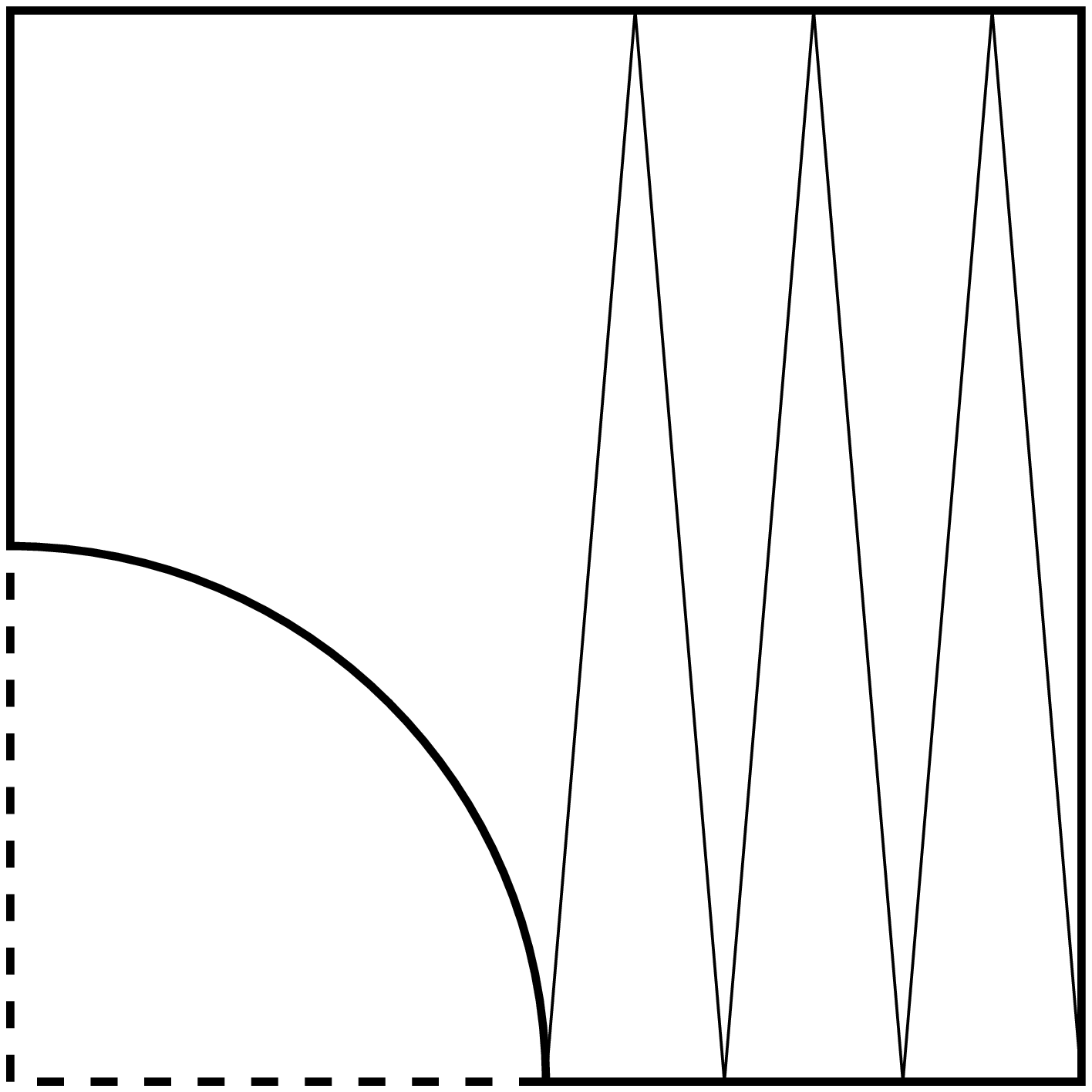,height=25mm} \\
12.04159 \\ 0.58E+03
\end{tabular}
\\[5mm]

\begin{tabular}{c}
\psfig{figure=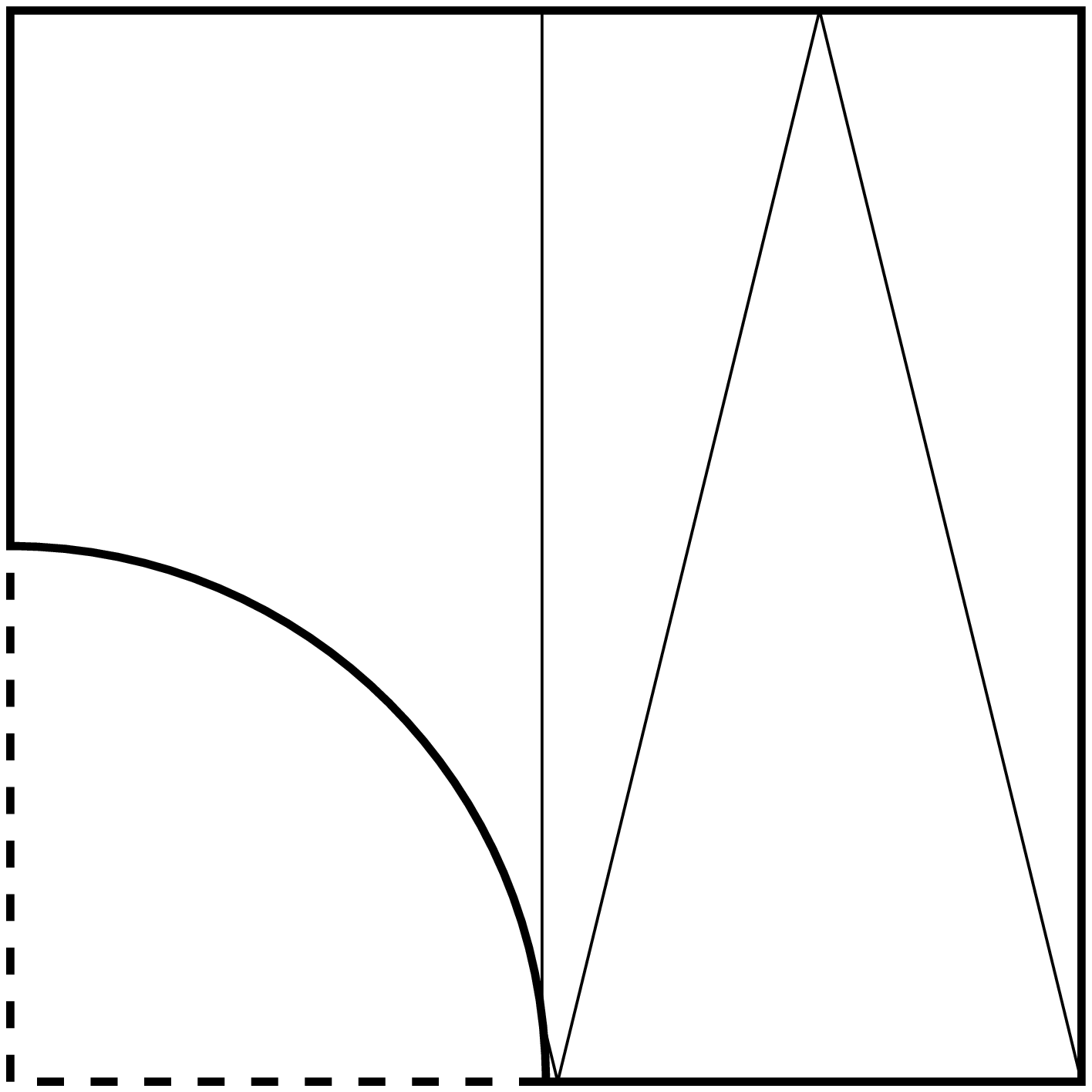,height=25mm} \\
 6.12133 \\-0.93E+04
\end{tabular}
&

\begin{tabular}{c}
\psfig{figure=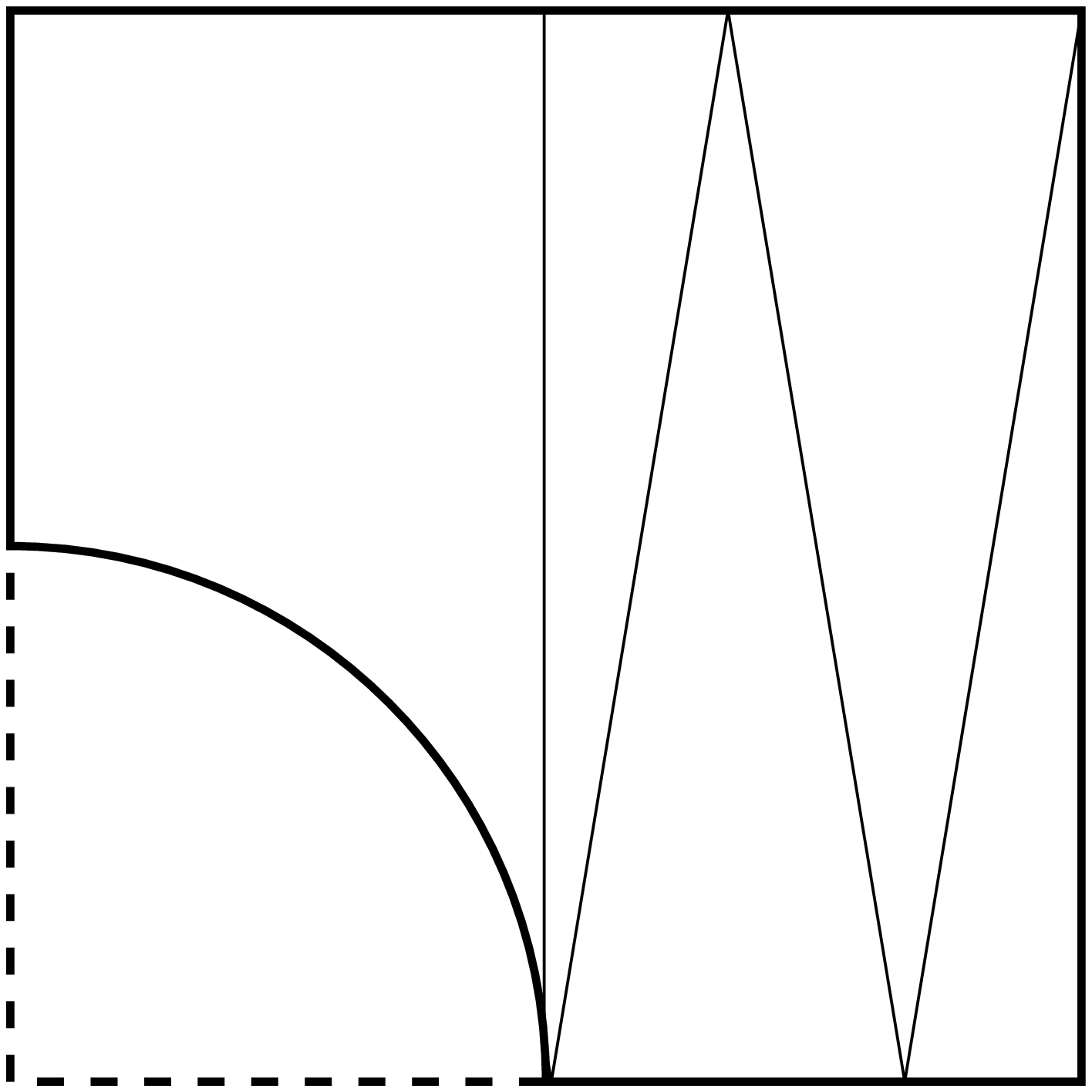,height=25mm} \\
 8.08221 \\-0.29E+05
\end{tabular}
&

\begin{tabular}{c}
\psfig{figure=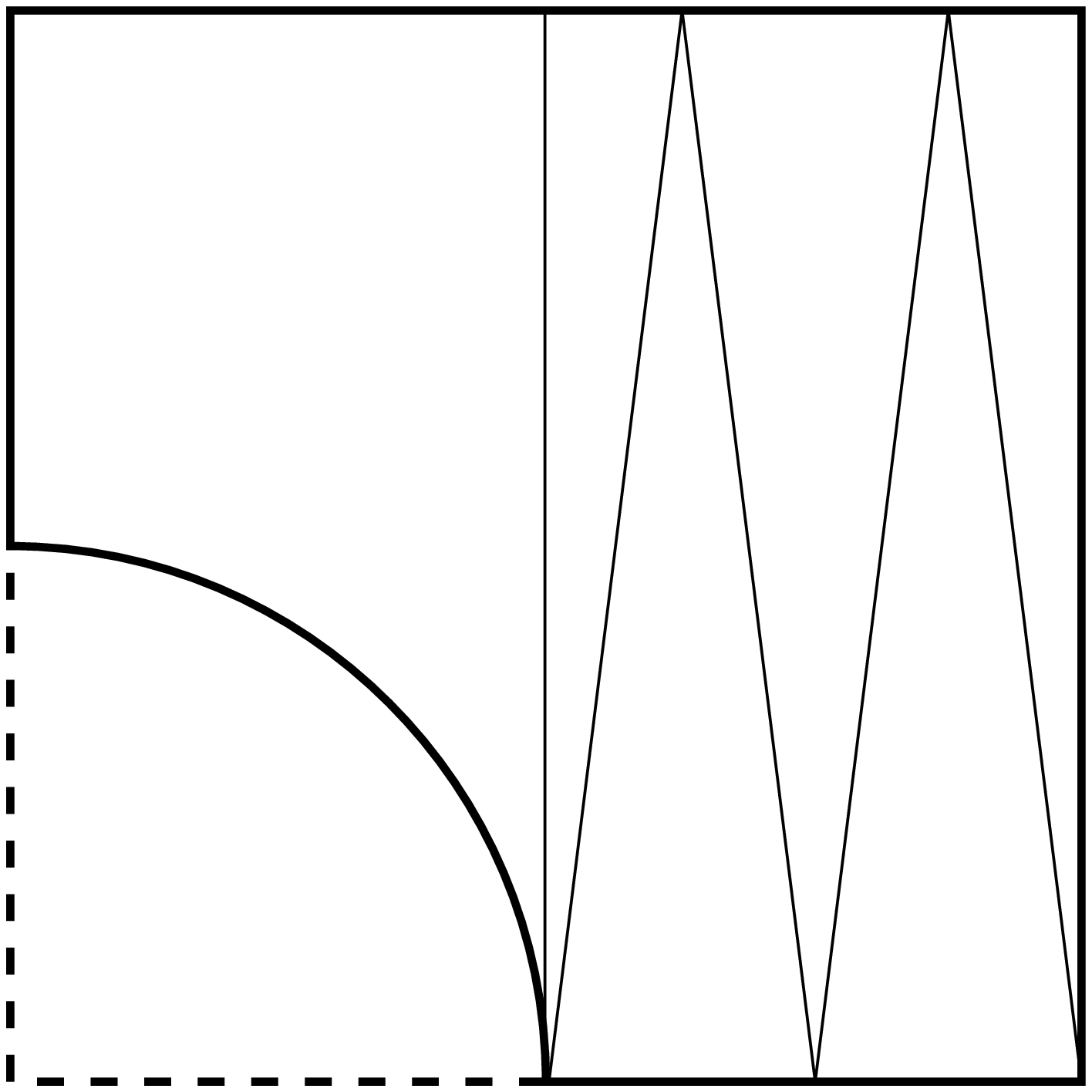,height=25mm} \\
10.06202 \\ -0.67E+05
\end{tabular}
&

\begin{tabular}{c}
\psfig{figure=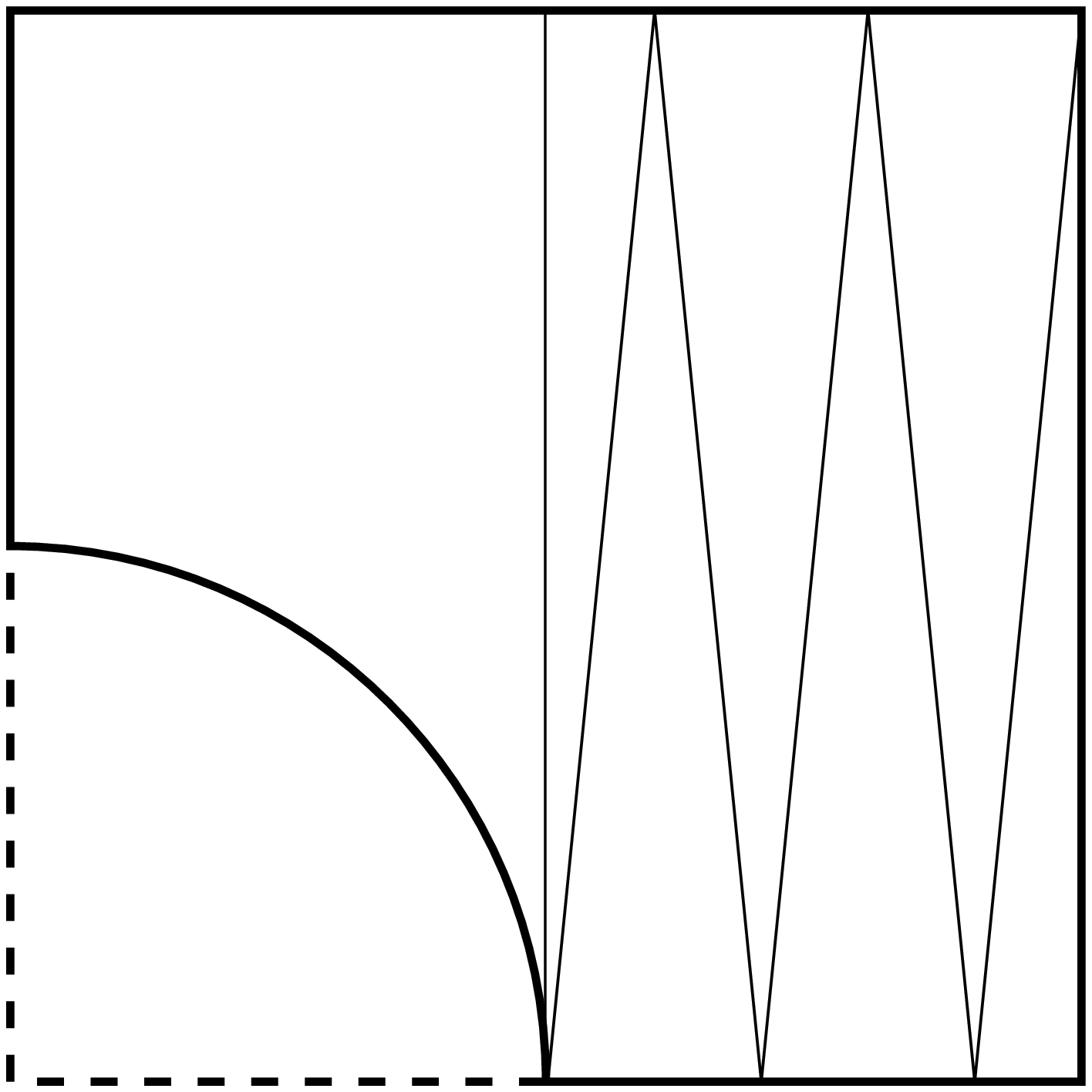,height=25mm} \\
12.04975 \\-0.13E+06
\end{tabular}
\end{tabular}
\end{center}
\caption{\label{lorb-trs} Some of the orbits which give rise to the peaks at large x
  in the length spectrum of ${\rm Tr}\, S$. Below the orbits the length and
  the dominating eigenvalue of the monodromy matrix are given. Due to
  diffraction the standard Gutzwiller formula fails to predict the
  contributions of these orbits correctly.}
\end{figure}

%% file: FIG/fig-pdelta.tex
\begin{figure}[tbp]
\begin{center} 
\leavevmode 
\psfig{figure=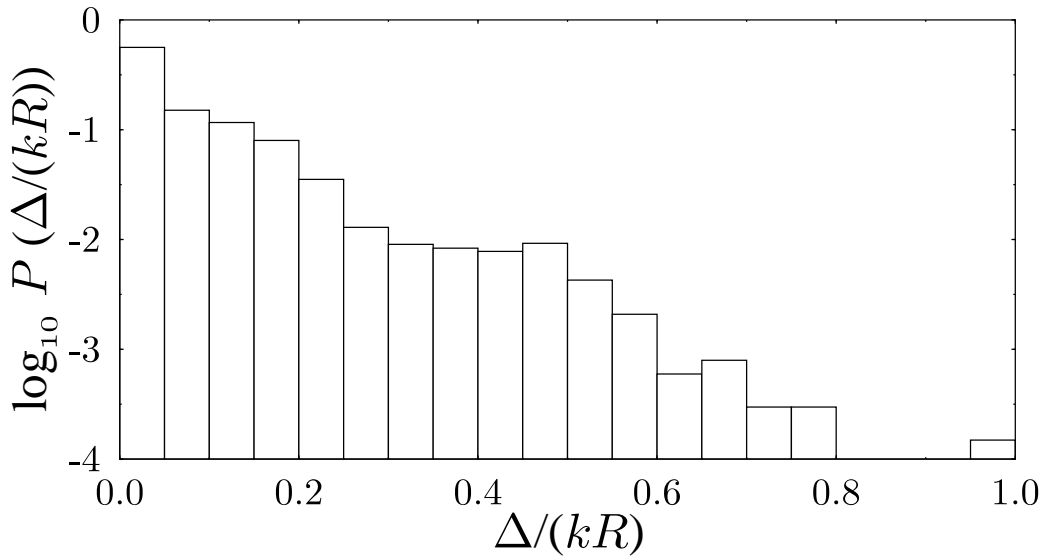,width=15cm}
\end{center}
\caption{The coarse--grained distribution of $\Delta/(kR)$. Note the
logarithmic scale.
\label{fig:pdelta}}
\end{figure}